# Final Report for SAG 22:
# A Target Star Archive for Exoplanet Science

**Natalie R. Hinkel,[1] Joshua Pepper,[2] Christopher C. Stark,[3] Jennifer A. Burt,[4] David R. Ciardi,[5] Kevin K. Hardegree-Ullman,[6] Jacob Lustig-Yaeger,[7] Ravi Kopparapu,[3] Lokesh Mishra,[8, 9] Karan Molaverdikhani,[10, 11, 12] Ilaria Pascucci,[13] Tyler Richey-Yowell,[14] E. J. Safron,[15] David J. Wilson,[16] Galen Bergsten,[13] Tabetha S. Boyajian,[17] J. A. Caballero,[18] K. Cunha,[6] Alyssa Columbus,[19] Shawn D. Domagal-Goldman,[3] Chuanfei Dong,[20] R. M. Elowitz,[21] Devanshu Jha,[22] Archit Kalra,[23] David W. Latham,[24] Jacob Luhn,[25] Carl Melis,[26] Navya Nagananda,[27] Eliad Peretz,[3] Sabine Reffert,[28] Kimberly Scarangella Smith,[29] Keivan G. Stassun,[30] Angelle Tanner,[31] Noah Tuchow,[32] Dimitri Veras,[33, 34, 35] and Jennifer G. Winters[24]**

*[1]Southwest Research Institute, 6220 Culebra Road, San Antonio, TX 78238, USA*

*[2]Department of Physics, Lehigh University, 16 Memorial Drive East, Bethlehem, PA 18015, USA*

*[3]NASA Goddard Space Flight Center, 8800 Greenbelt Road, Greenbelt, MD 20771, USA*

*[4]Jet Propulsion Laboratory, California Institute of Technology, Pasadena, CA 91109, USA*

*[5] Caltech/IPAC-NASA Exoplanet Science Institute Pasadena, CA, USA [6]The University of Arizona, Tucson, AZ 85721, USA [7] Johns Hopkins University Applied Physics Laboratory, Laurel, MD 20723, USA*

*[8]Geneva Observatory, University of Geneva, Chemin Pegasi 51b, 1290 Versoix, Switzerland*

*[9] Institute of Physics, University of Bern, Gesellschaftsstrasse 6, 3012 Bern, Switzerland [1 0]LSW: Landessternwarte, Zentrum für Astronomie der Universität Heidelberg, D-69117 Heidelberg, Germany*

*[11]MPIA: Max-Planck-Institut fur Astronomie, K¨onigstuhl 17, D-69117 Heidelberg, Germany*

*[12] LMU: Universit¨ats-Sternwarte, Ludwig-Maximilians-Universität München, D-81679 München, Germany*

*[13] Lunar and Planetary Laboratory, The University of Arizona, Tucson, AZ 85721, USA*

*[14]School of Earth and Space Exploration, Arizona State University, Tempe, AZ, 85207, USA*

*[15] Department of Physics and Astronomy, Louisiana State University, Baton Rouge, LA 70803 USA*

*[16]McDonald Observatory, University of Texas at Austin, 2515 Speedway, C1402, Austin, TX 78712, USA*

*[17]Department of Physics and Astronomy, Louisiana State University, Baton Rouge, LA 70803 USA*

*[18] Centro de Astrobiología (CSIC-INTA), ESAC, 28691, Villanueva de la Cañada, Madrid, Spain*

*[19]Johns Hopkins University, Baltimore, MD, 21218 [20]Princeton University, Princeton, NJ, 08544, USA*

*[21]Omitron, Inc. / NASA Contractor, Beltsville, MD 20705 [22]MVJ College of Engineering, Channasandra, Kadugodi, Bengaluru, Karnataka 560067, India [23]Carmel High School, Carmel, IN*

*[24]Harvard-Smithsonian Center for Astrophysics, Cambridge, MA 02138, USA*

*[25] Department of Physics Astronomy, The University of California, Irvine, CA 92697, USA*

*[26] Center for Astrophysics and Space Sciences, University of California, San Diego, CA 92093-0424, USA*

*[27] University of Rochester, Rochester, NY 14627, USA [28] Landessternwarte, Zentrum für Astronomie der Universität Heidelberg, Germany [29] State University of New York (SUNY) Oswego, NY, 13126*

*[30]Vanderbilt University, Department of Physics Astronomy, Nashville, TN 37235, USA*

*[31]Mississippi State University, Department of Physics Astronomy, Hilbun Hall, Starkville, MS 39762, USA*

*[32] Department of Astronomy Astrophysics and Center for Exoplanets and Habitable Worlds and Penn State Extraterrestrial Intelligence Center, The Pennsylvania State University, University Park, PA 16802, USA*

*[33] Centre for Exoplanets and Habitability, University of Warwick, Coventry CV4 7AL, UK*

*[34] Centre for Space Domain Awareness, University of Warwick, Coventry CV4 7AL, UK*

*[35] Department of Physics, University of Warwick, Coventry CV4 7AL, UK*

## ABSTRACT

Present and upcoming NASA missions will be intensively observing a selected, partially overlapping set of stars for exoplanet studies. Key physical and chemical information about these stars and their systems is needed for planning observations and interpreting the results. A target star archive of such data would benefit a wide cross-section of the exoplanet community by enhancing the chances of mission success and improving the efficiency of mission observatories. It would also provide a common, accessible resource for scientific analysis based on standardized assumptions, while revealing gaps or deficiencies in existing knowledge of stellar properties necessary for exoplanetary system characterization. To view report online, please go to: https://tinyurl.com/SAG22-FinalReport

### Executive Summary

A significant number of present and upcoming NASA missions will be targeting relatively small numbers of high-value stars to investigate the detailed properties of exoplanets. The ability to interpret and utilize the observational data from these missions will require detailed knowledge of the host star and planet. To date, there does not exist a central repository or archive of comprehensive stellar *and* exoplanetary data. A complete understanding of an exoplanetary system generally requires the knowledge not just of a single spectrum and a few bulk properties of the star, but a wealth of additional contextual properties supported by intensive observations and modeling.

- A target star archive to support community characterization of targets would enable NASA to be prepared for and better utilize exoplanet missions; such an archive would contain information on the stellar parameters, any companions, and any information about known planetary companions.
- The most highly targeted stars for exoplanet observations, where most detection methods overlap, will be the nearest FGK-type and early-M-type stars. While some of those stars are known to host planets, most of them (primarily the targets of direct imaging missions) are not. As a result, the stellar information necessary for target prioritization and planet characterization is sparse or non-existent for a large fraction of these stars. Additional observations will be critical to fill some of these gaps and may require support.
- We surveyed the interdisciplinary exoplanet community to determine the data most desired for characterizing stellar and planetary systems. The most common data products that all disciplines need are:

- Stellar physics parameters, including effective temperature, luminosity, mass, radius, age, chromospheric activity, and spectral energy distributions (SEDs)
    - Orbital properties of known planetary companions, including eccentricity
    - Physical properties of known planetary companions, including mass and radius
    - Spectroscopy of both the star and planet, with accompanying compositional information, such as stellar elemental abundances (Fe, Ca, K, Mg, P, S, Si, Al) as a proxy for planet composition, as well as direct planetary atmospheric composition (C, O), where available
    - More focused observations of low-mass stellar systems, including rotation rates, EUV flux, long-term variability across UV, optical, and IR, including measurement of activity cycles.

  A number of the above properties are not currently well-measured:
    - Both planetary mass and radius measurements are known for only ~20% of current exoplanets,
    - A total of only ~70 exoplanet host stars have Fe, Ca, K, Mg, S, Si, and Al abundance measurements important for building planets, while only 9 (of the 70) also have P measurements -- which is necessary for life.

  Unfortunately, some of the above host star properties are *not commonly measured* -- i.e. only exist for a small percentage of stars -- namely:
    - stellar chromospheric activity,
    - stellar magnetic activity,
    - stellar abundances for major components of planets,
    - element isotope abundances,
    - spectroscopy of planetary atmospheric gases, and
    - SEDs.

  It is important to the greater interdisciplinary exoplanet community that these rare or missing properties be specifically pursued in order to advance our understanding of exoplanets and their potential for habitability.
- A future target star archive would benefit a wide cross-section of the community. Such an archive would need to connect to and be interoperable with other current data archives, making it easier to cross-correlate target data from different resources -- whether active databases or telescope archives. Data within the archive should be presented in a way that users are able to assess the quality of data from a given survey or catalog. It would also need to be a "living" archive where archival or community members could update data content and information.

## 1. Motivation

A number of current, planned, and possible NASA missions are focused on discovering and measuring specific properties of exoplanetary systems. While these missions range from cubesats to flagships, all could substantially benefit from detailed information about the targeted star systems. This information could aid in the analysis of scientific results (e.g., accurate stellar radii are necessary to estimate the radius and bulk composition of transiting planets), observation planning (e.g., stellar activity priors could guide target selection), and even mission design (e.g., the occurrence rate of potentially Earth-like planets constrains the required aperture size of direct imaging missions). Accurately curated target information would ultimately enable and/or enhance mission success and improve the efficiency of scientific inquiry.

Precise exoplanet characterization is observationally intensive. Therefore, the target lists of these missions are relatively small, usually tens to hundreds of stars. Some of these target lists overlap significantly with one another. As a result, the total set of stars expected to be targeted by all such missions is fairly small, on the order of a couple thousand. The archive needs for targeted missions observing small numbers of stars is fundamentally different from the archive needs of large surveys like TESS and PLATO. While it would be ideal to have an archive that is both deep (all the stellar properties described in the report) and wide (millions of stars), that is not feasible in the near term. This SAG is focused on the archive needs of the "narrow" missions that are observing small numbers of stars to ensure that these observations, which are expensive on a per-star basis, are supported as thoroughly as possible.

**To date, there does not exist a central repository or archive of *comprehensive* stellar and exoplanetary data.** There are published results of large stellar surveys, such as Gaia, SDSS, Tycho, etc. Such surveys generally report broadband magnitudes, position and proper motion, and other bulk properties, for millions or billions of stars (see Appendices E and F). There are also specialized, targeted surveys of a few thousand or tens of thousands of stars with more intensive observations reporting spectra in a given wavelength regime, the abundances of a few elements, or time series observations of a given property. Tying together these various data sets are central archives that serve the cross-matched results of various surveys, such as Vizier, MAST, etc. *What these resources do not provide is a complete picture of many stars*. **A comprehensive understanding of an exoplanetary system goes beyond planetary spectra and a few bulk properties of the star--it requires a wealth of additional contextual properties supported by intensive observations and modeling.** The NASA Exoplanet Archive addresses some aspects of this work, but crucially, that archive only includes some information of known planet host stars, and a significant fraction of target stars of upcoming NASA missions do not yet have known planets and are thus not included in that archive.

The absence of a comprehensive target star archive results in wasted time spent duplicating the acquisition of target information, confusion as different groups adopt different parameters that inform analysis, and duplicated mission resources. In light of this, the ExoPAG Study Analysis Group 22 (SAG 22) was formed to examine the attributes for a comprehensive archive of stellar and planetary systems relevant to NASA exoplanet science, including the stellar and planetary properties themselves.

The goals of SAG 22 were as follows:

- Define the attributes of the sample of high priority stars to be included in a future archive.
- Survey the broad exoplanet community (e.g., including planetary scientists, geologists, heliophysicists, and biologists) to determine data required for characterizing stellar and planetary systems.
- Prioritize what stellar and planetary properties would be most useful to include in a future archive.
- Identify useful front-end and back-end features of such an archive.

While exoplanet science is a worldwide endeavor, as a SAG established by the NASA Exoplanet Exploration Program Analysis Group, we specifically limited our investigations to the needs of NASA missions. We discuss these missions in the context of the larger exoplanet field (e.g. non-NASA space missions, ground-based investigations), but focus on the specific benefit to NASA projects.

## 2. Approach

To accomplish the above goals, the SAG 22 co-leads reached out to a variety of communities so that membership could be as diverse and multidisciplinary as possible. The co-leads established a website with a member sign-up and advertised the SAG at a variety of conferences, posted on social media, and directly messaged AASWomen, AAS Committee on the Status of Minorities in Astronomy, AAS Committee for Sexual Orientation and Gender Minorities in Astronomy, Planetary Science Newsletter, Earth Science Women Network (ESWN), 500 Women Scientists, American Geophysical Union, as well as SACNAS to garner members. Early career scientists were particularly encouraged so that they could be part of the archive decision making from the beginning. As a result, SAG22 consisted of 3 co-leads and 34 members, which is one of the largest SAGs to-date.

The SAG members were assembled into four task forces (TFs) designed to gather the requisite information on existing and future exoplanet missions, the needs of the community, and existing exoplanet/stellar catalogs:

- TF1: Mission Observables and Deliverables.
  *Leads: David Wilson (U. Texas at Austin) & Karan Molaverdikhani (Landessternwarte, Zentrum für Astronomie der Universität Heidelberg)*
  This TF systematically gathered information about science drivers and data products for active and future exoplanet-related NASA missions. This information was used to help define the science needs of the archive.
- TF2: Target Lists and Target Criteria.
  *Leads: Ilaria Pascucci (U. Arizona) & David Ciardi (Caltech/IPAC-NExScI)*
  This TF examined the planned observations by NASA exoplanet missions. It assembled lists of individual stars slated for observation, and also sets of stellar criteria for missions without preliminary or published target lists.
- TF3: Interdisciplinary Use Cases.
  *Leads: Ravi Kopparapu (NASA GSFC) & Jacob Lustig-Yaeger (JHU APL)*
  While current exoplanet missions tend to be designed by astronomers, key insights into the physics of exoplanets are studied by a wide range of scientists, including astrobiologists, geologists, biologists, planetary scientists, heliophysicists, and others. This TF communicated with a number of such scientists to assemble a list of use cases for how the mission data products identified by TF1 could be used to investigate exoplanets, and to determine what information about the host star would be needed to interpret that data.
- TF4: Existing Catalogs.
  *Leads: Jennifer Burt (JPL) & Kevin Hardegree-Ullman (U. Arizona)*
  This SAG was formed partly through the realization that many properties of likely exoplanet mission target stars are not reliably or completely listed in existing astrophysical catalogs. Of course, many such properties are included, and we need to quantify their reliability and completeness. This TF outlined the parameters of that investigation by assembling a list of known catalogs of stellar properties, with a breakdown of the catalog contents, including estimates of the range of values of each parameter and the typical precision.

These TFs organized community surveys, conducted independent literature searches, and reached out to relevant experts. TFs worked together to address larger, overarching questions that spanned multiple TF topics. Finally, each TF assembled a detailed report addressing their goals, included in Appendices E-H of this report.

In addition to the research conducted by the TFs, SAG 22 directly invited representatives of many current, upcoming, and proposed NASA missions to discuss their missions at SAG-wide virtual meetings. SAG 22 held a total of nine such meetings (see Table 2.1), at which we discussed details of the target lists, as well as how a future target star archive could benefit the design, implementation, and scientific analysis associated with these missions.

| Mission Name | Contact Person |
|---|---|
| JWST | Jacob Bean |
| HabEx & LUVOIR | Shawn Domagal-Goldman |
| OST | Kevin Stevenson |
| ARIEL/CASE | Billy Edwards |
| Roman CGI | Vanessa Baily |
| CUTE | Kevin France |
| NEID | Arvind Gupta |
| SPARCS | Evgenya Shkolnik |
| Pandora | Jessie Dotson |

**Table 2.1:** Overview of the NASA mission representative who met with the SAG 22 members to discuss target lists and the ways in which a target star archive could be supportive.

Below we summarize and synthesize the findings of SAG 22. In Section 3, we discuss the interdisciplinary user base of a future archive and their desired data sets based on their use cases. In Section 4, we present the target lists of current and future exoplanet-related missions and describe the population of high priority stars. In Section 5, we present findings on the implementation of such an archive. Section 6 describes missing information from a variety of perspectives, including user feedback on data that does not yet exist but is critical for key science questions, as well as the results of an exercise the SAG employed to understand the difficulty in gathering critical properties for a small set of stars. Finally, in Section 7, we provide some concluding remarks.

## 3. User Base & Use Cases

With upcoming NASA exoplanet missions delivering data applicable to a broad range of the scientific community, we investigated how scientists across a variety of disciplines intend to use the data. Our focus was on what target star data would be most valuable to allow them to conduct their scientific research using mission data via an archive. We approached a wide interdisciplinary user base, from those designing future missions to those conducting observations, analyzing observations, and informing models.

TF3 reached out to hundreds of scientists from a variety of fields, including geology, (astro)biology, planetary science, astrophysics, heliophysics, and others, to determine how they would prefer to interact with a future target star archive, including the necessary observables and properties. Each scientist was asked:

1. What are your stellar or planetary science questions?
2. What stellar or planetary information will you need to interpret your data/models/output (including primary and secondary use case)?
3. What is ideal vs. threshold precision?
4. What wavelength range might these observations cover?
5. How have you used stellar databases/information in the past? What was missing?

A total of 66 responses were received, which were then categorized by TF3 into interdisciplinary groups (Fig 3.1) as well as larger sub-disciplines: Host Stars, Planetary Systems, Exogeology, and Habitability/Astrobiology. The answers, many long-form sentences, were processed into more easily parsable and uniform short phrases (Appendix A).

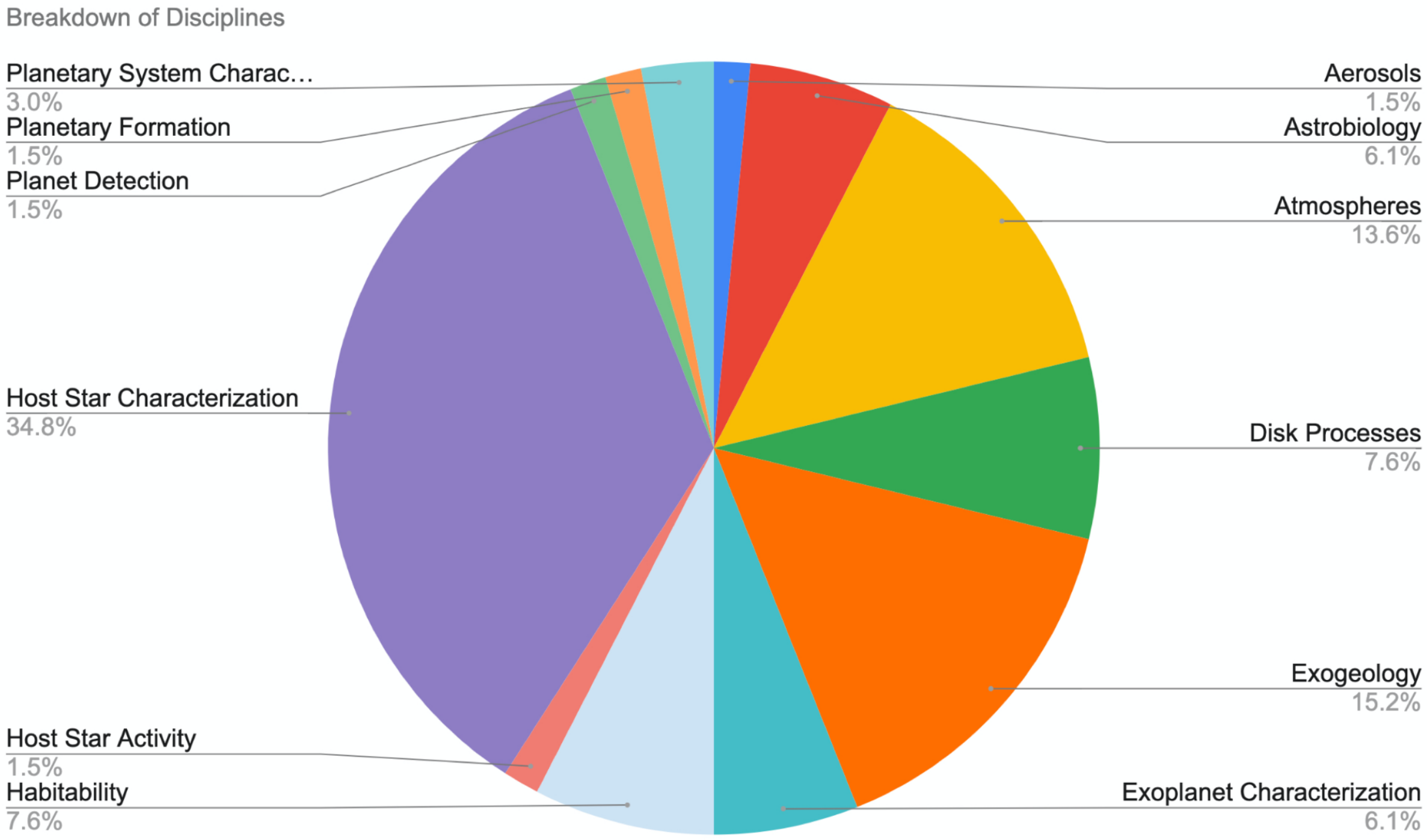

**Fig 3.1:** Pie chart showing the breakdown of the different discipline responses as labeled by TF3 (i.e. labels were not assigned by those who responded to the set of questions).

Table 3.1 lists the top 20 most frequent property responses for the full and individual disciplines as reported from the interdisciplinary science survey (Appendix A). The number in parenthesis indicates how often this property was cited within the survey responses. The most noticeable difference between the four example sub-disciplines is the variation in their important properties -- namely, characterizing host stars and planetary systems require a wide variety of properties, while exogeology, habitability, and astrobiology are mostly limited to the physical properties of the planet (mass, radius, and spectra) as well as the chemical composition (as typically inferred from the stellar abundances). This may be because the scientists who study astrobiology/habitability take a number of properties on assumption: rocky planets, stable star, stable climate, etc. Namely, by assuming an Earth-like exoplanetary orbit, they are able to study other kinds of variable conditions on the planet itself. In addition, the study of habitability and astrobiology requires a much more diverse set of characteristics, as noted by the presence of a large number of unique properties, i.e. ~120 properties as compared to the other 3 disciplines who had < 60 unique properties. Also, while it may be useful for the host star and planetary system communities to observe their targets across a wide range of wavelengths, exogeology and habitability/astrobiology are preferentially interested in the UV, optical, and IR bandpasses (Appendix A and G). It should be noted that responses may have been biased towards specific proposed or upcoming instruments/telescopes, such as HabEx or LUVOIR. Finally, we note that there are properties in every sub-discipline that are standalone because they do not quite fit; however, this really illuminates the point that exoplanet science is very interdisciplinary and that these sub-discipline "silos" do not work across the board.

Overall, while many of the different sub-disciplines have individual foci, common themes can be identified that would be useful for an interdisciplinary investigation; i.e, information/data collected from one discipline could flow into another discipline. **The most common data products that all disciplines need are:**

- **stellar effective temperature and luminosity,**
- **planetary mass and radius,**
- **spectroscopy of both the star and planet,**
- **stellar mass, radius, age, and activity,**
- **planetary atmospheric composition (C, O), and**
- **stellar elemental abundances (Fe, Ca, K, Mg, P, S, Si, Al) as a proxy for planet composition.**

There are other key properties that are important for providing a context for the stellar and planetary system, such as **planetary eccentricity, stellar SEDs, as well as more focused observations of low-mass stellar systems**, but they are more specific to the particular science application.

| All Frequency | Host Star Frequency | Planet & System Frequency | Exogeology Frequency | Astrobio Frequency |
|---|---|---|---|---|
| (22) Spectroscopy | (6) EffectiveTemp | (4) Luminosity | (5) Ca | (12) PlSpectra |
| (17) EffectiveTemp | (6) Spectroscopy | (4) Spectroscopy | (5) O | (9) Spectroscopy |
| (15) PlMass | (6) StMass | (3) EffectiveTemp | (5) PlMass | (8) PlAtmospheric Composition |
| (15) PlSpectra | (5) Luminosity | (3) PlMass | (4) Al | (6) PlMass |
| (15) StMass | (5) StActivity | (3) StMass | (4) Fe | (6) PlRadius |
| (14) Luminosity | (5) StRadius | (3) StRadius | (4) K | (5) EffectiveTemp |
| (12) PlRadius | (4) Age | (2) Eccentricity | (4) Mg | (5) SED |
| (10) Age | (3) Asteroseismology | (2) Inclination | (4) S | (3) C |
| (9) PlAtmospheric Composition | (3) Photometry | (2) Photometry | (4) Si | (2) AerosolOptical Properties |
| (9) StRadius | (3) Spectrophotometry | (2) PlCompanions | (4) StMass | (2) Age |
| (8) O | (3) StMagneticActivity | (2) PlRadius | (3) Age | (2) EvolutionaryTracks |
| (7) C | (3) StMagneticField | (2) StSpectra | (3) C | (2) HabitableZone |
| (7) SED | (3) StRotationPeriod | (2) TimeVariability | (3) EffectiveTemp | (2) LowMassStars |
| (6) Fe or [Fe/H] | (3) StRotationPeriods | (1) Abundances | (3) Luminosity | (2) Luminosity |
| (6) Ca | (2) flux | (1) Age | (3) P | (2) N |
| (6) K | (2) Granulation | (1) C | (3) PlRadius | (2) O |
| (6) LowMassStars | (2) Images | (1) Disk | (3) Spectroscopy | (2) PlSurfaceComposition |
| (5) Mg | (2) LowMassStars | (1) DiskComposition | (2) Cr | (2) StFlux |
| (5) P | (2) Metallicity or [Fe/H] | (1) DiskRadii | (2) Mn | (2) StMass |
| (5) Photometry | (2) SpaceVelocities | (1) Distance | (2) Na | (1) Ca |

**Table 3.1:** The top 20 most frequently referenced properties for the full and individual disciplines as reported from the interdisciplinary science survey. The number in parenthesis indicates how often this property was cited within the survey responses. Note that “Pl” is short for “planet” and “St” is short for “star” as a way to qualify the properties without impacting the results of the frequency count.

The most common properties identified above are not entirely surprising considering that fundamental stellar and planetary information is crucial for studying planetary system properties such as disk formation, volatile/ice-line location, atmospheric escape and retention, and habitability. These topics span all the disciplines discussed above indicating a common need, although we note that this set of data products alone would be insufficient to wholly address many of the science topics.

We note that many of the properties important to characterizing an exoplanetary system are time-series observations. While the parameters in the bulleted list above are physical parameters of the combined star/planet system, there are a variety of observational data that are crucial for characterizing the system. These include photometric variability in key bandpasses over different timescales including stellar rotation and planetary orbits, and spectroscopic trends encompassing chromospheric activity indices and radial velocities. Other valuable time domain data sets are spectroscopic signatures of planetary atmospheres as well as debris disks. Overall, time-series observations, particularly those related to the dynamical and orbital system properties, will need to be accounted for in an archive.

Part of the functionality of the survey of scientists across multiple disciplines by TF3 was to better understand the range of science use cases pertaining to current and upcoming NASA missions. While the full list of responses can be found in Appendix A, here we present a few example use cases:

- There are a variety of astrobiological science questions that will utilize JWST, TESS, and ground-based RV surveys in order to better understand liquids present on an exoplanet surface at 200-500 K (e.g. $H_2O$, $CO_2$, and $H_2SO_4$) as well as detection of gases, both “expected” and “unexpected”. Understanding these properties -- via planetary atmospheric compositions, instellation flux on the planet, planetary interior composition, planetary dynamics and tectonics, planetary orbital properties, stellar effective temperature, stellar ages, and SEDs -- will enable the interpretation of possible biosignatures within the context of their environment (including energy sources, availability of elements like CHNOPS, and clement conditions).
- JWST, Roman, HST, and future HabEx/LUVOIR high contrast imaging data are crucial for debris disk characterization that can inform disk composition and dynamics, volatile delivery, and disk variability, as well as the overall planetary system architecture. These sorts of investigations require spatially resolved disk photometry and spectra, planetary spectra, planetary mass, and stellar luminosity, as well as the ability to observe over long temporal baselines.

- Studying host star winds and activity -- which are essential for understanding the planet with respect to the stellar environment -- will utilize data from TESS, HST, ALMA, SPARCS, Chandra, and XMM-Newton in addition to Gaia and PLATO.

In terms of measurement precision, obtaining information about stellar spectra (~10% precision) from UV-NIR (0.2 - 25μm), spot temperatures (+/- 100K precision), and stellar masses (~20%) could be useful in accurately characterizing exoplanet atmospheres. UV and X-ray wavelengths allow for the study of a wide variety of areas, such as the radiation environment around a star, stellar activity, and activity cycles that could impact a planet's atmospheric composition and habitability. More specific responses can be found in Appendix A.

**A future target star archive would benefit a wide cross-section of the community.** It would enhance the chances of mission success, improve the efficiency of ground- and space-based observatories, provide a common and accessible resource for scientific analysis based on a standardization of assumptions, and could reveal any observational gaps or deficiencies for those properties necessary for exoplanetary system characterization. To this end, all of the responses in Appendix A were analyzed in conjunction with TF1 with respect to current upcoming NASA and non-NASA missions, whether they needed supplemental information not provided by the main mission, and whether ultimately the use case was in support of NASA missions.

## 4. Targets

Many future exoplanet studies and missions will focus predominantly on nearby stars. While some stars may be well-studied with multiple exoplanet observing methods, others may be limited to a single method. As a result, the sample of "high priority" stars accessible with multiple techniques/missions will be constrained to a relatively small number.

To better understand the overlap of the target lists for current and future exoplanet observations, SAG 22 established two TFs. TF1 created an exhaustive list of current and future exoplanet observatories, along with their science goals, wavelength ranges, and anticipated data products. As noted in the introduction, this investigation was limited to NASA missions to stay within the purview of the SAG 22 Terms of Reference. A simplified version of this list is shown in Table 4.1. TF2 then retrieved the exoplanet target lists for as many of the observatories listed in Table 4.1 as possible and cross-matched them with the 2MASS catalog. Finally, the 2MASS IDs were cross-matched with the TESS Input Catalog (TIC-8.1) to retrieve stellar parameters for >95% of the targets. While the TIC is not fully complete nor reliable, it serves as a convenient, single catalog that encompasses all stars in the combined target list. Table 4.1 color-codes the target lists used in this report by method: red = transit spectroscopy, blue = extreme precision radial velocity (EPRV), green = ground-based direct imaging, orange = space-based direct imaging. The target lists of future observatories should be interpreted as current best guesses based on the available understanding of each observatory at the time of this report.

| Observatory | Science Case | Wave-length | Source of Target List Used in this Report |
|---|---|---|---|
| *NASA Observatories* | | | |
| Hubble Space Telescope | Transit Spectroscopy | UV/VIS/NIR | |
| | Transit Photometry/Phase Curves | VIS/NIR | |
| | Host star SEDs | UV/VIS | |
| | Direct Imaging | VIS/NIR | |
| Keck Observatory | Transit Spectroscopy, RV observations, Direct imaging | VIS/NIR | |
| James Webb Space Telescope | Transit and secondary eclipse spectroscopy | NIR/MIR | GTO, ERS, and Cycle 1 GO Programs |
| | Direct Imaging of planets & disks | NIR/MIR | GTO, ERS, and Cycle 1 GO Programs |
| HabEx | Direct Imaging of planets & disks | UV/VIS/NIR | HabEx Final Report and private comm. (C. C. Stark) |
| | Transit Spectroscopy | UV/VIS/NIR | |
| LUVOIR (A & B merged) | Direct Imaging of planets & disks | UV/VIS/NIR | LUVOIR Final Report and private comm. (C. C. Stark) |
| | Transit Spectroscopy | UV/VIS/NIR | |
| | Astrometry | VIS | |
| Origins | Transit and Eclipse Spectroscopy | NIR/MIR | Origins Final Report and private comm. (K. Stevenson) |
| | Direct Imaging of protoplanetary disks | NIR/MIR | |
| ARIEL/CASE | Transit Spectroscopy | NIR | Edwards et al. (2019) |
| Roman Space Telescope | Microlensing | VIS/NIR | |
| | Transit photometry | VIS/NIR | |
| | Direct Imaging of planets & disks | VIS/NIR | SIOS Lab Imaging Mission Database (D. Savransky) and private comm. (V. Bailey) |
| CUTE | Transit Spectroscopy | NUV | Private comm. (K. France) |
| NEID | EPRV | VIS/NIR | Gupta et al. (2021) & GTO Programs |
| *Non-NASA Observatories* | | | |
| GPI | GPIES survey - direct imaging of debris disks and self-luminous exoplanets | NIR | Nielsen et al. (2019) |
| SPHERE | SHINE survey - direct imaging of debris disks and self-luminous exoplanets | NIR | Desidera et al. (2021) |
| MAROON-X | EPRV exoplanet detection around M dwarfs | VIS/NIR | Private comm. (J. Bean) |
| ESPRESSO | EPRV exoplanet detection around G dwarfs and later | VIS | GTO Programs (http://www.eso.org/sci/observing/teles-alloc/gto.html) |
| CARMENES | EPRV exoplanet detection around M dwarfs | VIS/NIR | Reiners et al. (2018) |
| California Planet Search | EPRV | VIS/NIR | Private comm. (B.J. Fulton) |

**Table 4.1:** List of current and future exoplanet-observing telescopes. The color-coded rows were included in the SAG 22 target list analysis.

We note that some types of missions are deliberately not included in this table. For instance, TESS and PLATO are both exoplanet missions, but are large surveys of hundreds of thousands of stars. Incorrect stellar information for a small subset of those stars does not present the kind of mission impact as it does for highly targeted missions observing a few dozen stars. TESS and PLATO use large compiled star catalogs for selecting targets that do not require such highly individualized stellar information as do the other missions listed in the table. Similarly, although Gaia is expected to discover many massive planets via astrometry, the core science of that effort does not depend on detailed stellar information requiring a new archive. Since those missions do not require a highly curated archive of the sort discussed in this report, we are deliberately leaving them out of Table 4.1.

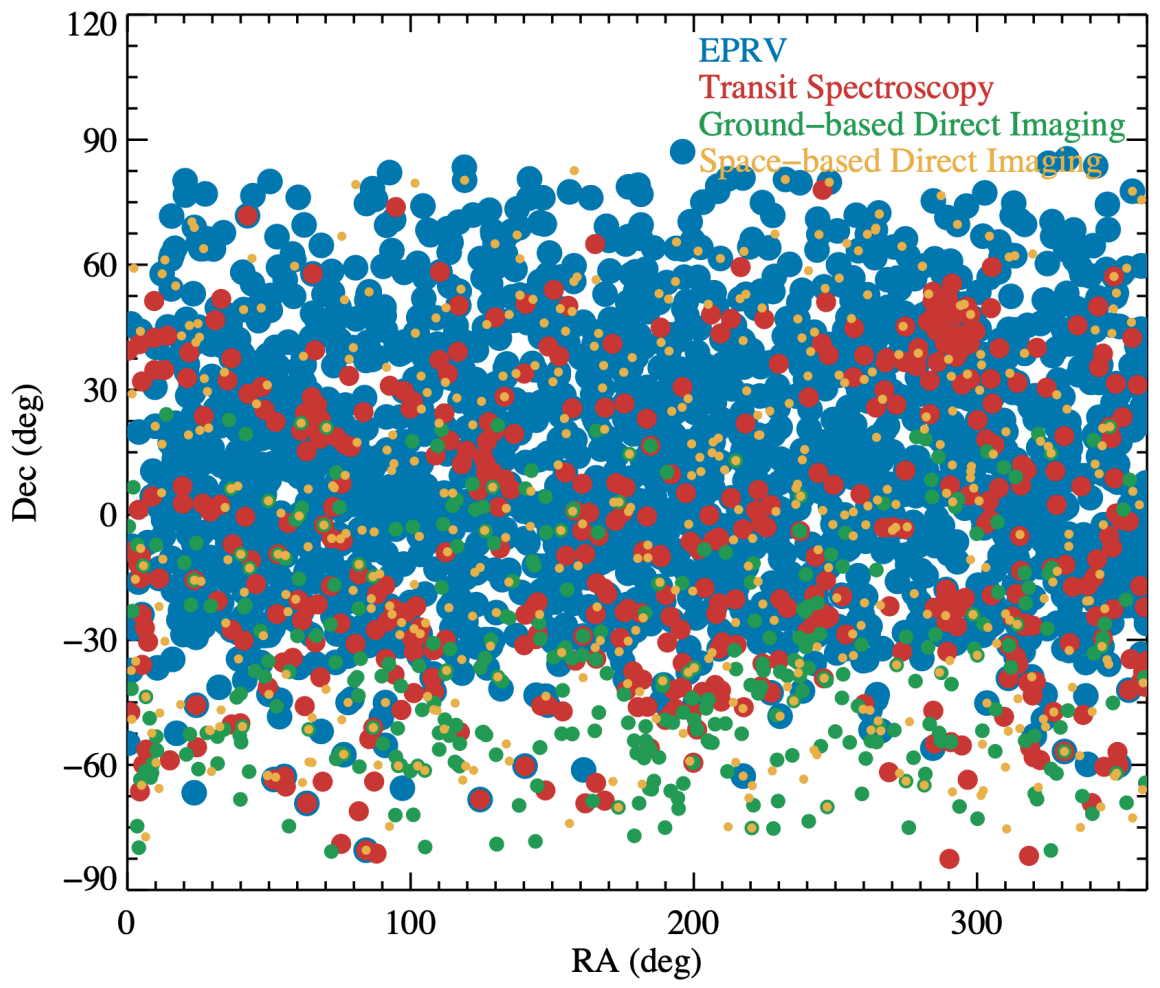

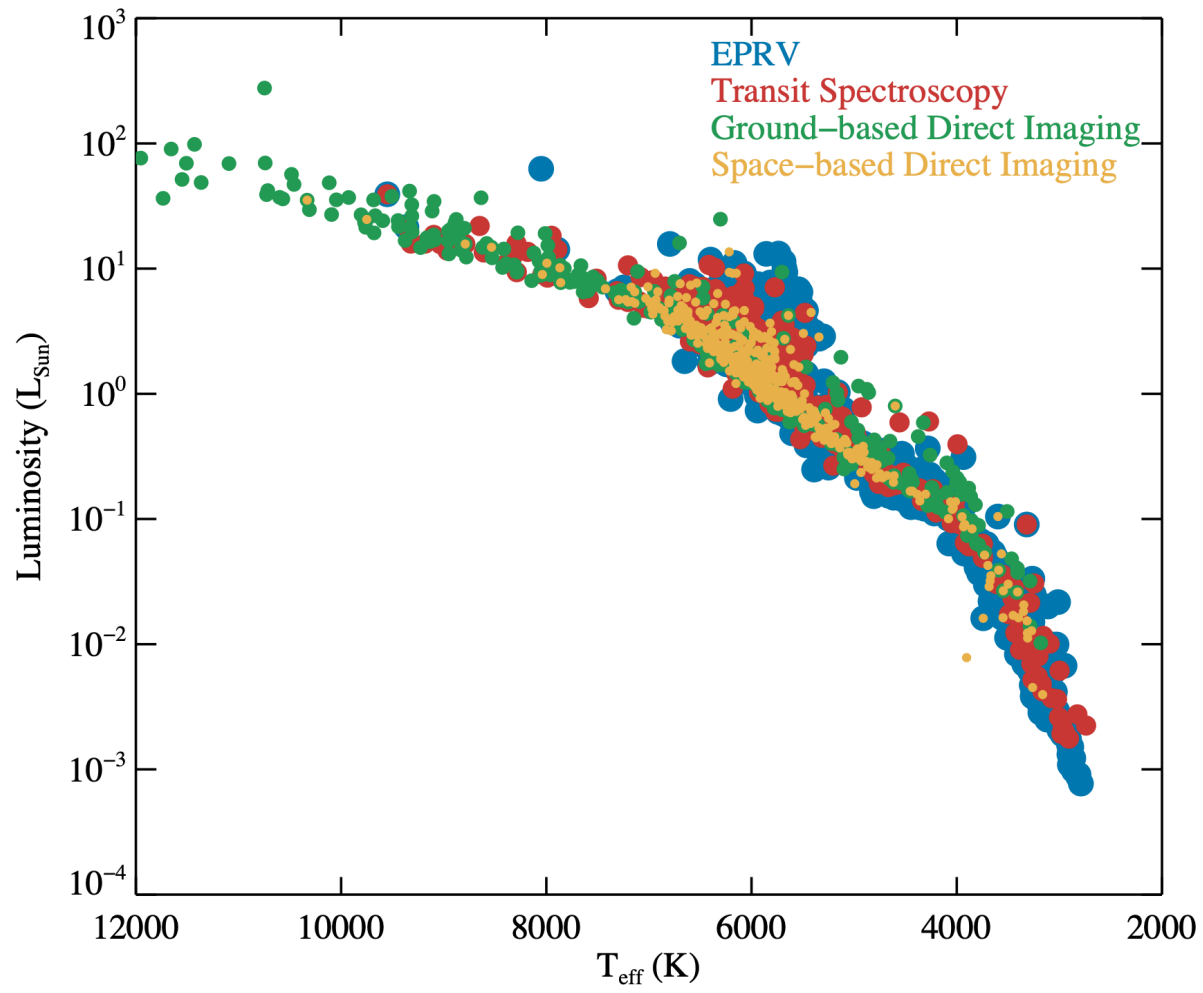

**Figure 4.1:** Distribution of targets color-coded by observing method. *Left:* Distribution of targets on the sky. *Right:* Distribution of targets relative to the main sequence.

The combined targets, given in Appendix B, are predominantly main sequence stars, covering a broad range of spectral types. Temperatures range from ~3000 to 12,000 K and luminosities range from ~0.001 to 100 times solar, as shown in the right panel of Figure 4.1. The left panel of Figure 4.1 shows the distribution of targets on the sky, color-coded by technique. While the targets are distributed broadly uniformly, a clear northern and southern bias is shown in the EPRV and ground-based DI target lists, respectively. The Kepler field of view is noticeable as a small cluster of targets for transit spectroscopy at RA ~290°, Dec ~44°. No clear bias toward the ecliptic poles is evident.

Figure 4.2 plots the targets by luminosity and distance. The left panel, color-coded by method, shows that the EPRV method largely overlaps with the other methods, but is primarily limited to mid-G and later-type stars. Transit spectroscopy targets tend to be more distant, as the chance alignment of an edge-on planetary orbit has a small probability and the number of stars increases as distance cubed. Ground-based direct imaging focuses primarily on still-cooling gas giants beyond a few AU orbiting younger stars, biasing the sample to earlier-type stars at distances beyond 20 pc (see Figure 4.1, right panel). Finally, future space-based direct imaging missions primarily focus on detecting Earth-like planets, biasing them toward the nearest stars with the largest projected habitable zones.

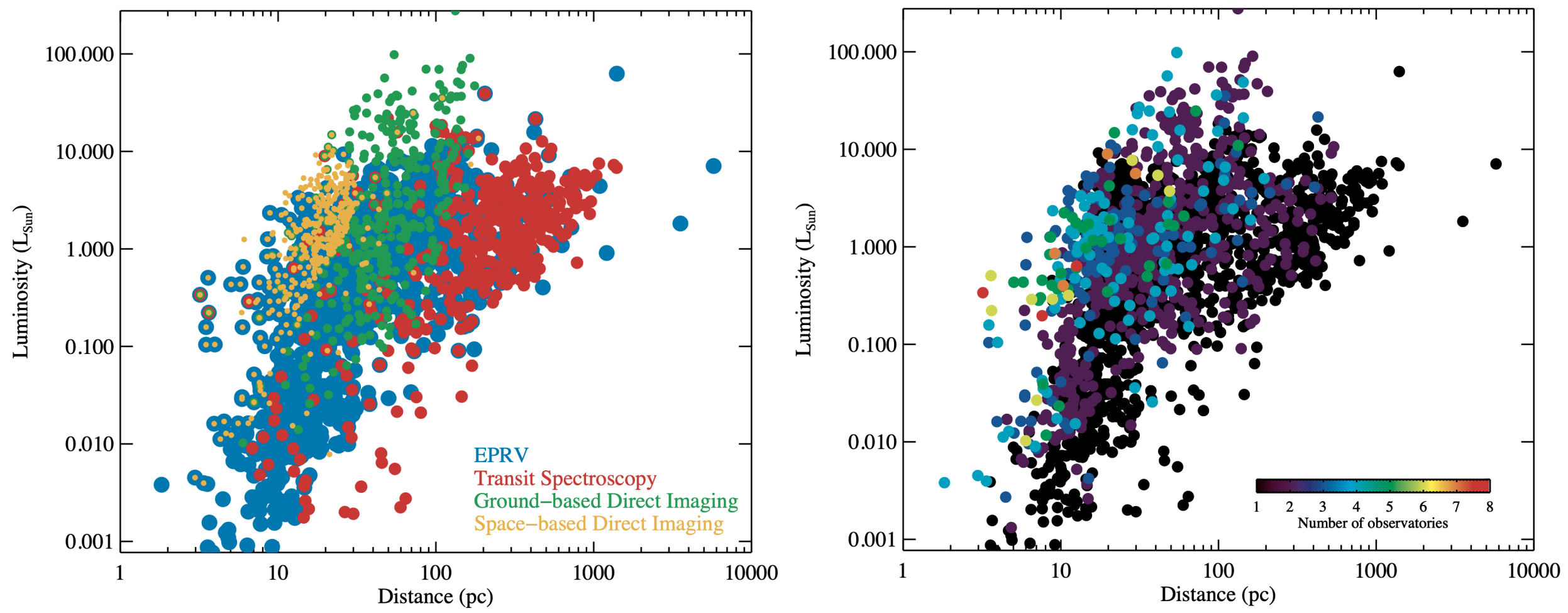

**Figure 4.2:** Distribution of targets in luminosity-distance space. *Left:* Color-coded by observing method. *Right:* Color-coded by the number of observatories potentially targeting the star.

The right panel of Figure 4.2 shows the same set of targets, color-coded instead by the number of observatories that may target the system. While the right panel shows a fair amount of scatter due to the specificity of each observatory's target list, a trend is clear: **the most highly targeted stars for exoplanet observations, where most detection methods overlap, will be the nearest FGK-type and early-M-type stars**. This region of greater overlap is largely driven by space-based direct imaging missions, including the upcoming James Webb and Roman Space Telescopes, as well as future mission concepts like HabEx or LUVOIR. The full target list can be found in Appendix B.

## 5. Archive Functionality

Here we consider the archive functions that would benefit NASA missions and the scientific use of their data. We note that while this SAG has compiled a notional target list for such an archive, that preliminary list is incomplete. A more comprehensive and reliable target list that best promotes NASA's exoplanet science goals could be formed at the outset by future archive curators. Furthermore, new observations, discoveries, and missions will continue to emerge while such an archive is implemented and used. **As such, we find that the primary function of such an archive is to serve as a living data source, with ongoing curation to include new target stars, incorporate new stellar and planetary data, and update the archive accessibility as web pages, APIs, and query-tools evolve.** To promote a living archive, archive curators could regularly liaise with scientists participating in relevant NASA missions, conduct surveys of its user base, and consult with the future user base of NASA missions.

We consulted with the scientific community to learn what scientists would like to see for the front-end user experience of such an archive. The full list of such requests categorizes items as “Needs” (the archive will be significantly deficient without the capability), “Wants” (the archive can serve its core purpose without the capability, but the community would significantly benefit with its inclusion), and “Ideas” (capabilities that are not immediately needed, but would provide benefits and should be considered when development resources are available). While the full list is provided in Appendix C, the “Needed” capabilities, in no particular order, are:

- Resolve common star names for a specific object or list of objects.
- Distinguish between observed and derived stellar parameters.
- Search, filter, and download subsets of data, based on provenances or multiple parameter values or ranges or measurement dates, for groups and/or individual targets, allowing discontinuous range, via both web interface and scripting.
- Filter by stars with known or confirmed planets and the properties of those planets.
- Allow users to select subsets of data with enforced consistency between parameters.
- Provide examples of and tutorials on constructing queries. Allow users to provide their own examples via a shared wiki or github repository.
- Permit interactive tabular searching/filtering and plotting of properties or data products through a browser.
- Retrieve and compare all published values of a specific parameter for a given target. Include tools for determining mean/median/etc. of a given parameter.
- Summary/overview of system information for each star (planetary and stellar companions, disks, cluster/association membership).
- Maintain user accounts to save queries, plots, or download scripts.
- Include as much cross-correlated data as possible (e.g. survey IDs and associated data).
- A single resource for all publicly available broadband photometry, high resolution spectra, etc. including references/sources.
- Cross-link individual targets and parameters to external databases, archives, and ADS.

One key archive function will be the strategy for identifying and ingesting new data for all target stars. The initial build and ongoing maintenance of the archive will require regular perusal of existing literature, review of new papers, and connection to NASA mission development plans. Archive scientists will make ongoing judgements about the value and utility of incorporating new data on these stars. While guidelines for making those judgements should be developed at the establishment of the archive, SAG 22 consulted with representatives from NASA archives, specifically David Ciardi (NExScI) and David Rodríguez (MAST), who offered the following thoughts on running a NASA archive from the survey of interdisciplinary scientists:

- Archives usually use relational databases. In such cases, it is important to create a thorough, carefully constructed data model that describes the data to be captured as explicitly as possible. Planning for future changes and creating a flexible data model is essential.
- There is always a decision that needs to be made between providing all the sources of parameter information, versus the archive making a decision on the "best" parameter.
- Curation, including selection of best parameters, is hard, time-consuming, and people-power-intensive. Make sure that ongoing curation activities are closely connected to the core use case for the archive.
- There is a trend towards databases that accept many data types for a given record. The archive should consider Uniform Resource Identifiers (URIs) to connect to other data sources on the cloud that do not rely on hardcopy versions of all data.
- **It is important to provide proper connections between the catalog data and the source catalogs or archives, with code too**; especially other NASA archives, but also other non-NASA archives. It is also important to provide an Application Programming Interface (API) with as much code documentation as possible.
- Expect exceptions to all data rules and that not all data can be captured. Some things will not fit, and the archive managers have to plan for such cases.
- For some quantities, the utility of a parameter changes over time (i.e. transit times growing stale), or the fundamental information changes over time (i.e. star coordinates changing, based on precession, proper motion, and even secular acceleration).
- Provenances can be tracked through doi's, linked to ADS. If derived parameter values that depend on other parameters are provided, it is necessary to clarify and provide the links from those other parameters.
- Ticket systems that allow the community to provide direct feedback, with additional information, clarification, and fixes are useful tools. Such systems require commensurate staffing levels.

## 6. Missing Information

During the course of SAG 22, we found that there are many instances of “missing information”, namely: A) data that can be found vs. data that exist but may be difficult to find, B) data that does not exist but is needed by members of the community, and C) metadata that is not presented but is necessary for overall comprehension and utilization. In this section, we review these three cases of missing information.

A. Finding important stellar information

As an illustrative exercise, SAG 22 members attempted to collate as many stellar properties as possible for a handful of quasi-random stars from the TF2 target lists to test for completeness and reliability of existing star catalogs. Four representative stars were chosen: HIP 36366 (F1V), HIP 55846 (G9), HIP 57939 (K1), and HIP 33226 (M3V), which covered a wide range of stellar types. These stars are fairly average, i.e. not deeply and thoroughly characterized like more popular stars, such as TRAPPIST-1. The SAG22 membership was polled during multiple Zoom calls to determine the list of standard stellar parameters to be found during the exercise (listed in Fig 6.1). In an attempt to replicate reasonable real-world time constraints, volunteers were asked to spend ~2 consecutive hours to find as many of the standard stellar properties as possible. This exercise was not intended to be a deep dive into the literature, but instead illustrates the ease or difficulty in finding common stellar properties by community members. This exercise was performed collaboratively, over Zoom, in order to take advantage of as much expertise as possible. Members used the TF4 list of current  datasets, catalogs, and databases (found in Appendix D), in addition to other online resources. The SAG 22 volunteers included Joshua Pepper, Kevin Hardegree-Ullman, David Wilson, and Natalie Hinkel.

The results of this exercise are shown in Fig. 6.1, where the findings have been abstracted to indicate whether property or error information was found (orange) or whether it was not found (purple). Of the 288 total values and errors for the 4 stars, 100 (~35%) were not found during the allotted time. This is not to say that all these measurements do not exist, only that they could not be found by 4 members of the community within a limited amount of time. For example, in the case of HIP 57939, there are apparently hundreds of literature references for this star when searching for it by another name, Gmb 1830. However, only a few of these references came up when searching via the HIP and HD (Simbad default) names since (1) stars appear in papers with the different IDs, complicating the search; and (2) not all useful data is in easily queryable electronic tables, showing an important lack of inaccessible, cross-correlated data. In addition, the quality of properties and their uncertainty, i.e. whether a single epoch of data is sufficient or whether a time series is necessary, were not assessed since the data/precision needs are dependent on the science use case. Overall, though, we find that Fig. 6.1 maps out missing data that could be influential for upcoming missions/surveys.

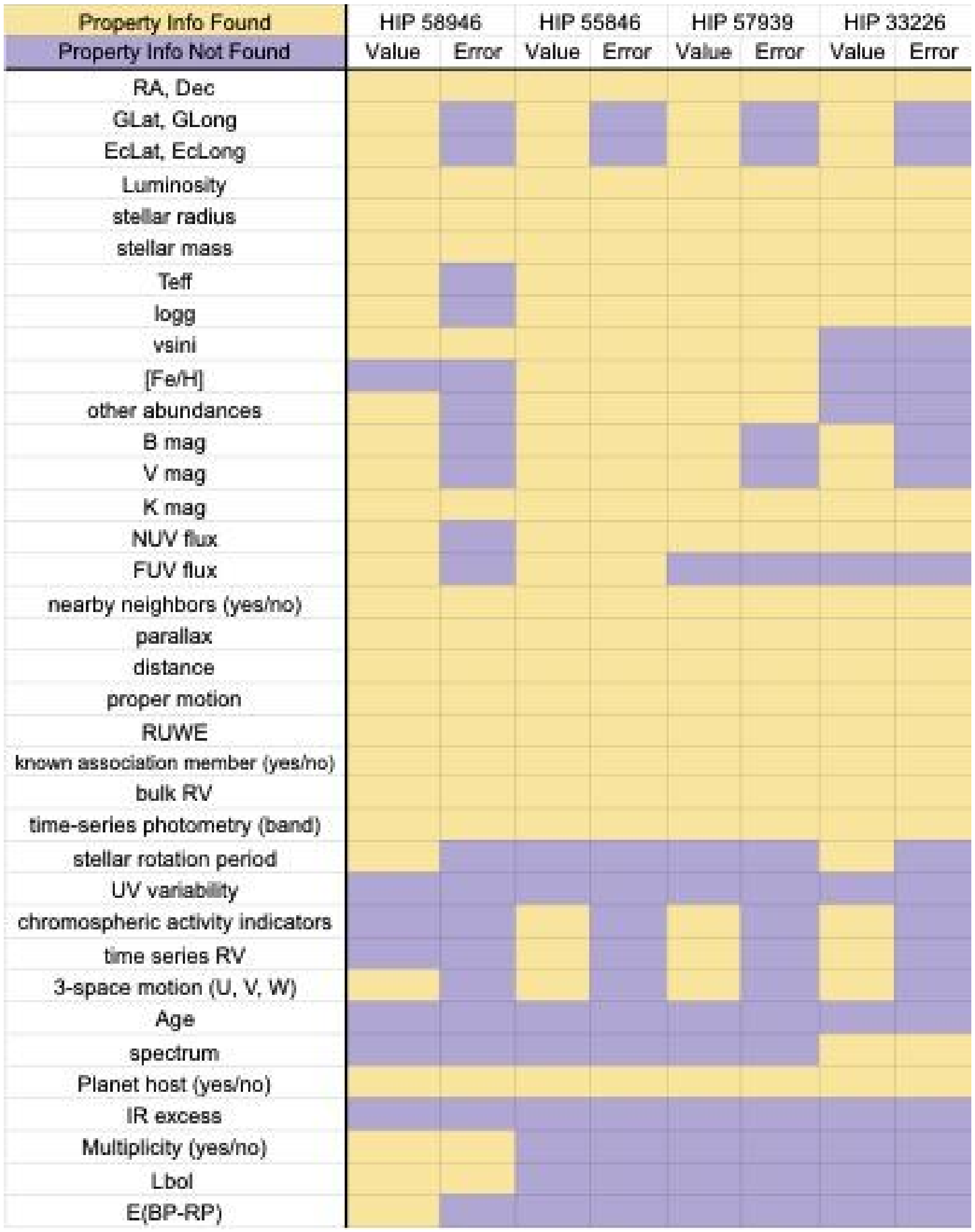

**Fig. 6.1:** An illustrative exercise showing which stellar information is available or difficult to find. The abstracted image indicates which standard stellar properties could be found for four average stars from the TF2 target lists of varying stellar types. This exercise was performed by four SAG members during a single 2-hour session to find as much information as possible. Orange indicates that the property value or uncertainty were found, purple shows that the property or error was not easily found. We will note that for those properties that were binary in nature (e.g. planet host, multiplicity, association membership, etc.), both the value and error were coded with the same color for simplicity.

## B. Data that does not exist but is needed by the community

While some of the stellar properties that could not be found during the Fig. 6.1 exercise actually do exist, there are also values that are needed by the community that have not yet been measured. As mentioned in Section 3, within the set of questions presented to interdisciplinary scientists, they were asked to identify what was missing within currently available databases or archives. Table 6.1 shows the abbreviated responses from those people who responded to the question (10 out 66 people), where the full responses can be found in Appendix A. **The most common concern across the sub-disciplines was that it was difficult to cross-correlate target data from different resources, whether active databases or telescope archives.** There are also a variety of properties that are necessary for exoplanet science but are lacking in current databases, either because they have not been amalgamated or because they do not exist: spectroscopy of atmospheric gases (especially high resolution), broadband spectroscopy, stellar abundances for elements important to planet formation, high precision UV data, and information on stellar activity.

| Frequency of responses regarding missing data in current databases |
|---|
| (7) cross-correlated target data |
| (2) stellar activity indicators (any) |
| (2) stellar abundances for major components of planets |
| (2) spectroscopy of atmospheric gases |
| (2) broadband coverage |
| (1) accurate measurement uncertainties |
| (1) cross-correlated planet and star data |
| (1) ease of use for multi-parameter searches |
| (1) planetary surface compositions |
| (1) CMEs |
| (1) realistic stellar models for a larger range of stars |
| (1) UV SEDs |
| (1) predictions of direct imaging observations of Earth like exoplanets |
| (1) optical constants for organic aerosols |
| (1) database of exoplanet flux |
| (1) Zeeman doppler imaging |
| (1) high spectral resolution |
| (1) cross-referenced archival photometric data |
| (1) complete survey information |
| (1) oscillation modes |
| (1) X-ray SEDs |
| (1) continuum opacities |
| (1) list of planets in the HZ |
| (1) asteroseismology data |
| (1) elemental isotope abundances |
| (1) UV data |
| (1) high precision EUV data |
| (1) stellar magnetic activity |

**Table 6.1:** Similar to Table 3.2, the frequency (in parenthesis) of properties missing from current databases as indicated by a survey of interdisciplinary scientists (see Section 3). Those properties highlighted in yellow were also listed as essential to characterizing stellar and planetary systems (Section 3).

Some of this missing information was made more obvious when putting the responses from the interdisciplinary scientists, Appendix A, in context with those stellar and planetary properties they said were most important for their science (see Section 3) -- which have been highlighted in Table 6.1. For example, measuring both planetary mass and radius is crucial to properly classify and characterize a planet, yet -- at the time this report was written -- only ~20% of the planets within the NASA Exoplanet Archive have both properties measured. This is to say nothing of the associated uncertainty. In addition, there are currently only ~75 exoplanet host stars who have O, C, Fe, Ca, K, and Mg abundance measurements, only 7 of which also have P measurements -- which is necessary for life. While it's clear that these measurements are difficult to obtain, it's important for overall exoplanetary characterization that the community make a concerted effort to obtain these crucial properties for as many stellar and planetary systems as possible. **Stellar activity indicators were listed as being key for interdisciplinary science, but they are not available for a wide range of stars. The same is true for stellar elemental abundances for the major component of rocky planets (i.e. Fe, Mg, Si, Al, and Ca). It is also apparent that the measurement of organic aerosol and clouds, as well as lab/theoretical studies of exoplanet atmospheres, are important for the characterization of exoplanets.** From these responses, and in light of current/upcoming NASA and non-NASA missions, it is clear that there are a number of gaps in our current understanding of stellar and planetary systems that should be addressed as quickly as possible, in order to capitalize on upcoming mission data.

C. Data information useful for comprehension

An issue expressed by scientists on the SAG and those we communicated within our investigations was the **difficulty in quantitatively assessing the quality of data from a given survey or catalog.** For many science goals, it is necessary to identify and filter out low-quality data. It is also unclear how to compare uncertainties in measurements that utilized different approaches to estimate or characterize errors. Instead, it was suggested that, where available, summaries of the likelihood and/or the posterior distribution computed with a well-chosen interim prior with broad support would be useful for determining true uncertainties. In addition, complete information about the entire sample surveyed, as well as what observations were taken (or not taken) for each object and how it was decided to take those observations (namely target of interest vs. homogenous survey), would help reveal potential biases within the data. Overall, there is a wide variety of data and accompanying observing decisions that are important for users to understand archival measurements within a proper context. That information should be provided, in a consistent and comparable manner, whenever possible.

## 7. Conclusions

SAG22 performed a comprehensive study of the utility of an archive of stellar and planetary system properties focusing on current and future NASA missions that will be intensively observing specific systems for exoplanet studies. We conducted a broad survey of scientists that could benefit from such an archive to understand their needs, generated an exhaustive list of exoplanet-observing missions, examined the target lists of those missions, and studied the implementation of such an archive. The findings of this study also directly incorporate the suggestions and opinions from representatives of many current and upcoming NASA missions as well as personnel operating NASA archives. Overall, this report encapsulates the knowledge and experience from a wide range of stellar and exoplanet scientists, as well as their recommendations for data, properties, and tools, to ultimately create a useful target star archive.

**This study found that there is a large need for a centralized repository of curated, detailed stellar and planetary system data for a relatively small collection of hundreds of nearby FGKM stars**. It is clear that a target star archive would greatly benefit current and future NASA missions as they define their target lists and simulate future surveys, ultimately increasing the chances of mission success. Further, such an archive would be of use to a broad variety of interdisciplinary stellar and exoplanet scientists. It would support greater efficiency and accuracy of scientific analyses, and the curation of data would highlight any notable gaps in stellar/system parameters that could motivate additional observations.

This study and its appendices provide detailed findings of the stellar data desired by scientists and missions, methods for archive implementation that would enhance usability, and the population of high priority stars.

## Acknowledgements

We appreciate the contributions from the SAG 22 leads, members, speakers, and advisors in the creation of this report and hope that it serves as a useful starting point for future NASA considerations regarding archive needs for exoplanet science. The authors would like to acknowledge Lynne Hillenbrand as having worked with the group and participated in the process. Part of the research by JAB was carried out at the Jet Propulsion Laboratory, California Institute of Technology, under a contract with the National Aeronautics and Space Administration (NASA). JAC acknowledges PID2019–109522GB–C51. IP acknowledges partial support from a NASA/ADAP grant. DV gratefully acknowledges the support of the STFC via an Ernest Rutherford Fellowship (grant ST/P003850/1). The research shown here acknowledges use of the Hypatia Catalog Database, an online compilation of stellar abundance data as described in Hinkel et al. (2014, AJ, 148, 54), which was supported by NASA's Nexus for Exoplanet System Science (NExSS) research coordination network and the Vanderbilt Initiative in Data-Intensive Astrophysics (VIDA).

# Appendix A: Science Use Cases & Missions (TF3)

To view table online, please go to: https://tinyurl.com/SAG22-ScienceUseCasesMissions

| Discipline | Science Case | Parameter | Parameter (Short) | Ideal Precision | Minimum Precision | Wave-length Range | Doable? | Mission / Telescope / Resource (Required to make doable) | Non-NASA Missions (Required to make doable) | Supplemental (Info not provided by main mission) | Supports NASA Missions ? | Notes |
|---|---|---|---|---|---|---|---|---|---|---|---|---|
| Host Star | | | | | | | | | | | | |
| Host Star Activity | Stellar wind | Flare rate, energy distribution, CME rate/properties? | Stellar Activity, SED, CME | | | X-ray, FUV/NUV, optical, radio | Yes (radio, optical, UV, Xray) | HST -- targeted (UV, optical), TESS -- survey (optical), Alma, Sparcs, Cute | VLA | | Yes | |
| Host Star Characterization | stellar flare spectra and stellar UV irradiances | (1-2000A) | Stellar Spectra, Spectroscopy, UV | | | UV | Yes | HST -- targeted, Sparcs (small set of stars) | | | Yes | |
| Host Star Characterization | Observational Exoplanet Characterization | High resolution M-dwarf Stellar spectra and activity levels | Low-mass Stars, Spectroscopy, Stellar Activity | High, medium, low activity for early, mid, late M dwarfs would be a good start | < 10% | UV, optical, MIR | Yes | APOGEE (in the works) | CARMENES, maybe PLATO | | No | |
| Host Star Characterization | Trying to go from stars to planets to life, focusing on bio-essential elements that are not well measured in stars | Abundances of N, F, P, Cl, and K | N, F, P, Cl, K, Spectroscopy, Planetary Mass, Planetary Radius | < 0.1 dex | 0.2 dex | optical, NIR | No | | | | No | |
| Host Star Characterization | Activity | Effective temperature, spectral type, metallicity via spectroscopy | Effective Temperature, Metallicity, Spectroscopy | High-resolution (~100,000) | (>30,000) | optical | Yes | NASA ground-based surveys | | | Yes | |
| Host Star Characterization | Activity | Magnetic activity, magnetic field strength and/or topology (polarimetry required), activity cycles, correlations with activity indicators via spectroscopy | Stellar Magnetic Activity, Stellar Magnetic Field, Stellar Activity, Spectroscopy | High-resolution (~100,000) | (>30,000) | optical | Yes (except activity cycles) | TESS (except polarimetry) | | | Yes | Activity cycles will depend on lifetime of TESS relative to unconstrained M dwarf activity cycles duration |
| Host Star Characterization | Activity | Jitter / Granulation via spectroscopy | Stellar Jitter, Stellar Granulation, Spectroscopy | High-resolution (~100,000) | (>30,000) | optical | Yes | | Ground-based spectroscopy | | No | |
| Host Star Characterization | Activity | Flicker / Granulation via photometry | Stellar Flicker, Stellar Granulation, Photometry | High-resolution (~100,000) | (>30,000) | optical | Yes | TESS | | | Yes | Limited to the set of brightest stars |
| Host Star Characterization | Activity | Mass, radius, rotation period, age via photometry | Stellar Mass, Stellar Radius, Stellar Rotation Period, Age, Photometry, Effective Temperature, Stellar Mass, Luminosity | High-resolution (~100,000) | (>30,000) | optical | Yes | Gaia (distances, ~radii), TESS (rotation period), | | Model inferences | Yes | |
| Host Star Characterization | Activity | Radiation environment via spectrophotometry | Stellar Radiation, Spectrophotometry | | | X-ray, UV | Yes | Chandra, HST, XMM | | | Yes | |
| Host Star Characterization | Activity | Activity / activity cycles / activity indicators via spectrophotometry | Stellar Activity, Spectrophotometry | | | X-ray, UV | Yes | Chandra, HST, XMM | | | Yes | Same activity cycles issue as above |
| Host Star Characterization | Activity | Coronal Heating via spectrophotometry | Spectrophotometry | | | X-ray, UV | Yes | Chandra, HST, XMM | | | Yes | |
| Host Star Characterization | What is the amplitude and impact of EUV radiation and CMEs on habitable zone planets as a function of stellar mass and age | stellar 100-911 Ang emission and variability on flare, rotational, and evolutionary time scales. A secondary want (but not need) is B-field estimates for rocky exoplanets.<br><br>EUV: 100-911 Ang (+ FUV: 1020 - 1650 Ang as supporting data; important to note that FUV data is not a substitute for EUV data) | Stellar Variability, Stellar Rotational Period, Stellar Magnetic Field | EUV absolute flux accuracy of 10% (ideal), 50% (threshold). [ Note the state-of-the-art is somewhere between 300% and 10000%. ]. Spectral resolution of 0.5 Ang (ideal), 5 Ang (threshold). Temporal resolution of 20s (ideal), 300s (threshold). | | EUV, FUV | No | | | | No | Possible upcoming mission: ESCAPE |
| Host Star Characterization | Luminosity, Temperature evolution | L, T_eff, log(g), [Fe/H] etc. | Luminosity, Effective Temperature, Surface Gravity, Metallicity, Multiband Magnitudes, Stellar Radius, Asteroseismology, Age | Subject of future sensitivity analysis. Currently in progress | 10% ? subject of future study | optical | Yes | TESS (age via asteroseismology) | Gaia (Luminosity), PLATO (asteroseismology) | | Yes | Inferred from large sample |
| Host Star Characterization | Stellar/ Planetary Ages | Stellar rotation period | Stellar Rotation Period | <10 % | subject of future study | | Yes | TESS | | | Yes | Likely only for young stars |
| Host Star Characterization | Stellar/ Planetary Ages | Asteroseismic modes | Asteroseismology, Age | | | | Yes | TESS | PLATO | | Yes | Small number of nearby, main-sequence stars (hotter than G) apply |
| Host Star Characterization | Convective zone radii | from Asteroseismology/ fitting stellar models | Asteroseismology | Model dependent, models often give high precision estimates, but may be different from empirically determined measurements from asteroseismology | | | Yes | TESS | PLATO | | Yes | Few stars have precise enough measurements of asteroseismic modes to make this feasible |
| Host Star Characterization | UV evolution, stellar winds, CMEs | rotation periods, magnetic field strength and topology, stellar winds<br><br>Lyα in the UV. (Radio ~ 10s GHz) | Stellar Rotation Periods, Stellar Magnetic Field, Stellar Winds, Stellar Mass, Stellar Radius | <10% | <50% | optical, IR, radio, UV | Yes | HST (UV), TESS (rotation) | VLA, PLATO (solar age stars) | | Yes | Some properties are non-UV specific |
| Host Star Characterization | proxies for stellar ages | rotation periods, magnetic activity indicators (e.g. flare rates, H alpha emission , Ca HK, UV/X-ray flux, LIthium), 3D space velocities | Stellar Rotation Periods, Stellar Magnetic Activity, Ca H&K, UV flux, X-ray flux, Li, Space Velocities, Effective Temperature, Stellar Mass, Luminosity | Didn't have enough time to provide more details | | X-ray, UV, optical | Yes | TESS (flare rates), Keck and similar telescopes (Ca HK, H-alpha, Li), Chandra, XMM, HST (UV) | Gaia (kinematics) | | Yes | |
| Host Star Characterization | UV evolution, activity | Rotation periods, ages, Teff, EUV models, long-term monitoring of low-mass stars<br><br>0.1 A - 3000 A | Stellar Rotation Periods, Age, Effective Temperature, EUV models, Stellar Activity, Low-mass Stars, Effective Temperature, Stellar Mass, Luminosity | <10% | | UV | | | | | | Skipped in interest of time/repeat of previous lines |
| Host Star Characterization | Stellar gravity darkening | R, L, Teff, log(g), stellar oblateness | Stellar Radius, Luminosity, Effective Temperature, Surface Gravity, Stellar Oblateness | | | | Yes | TESS, Kepler, ground-based interferometry (CHARA) | | | Yes | Only transiting planets, very small numbers (~10) |
| Host Star Characterization | Use multistar, particularly true binaries/co-moving pairs, as a backdrop to better characterize the formation, dynamics, and evolution of exoplanet hosts and their planets. | high-resolution imaging, astrometry, Gaia | Images, Astrometry, Coordinates, Space Velocities | best typically values today are fine | | optical, IR | Yes | Keck AO, ground-based spectroscopy/RV | Gaia (astrometry), ALMA (non optical/IR) | | Yes | |
| Host Star Characterization | Spectropolarimetry | ZDI images and fB | Stellar Magnetic Activity, Zeeman Doppler Imaging | <10% | | optical | Yes | | HARPS and many other ground-based | | No | LUVOIR A polarimeter would help |
| Planet Detection | Circumbinary planets / EBs | Spectroscopic and photometric observations of the host binary systems, including precision radial velocity measurements, high-contrast imaging and archival time-series photometry. | RV, Images, Photometry, Stellar Mass, Stellar Radius, Time Variability | Spectroscopic observations with R ~ 5,000-10,000+ and archival photometry with timing uncertainty on the order of minutes. | Threshold spectroscopic resolution is R ~ 5,000, ideal would be ~20,000. Threshold timing precision on the archival photometry is ~10 min, ideal is ~1 min. | optical | Yes | Kepler, TESS, ground-based RV | PLATO | | | Future direct imaging could do binary planets |
| **Planet & System** | | | | | | | | | | | | |
| Disk Processes | Debris disk characterization | Habitability, Volatile delivery, time variability of debris disks, planetary system architectures and whether ours' is common, debris disk composition and dynamics, observation planning | SED, Planetary Spectra, Planetary Mass, Orbit, Stellar Radius, Inclination, Distance, Planetary Radius, Disk, Photometry, Luminosity, Time Variability | Stellar spectrum (UV-NIR), stellar variability (FUV-NIR), direct imaging spectrum of planet (UV-NIR), direct image of planetary system from 0.5-50 AU (VIS-IR), planet mass, planet orbit, high spatial-resolution multi-color images of faint debris disks from VIS-IR, stellar diameter, compact stellar binary information | | UV, optical, NIR, NUV, FUV | Yes | HST, Roman, JWST | Alma | | Yes | HabEx, LUVOIR... |
| | | | | (distance and dmag), system inclination, stellar inclination, distance to system, bolometric stellar luminosity | | | | | | | | |
| Disk Processes | Debris disk characterization | photometry | Photometry | 1-10% accuracy | | optical | Yes | Roman, HST | | | Yes | Future coronagraphs |
| Disk Processes | Lots of circumstellar disk stuff | Infrared excess and ideally a L_IR/L* value for a given disk | IR Luminosity, Luminosity, Infrared Excess | not sure | | MIR, FIR | Yes | JWST | Alma | | Yes | Really just for the massive disks |
| Disk Processes | Disk fractionation Processes | stellar ages | Age, Effective Temperature, Stellar Mass, Luminosity | <10% | | | | | | | | Only for a very small number of distant stars and to a very < 10% high precision, emailed 6/15, follow up 6/22 |
| Disk Processes | Disk fractionation Processes | chemical abundances (Mg/Si, C/O) | Disk Composition, C, O, Mg, Si | ~5%, 0.02 dex | | UV, optical, IR | Yes | | Alma (gas phase e.g. C/O) | | | How will Mg and Si measurements be made? Since it requires dust (and not gas) it was not clear how this could be measured -- emailed 6/15, follow-up 6/22 |
| Exoplanet Characterization | Exoplanet atmosphere characterization, contamination of stellar active regions to transmission spectra, exoplanet characterization in general. | Accurate model stellar spectra, especially in strong atomic lines like Na I D and the Balmer lines. Accurate planetary radii. | Stellar Spectra, Spectroscopy, Na, Planetary Radii | 10% | 10% | optical, NIR | Yes | TESS, Kepler, ground-based spectroscopy | Gaia (stellar params), ground-based SEDs, PLATO, CoRoT | | Yes | |
| Exoplanet Characterization | Planetary magnetic fields | high resolution optical spectra, more observations of planets/brown dwarfs at lower radio frequencies (< 100 MHz) | Low-mass Stars, Planetary Magnetic Fields | <10 mJy | | optical, radio | Yes and No | ground-based spectroscopy | LOFAR, SKA | | | Optical: Ca II K modulation from star-planet interaction (e.g. Cauley et al. 2019); Radio: **can only observe very strong planetary magnetic fields from Earth** (>= 10 MHz limit means planet must have >= 28 G magnetic field [Jupiter is 4G]) |
| Exoplanet Characterization | Detect and characterize the atmospheres of terrestrial exoplanets for status of habitability and presence of biosignatures | Stellar spectra and its time dependence, spot covering fractions and temperatures, stellar radii and mass (0.2 - 25 μm) | Stellar Spectra, Spectroscopy, Effective Temperature, Stellar Radius, Stellar Mass, Time Variability | For stellar spectra, 10% precision. For spot temperatures, +/- 100K. Masses < 20% | | FUV, NUV, optical, NIR, MIR | Yes | JWST, Pandora | | | Yes | |
| Exoplanet Characterization | Activity | Planetary companions and multiplicity via spectroscopy | Planetary Companions, Spectroscopy | High-resolution | | optical | Yes | NEID, Keck | | | Yes | |
| Planetary Formation | Formation & Evolution of Planetary Systems | Planet mass, orbital period & phase, eccentricity & pericenter direction; ideally inclination and ascending nodes. Host star bulk properties (e.g., density, mass, radius, temperature, abundances, kinematics) would be nice, too. | Planet Mass, Planet Period, Eccentricity, Pericenter, Inclination, Stellar Mass, Stellar Radius, Effective Temperature, Stellar Velocity, Abundances, Spectroscopy | Ideal is 0. :) In practice, ~1-2% for masses and 0.01 for eccentricity or 2\pi/100 for angles would be great. I realize that's probably not realistic. I really hope we can get to 10% mass, but could at least do some science since 20%. Eccentricities to 0.1 and inclinations to ~15 degrees would be usable. | | optical, NIR | Yes | NEID, Keck, TESS (subset) | Gaia, CHEOPS/PLATO (subset) | | Yes | |
| Planetary System Characterization | Planetesimal belts | Debris disk inner/outer radii, extents (wavelength best for big grains) | Disk Radii | sigma_Rdisk/Rdisk < 0.1 (inner and/or outer disk radius) | | sub-mm/mm | Yes | JWST | ALMA | | Yes | |
| Planetary System Characterization | Dynamical state/stability | Existing planet semi-major axes, eccentricities, masses | Planetary Companions, SMA, Eccentricity, Planetary Mass | <10% | | | Yes | TESS, RV surveys | CHEOPS, PLATO, Gaia | | Yes | LUVEx |
| **Exogeology** | | | | | | | | | | | | |
| Exogeology | To what extent might exoplanets be Earth-like? | Abundances of the "Major elements" (all on a given object), Si, Ti, Al, Fe, Mn, Mg, Ca, Na, K, Cr, P, S, O, plus key trace elements like U and Th | Si, Ti, Al, Fe, Mn, Mg, Ca, Na, K, Cr, P, S, O, U, Th | Comparable to estimates of Solar Photosphere | | UV, optical, IR | No | | | | | |
| Exogeology | diversity of planets as a | host star age and mass | Age, Stellar Mass, Effective | 1% | 10% | UV, optical, IR | Yes | TESS | CHEOPS, PLATO, Gaia | | Yes | Age is particularly hard -- easier in a moving |
| | function of host star properties | | Temperature, Luminosity | | | | | | | | | group or with asteroseismology |
| Exogeology | Planetary system architectures | planetesimal belts (mass and location), planets (masses and orbits) | Disk Mass, Disk Radii, Planetary Mass, Orbit | a few % | 10% | optical, IR, mm | Yes | JWST, TESS | CHEOPS, PLATO, Gaia | | Yes | See line 38 and 39 |
| Exogeology | What are the compositions of rocky exoplanets? How well does a stellar composition translate into a rocky planet composition? What are the consequences of variable composition on rocky planet evolution and habitability? | Stellar abundances for at least Mg, Fe and Si are key for understanding the bulk composition of a planet. Al and Ca are vital for understanding the melting behavior at the surface. C and O tell us about the oxygen fugacity during planet formation which can affect the chemistry of the planet's core and mantle. U, Th and K all power the interior dynamics of planets.<br><br>In addition to stellar abundances, stellar age is very important for modeling planets to know whether they contain sufficient heat to produce volcanism. | Mg, Fe, Si, Al, Ca, C, O, U, Th, K, Age, Planetary Mass, Planetary Radius, Effective Temperature, Stellar Mass, Luminosity | In some sense, having any abundances are useful. In terms of precision, factors of 10% are probably most useful. Even 50% can be used to explore end-member cases, but lower is better. For age, ~2 Gyr precision is more than enough. | | UV, optical, IR | Yes for some abundances (not U, Th) | ground-based spectroscopy | PLATO ground-based abundances | | Yes | Checked out paper and it's for Jupiter atmospheres -- not rocky planets or interior compositions. https://arxiv.org/pdf/2005.02528.pdf |
| Exogeology | Interior planets | Stellar Masses, | Stellar Mass | 1-2% | 1% | | | | | | | See 42 |
| Exogeology | Interior planets | Stellar Ages | Age, Effective Temperature, Stellar Mass, Luminosity | ~100 Myr | 2% ? | | | | | | | See 42 |
| Exogeology | Determine the surface compositions and possible volcanic volatile contribution to secondary atmospheres | mass, orbital distance, surface T, atmospheric composition of planet; composition (metals) of host star | Planetary Mass, Orbit, Surface Temperature, Atmospheric Composition, Spectroscopy, O, Si, Fe, Mg, Ca, Al, Na, K, Ti, Cr, Mn, Ca, Ni, S, Cl, F, S, Planetary Spectra | for atmosphere composition, right now a yes/no is ok | | UV, optical, IR | Yes | RV survey, TESS, ground-based spectroscopy | Gaia | | Yes | LUVOIR |
| Exogeology | composition/redox state of atmosphere | Abundance of two species each for the elements C, H, O, P, S. Ideally the same for any other major constituents.<br><br>0.2-25 um, but can definitely get away with subset of this | C, H, O, P, S, Isotopes, Spectroscopy, Planetary Spectra | OOM measurement of at least two species for each element. | upper/lower limits for at least one species for each element, and for two+ for at least one element. | UV, optical, IR | Yes | JWST, Roman?, ARIEL+CASE | ARIEL | | Yes | |
| Exogeology | Characterize the abundances of host stars (FGK-types and M-dwarfs) and their rocky planets | Abundances of: Fe, Mg, Si, Ca, Al, K, P, C, O<br>Mass and radius of planets | Fe, Mg, Si, Ca, Al, K, P, C, O, Planetary Mass, Planetary Radius, Spectroscopy, Low-mass Stars | < 0.1 dex, < 10% | 0.2 dex, < 50% | optical, NIR | Yes for FGK-stars<br>No for M-dwarfs | ground based spectroscopy, TESS | CHEOPS, PLATO, Gaia | | Yes | |
| Exogeology | interior of rocky/icy objects; giant planet moons; habitability | interior structure (composition, density, plate tectonics), heat flow; dynamics (planetesimal delivery) | Planetary Interior Composition, Planetary Mass, Planetary Radius, Plate Tectonics, Planetary Heat Flow, Planetary Dynamics | none specified | | optical, IR | No | | | | | |
| **Habitability & Astrobiology** | | | | | | | | | | | | |
| Aerosols | super-Earths / mini-Neptunes | planetary mass, radius, temperature. atmospheric composition, bulk metallicity. stellar R, T, SED | Planetary Mass, Planetary Radius, Temperature, Atmospheric Composition, Metallicity, Stellar Radius, Stellar Effective Temperature, SED, Spectroscopy, Planetary Spectra | varies per planet, but 10 ppm for planetary atmospheres. better than 20% for planetary masses | | UV, optical, IR | Yes | JWST, Roman?<br>ARIEL+CASE, HST | | | Yes | |
| Astrobiology | Long-term habitability and biosignature emergence | Age, stellar evolutionary tracks, atmospheric models/ habitable zone formulations | Age, Evolutionary Tracks, Atmospheric Composition, Habitable Zone, Effective Temperature, Stellar Mass, Luminosity | see each of these properties individually below | | optical, IR | Yes | JWST, ground-based transits | | | Yes | Only for some specific systems, probably better for super-Earths |
| Astrobiology | Broadly I'm interested in life detection. More specifically, I'm interested in characterizing gaseous and surface biosignatures and interpreting them within the context of their environment. Because I'm a biologist, I focus on different elements of habitability rather than just the presence of liquid water. These include assessing energy sources, availability of elements (CHNOPS), clement conditions. | Because I'm interested in photosynthetic biospheres, I'd ideally like information on the light quality (wavelengths available) and quantity (photon flux to the top of the atmosphere as well at the surface of the planet). Knowing what wavelengths and flux are available will help constrain the possibility of photosynthetic biospheres, and their gaseous products (e.g., O2). For example, see Lehmer et al. 2018). This can potentially help select targets for e.g., JWST, especially when targeting O2 as a biosignature on M dwarf exoplanets. | Instellation, Atmospheric Composition, SED, SMA, Eccentricity, Low-mass Stars, Planetary Spectra | Higher resolution wavelength data (rather than broadband) is ideal...20-40 nm? | | UV, optical, NIR | Yes | JWST, TESS or ground-based RV, HST, Chandra | | | Yes | |
| Astrobiology | What are the abundances of bioessential and rock-forming elements in stars? How do those elements get distributed into planets? Can habitability be related to stellar abundances in a meaningful way? | Surface temperature, presence of water or water/ water vapor, day/night cycles, seasons (orbital tilt basically), molar ratios of elements to compare with chemical reactions such as: N, P, K, Ca, C, S, Gases in the atmosphere (CO2 mainly), Evidence for Geodynamics (plate tectonics or or some form of dynamism? Magnetic field?) | Surface Temperature, Stellar Abundances, Atmospheric Composition, Planet Rotation, Obliquity, Planet Magnetic Field, Plate Tectonics, N, P, K, Ca, C, S, Spectroscopy, Planetary Mass, Planetary Radius, Planetary Spectra | Stellar and atmospheric composition to within 15%, planet parameters to within 10% | | UV, optical, IR | Yes for some parameters Not sure about geodynamics | ground-based abundances, JWST, ARIEL+CASE, TESS, HST | CHEOPS, PLATO | | Yes | |
| Astrobiology | What are the compositional parameters of different worlds that may affect habitability? | Chemical species in the atmospheres of planets would be a great start. Relative abundances of any species, even better.<br><br>(To know the abundances of a species within a factor of 2 would be ideal. Knowing the abundances within an order of magnitude, one can still learn a lot.) | Atmospheric Composition, Spectroscopy, Planetary Mass, Planetary Radius, Planetary Spectra | 50% | | | Yes | JWST, ARIEL+CASE, HST | | | Yes | |
| Atmospheres | How much H2 can be generated through Fe photo-oxidation on anoxic planets with oceans like the Archean Earth? Astrobiology, photochemistry, exoplanet ecosystems | UV-Vis Spectra, whether exoplanets have oceans and are anoxic. (180-400 nm) | Habitability, Planetary Spectra, Spectroscopy, Planetary Mass, Planetary Radius | Photon flux of exoplanets as seen from their star | | UV, optical | Yes? | | | | yes? | LUVOIR-A, -B, HabEx |
| Atmospheres | atmospheric composition (gases, aerosols, and ices) | remote sensing spectroscopy of exoplanetary atmospheric gases and suspended particulates (aerosols and ices) | Atmospheric Composition, Planetary Spectra, Spectroscopy, Planetary Mass, Planetary Radius | | | submm, Radio, IR | Yes | JWST, ARIEL+CASE | | | Yes | |
| Atmospheres | Carbon isotope ratios | CO observations | C, O, Isotopes, Spectroscopy, Planetary Spectra | R ~ 100,000 | R ~ 50,000 | NIR | Yes | JWST, ground-based spectroscopy | | | Yes | |
| Atmospheres | organic aerosol and clouds | experimental studies on compositionally diverse organic aerosol and ice analogs to 1) determine their optical constants and 2) determine low temperature saturation vapor pressures | Aerosol Optical Properties, Saturation Vapor Pressure | | | UV, optical, IR, submm | Yes | individual NASA program calls | | | Yes | highlights a gap in what we know and what we want to know -- because there are no dedicated programs to answer these questions, which are crucial for understanding exoplanets |
| Atmospheres | Atmospheric retrievals | Stellar Effective Temperature | Stellar Effective Temperature | <100 K | <200K | NIR, MIR | | | | | | Follow up - emailed 6/15, follow-up 6/22 |
| Atmospheres | Atmospheric retrievals | absolute stellar fluxes | Stellar Flux | <10% | | NIR, MIR | | | | | | Follow up - emailed 6/15, follow-up 6/22 |
| Atmospheres | Habitability & Biosignatures | C, N, O | C, N, O, Spectroscopy, Planetary Spectra | 0.01 dex | 0.1 dex | optical | Yes? | | | | Yes? | LUVOIR |
| Atmospheres | Photochemistry | SED/UV data, UV range (1216 - 400 nm) | SED | | <1 nm resolution | UV | Yes | HST | | | Yes | |
| Atmospheres | Laboratory and theoretical studies of exoplanet atmospheres | Collisional parameters for a variety of atmospheric types (e.g., CO2, CH4, H2O, H2/He dominated); molecular opacity data at high spectral resolution; photoabsorption/photodissociation/reaction rates at exoplanet-like temperatures; optical properties of particulates with a wide range of compositions (a few %) | Molecular Collisional Parameters, Molecular Opacity Data, Photochemical Reaction Rates, Aerosol Optical Properties, CO2, CH4, H2O, H2, He | 1-5% | accuracy | UV, optical, IR | Yes | individual NASA program calls | | | | highlights a gap in what we know and what we want to know -- because there are no dedicated programs to answer these questions, which are crucial for understanding exoplanets |
| Habitability | What are the habitability conditions? How life has emerged on earth? | The ultimate outputs are chemical abundances to investigate the possibility of biogenic origin of chemical disequilibrium.<br><br>In general investigating molecules is more relevant to the mentioned questions than studying atoms/ions (although these are also relevant). Therefore, NIR and IR wavelengths would be more suitable for such studies. In addition, temperate planets should have a peak emission intensity in IR, which makes targeting these wavelengths more feasible. | Planetary Surface Composition, Planetary Interior Composition, Planetary Dynamics, Tectonics, Effective Temperature, Planetary Mass, Planetary Radius, Planetary Distance, Spectroscopy, Planetary Spectra | probably 1e-7 or lower in terms of relative abundances, but translating this into observables (e.g. ppm or S/N in transmission or emission spectra) would be highly non-linear and case dependent | | IR | Yes for some parameters No for others (direct planetary interior composition, dynamics, tectonics, etc) | JWST, TESS, ARIEL+CASE | | | Yes | |
| Habitability | i) Habitability, and ii) Inhabitance. (i) requires a surface liquid, moderate surface temperature ('moderate' defined by the liquid), and stable surface conditions. Note that the preferred liquid is water, but other substances liquid at 200-500K (CO2, H2SO4, sulfur, maybe others?) are of interest. Detection of (ii) will be by 'unexpected' gases in the atmosphere, for | Atmospheric density, T/P profile of atmosphere and hence surface Temp and pressure, planetary elemental abundances, ideally crustal abundances, atmospheric gas composition, including trace gases (CF methane on Mars, phosphine on Venus etc). To a limited extent, stellar flux at surface (all wavelengths), particle/radiation flux.<br><br>Wavelength: I don't know - others might, but the technology to detect planetary surface composition at 10 parsecs has | Planetary Atmospheric Density, Planetary Atmospheric Temperature-Pressure Profile, Planetary Surface Temperature, Planetary Surface Pressure, Planetary Surface Elemental Abundances, Planetary Atmospheric Composition, Stellar Flux, SED, Spectroscopy, Planetary Spectra | For surface temp, +/- 10K would be great, +/- 30K useful. Pressure depends entirely on what the pressure is (on Mar 1 millibar is significant, on Venus it is not etc). Atmospheric composition - down to ppm level good, ppt level ideal but scifi with any planned mission. Surface liquid can be inferred, but measurement (glint? polarization?) would be cool. | | optical, IR | Yes for some parameters Small sample for others (T/P profile, direct planet compositions) | JWST | | | Yes | |
| | which you need to be able to model what the expected gases are, and then detect what is there.. | not been invented yet anyway. | | | | | | | | | | |
| Habitability | SETI | ages, habitable zones, infrared excess, precise coordinates | Age, Evolutionary Tracks, Atmospheric Composition, Habitable Zone, Effective Temperature, Stellar Mass, Luminosity, Planetary Spectra | | | optical, IR, Radio | Yes | TESS, JWST | | | Yes | |
| Habitability | planetary habitability | larger coverage SEDs and surface albedo spectra | Stellar SED, Surface Albedo, Planetary Spectra | | | | Yes | JWST, HST, Chandra | | | Yes | LUVOIR |
| Habitability | surface modeling | hi-res UV data, mostly of M stars because of their variability (2-5 um) | Stellar UV, Low-mass Stars | Down to 120 nm | | UV | Yes | HST, UVscope | | | Yes | Follow up - emailed 6/15, follow up 6/22 |

## Appendix B: Target List (TF2)

To directly download, please go to: https://tinyurl.com/SAG22-TargetList

Columns: 2MASS_name, target_lists/ARIEL_targets_2MASS_matches.csv, target_lists/cute_target_list_with2mass.csv, target_lists/EPRV-CARMENES_GTOtargets_with2mass.csv, target_lists/EPRV-CPS_targets_with2mass-new.csv, target_lists/EPRV-ESPRESSO_GTO_targets_with2mass.csv, target_lists/EPRV-MAROON-X_targets_with2mass.csv, target_lists/EPRV-NEID_EarthTwinSurvey_Gupta2021_Table1_with2mass.csv, target_lists/EPRV-NEID_GTO_targets_with2mass.csv, target_lists/gpies_target_list_with2mass.csv, target_lists/habex_with2mass.csv, target_lists/jwst_target_list_with2mass.csv, target_lists/luvoirA_with2mass.csv, target_lists/luvoirB_with2mass.csv, target_lists/ost_exoEarths_with2mass.csv, target_lists/roman_CGI_blind_search_with2mass.csv, target_lists/roman_CGI_knownplanets_with2mass.csv, target_lists/roman_cgi_knowndisks_with2mass.csv, target_lists/sphere_shine_target_list_with2mass.csv, TIC, HIP, TYC, UCAC4, _2MASS, OBJID, WISEA, GAIA, APASS, KIC, RAJ2000, DEJ2000, GLON, GLAT, ELON, ELAT, VMAG, TMAG, JMAG, KMAG, TEFF, LUM, PLX, DIST, PMRA, PMDE, RAD, MASS

Example data (3000 stars in total):

08523579+2819509,1,0,0,1,1,0,0,1,0,1,1,1,1,0,0,1,0,0,332064670,43587,1949-02012-1,592-044667,08523579+2819509,-1,,2776704,-1,-1,133.14921
,28.330821,196.79465,37.698898,127.73245,10.367099,5.95100,5.20600,4.76800,4.01500,5250.00,0.63540000,79.427400,12.585500,-485.87200,-2
33.65100,0.96400000,0.900000

09061775+1924080,1,0,0,1,0,0,0,0,0,0,0,0,0,0,0,0,0,0,243244680,-1,1404-01186-1,548-047168,09061775+1924080,-1,J090617.78+192407.6,50048
0,43603131,-1,136.57399,19.402236,208.78176,38.127630,133.29634,2.6867598,10.1540,9.66900,9.14400,8.83700,6128.00,2.3623400,5.4753000,
181.69000,45.492000,-41.791000,1.3640000,1.15800

05154075+1616435,1,0,0,0,0,0,0,0,0,0,0,0,0,0,0,0,0,0,411839167,-1,1283-00739-1,532-012276,05154075+1616435,-2138635738,J051540.75+1616
43.3,345216,43006541,-1,78.919807,16.278755,186.97526,-12.640859,79.294237,-6.7496933,11.2370,10.8240,10.1990,9.89300,6513.00,4.9042600
,2.6639000,372.63200,2.7170000,-12.983000,1.7390000,1.33800

## Appendix C: Archive Implementation Request

| Requests | Example | Need | Want | Idea |
|---|---|---|---|---|
| Ability to resolve stellar names or coordinates into a specific object (or list of objects) | Simbad, vizier | X | | |
| Ability to distinguish between observed and derived stellar parameters | | X | | |
| Provide access to underlying data used to determine stellar parameters (1D spectrum, wavelength solutions, etc). Not necessarily hosting, but would provide a link and instructions on how to access | ExoFOP | | X | |
| Provide tools to derive parameters from the underlying data stored on the website | EFv2 on the Exoplanet Archive | | | X |
| Full catalog download -- access to existing stellar catalogs from their home site, flag to let you know that you can download the whole thing if desired | LAMOST, APOGEE, GALAH | | X | |
| Download a significant chosen fractions of a given catalog, e.g. Get the distance and metallicity for all of the stars but not all of the additional parameters. Make this possible via web interface (check boxes) *and* scripting (SQL, python, etc) | Vizier | X | | |
| Tutorials on how to construct queries and a large variety of use examples. Potentially a sub-area where users can upload examples of their own use cases that can be upvoted for relevance/success. Maybe a shared wiki or github repository | Gaia archive | X | | |
| Quick links or zipped packages of reduced data products for each star classified by data type (e.g., stellar spectra, images of planetary scene, etc.) | ExoFOP, though it doesn't sort by data type | | X | |
| Web-based interface for searching ranges of parameters (not code-based) that supports both tabular and plotted results -- preferably on the same page | Vizier | X | | |
| Plotting capabilities: plotting stellar parameters from a search, plotting data products (light curves, spectra, etc) | filtergraph | X | | |
| Ability to filter by stars with known or confirmed planets. | Exoplanet Archive, Hypatia, Vizier -- "targets with | X | | |
| | children" | | | |
| Summary/overview of planetary system for each star (planets & basic parameters, disk location, etc.) | Exoplanet Archive | X | | |
| Ability to rederive planetary parameters using different stellar/planet parameter solutions (M star, Rp/Rstar, Teff, etc) | | | X | |
| Some kind of planet ephemeris info, e.g., a link to a transit finder tool for observers doing followup obs? | https://astro.swarthmore.edu/transits/transits.cgi | | X | |
| Ability to search by planetary parameters as well as stellar parameters (e.g., select planets of a given radius range) | Exoplanet Archive | X | | |
| A list of non-detections would help with the future mission plannings, maybe with some info on the completeness of HZ coverage in the observations, etc | No existing example in an archive, but some catalog papers provide this (e.g. Rosenthal+ 2021) | | | X |
| having the ability to retrieve all published values of a given target's stellar or planetary parameters, include tools for determining mean/median/etc of a given parameter. Have ability to remove certain entries before computing mean/median | Hypatia Catalog | X | | |
| Easy way to differentiate data sources, or to filter by specific source catalogs. | Hypatia Catalog | X | | |
| Ability to see/plot parameters from different sources (papers/catalogues/etc) for a given star, to look for agreement or lack thereof | Exoplanet Archive | X | | |
| Flag to check for consistency among different solutions for individual parameters for a given target (e.g. are all the Teff values within X-sigma?) | use of flags in Exoplanet archive, but not this exact implementation | | X | |
| Keep track of author notes (e.g. priors used) | basic version is Vizier's table notes | | | X |
| User account interface to save queries, or ability to export a setting-specific code to easily recreate searches in the future. | Simbad results, Gaia, Exoplanet Archive, etc | X | | |
| Ability to select a consistent set of stellar and planetary parameters (to allow for population studies, want stellar and planetary parameters that are consistent with one another) | Exoplanet Archive | X | | |
| Make connections to other databases (whether a star has a TESS lightcurve, X-ray detections, additional paper references, in telescope archives) | MAST for observational data, simbad/vizier for literature references | X | | |
| Easy to implement multiparameter searches (can specify multiple ranges, criteria, etc -- planet radii and stellar Teff and orbital period) | Exoplanet Archive, MAST, most catalogs | X | | |
| One-stop-shop containing links to all the publicly available time series photometry | AAVSO | | X | |
| One-stop-shop containing all the publicly available broadband photometry including references/sources | Simbad, though often not for the bright stars we're interested in | X | | |
| Easy to access list of planets that were in the habitable zone (with the ability to choose from a list of HZ definitions: conservative, optimistic, etc) to make it easier to access some of the most important planets without too much cross-correlation. | Habitable Zone Gallery | | X | |
| Provenance of entries that are filterable (require references -- need to be able to track where values in the archive come from) | Exoplanet Archive | X | | |
| Accurate summaries of the likelihood and/or the posterior distribution computed with a well-chosen interim prior with broad support. | has not yet been implemented | | | X |
| Complete information about the entire survey, and not just the targets with detections. That means we need information (ideally nearly uniform or at least homogeneous) about the entire sample surveyed, as well as what observations were taken for each object and _how_ it was decided to take those observations (e.g., part of a homogeneous survey versus extra observations because the target was interesting) | has not yet been implemented, beyond the scope of this archive | | | X |
| Transparency regarding the decisions that were made within the database (including data or not, including stars or not, including references or not, etc), either in general or for individual stars. | | | X | |
| As much cross-correlated target (star and planet) data as possible [e.g. cross matching survey IDs and show all data for the star across all surveys, should include the cross match methodology] | Simbad, Vizier, Gaia | X | | |
| Compile/link to publicly available, archival Low-resolution and High-resolution (adaptive optics or speckle) images | ExoFOP, DIVA | | X | |
| Allow for multiple discontinuous ranges within a single search parameter (ex. 0.5-0.7 AND 1.2-1.6 solar masses) so you don't need to make multiple searches. | Exoplanet Archive | X | | |
| Allow for in-engine calculations using multiple parameters (ex. calculate equilibrium temp given certain parameters). Make your own column functionality, where you can define your own equation that pulls in values from the archive and report/plot results | Filtergraph, IRSA | | X | |

# Appendix D: Existing Catalogs (TF4)

To view the entire multi-page spreadsheet online, please go to: https://tinyurl.com/SAG22-ExistingCatalogs

## Survey/Catalog Resources

| | Description | Description of Stellar Parameters | Wave length | # of stars | Max dist | Observational method (spec v. phot) | Magnitude range | Spectral type | SNR parameters | Sky coverage | Link to catalog | Link to seminal paper | Spectral resolution | Abundances (Fe/H, M/H, alpha/Fe, etc) | Year of most recent data release (live vs. finalized) | Interface to data (Simbad, MAST, vizier, SQL query, csv download, etc) | Link to info on how to query catalog |
|---|---|---|---|---|---|---|---|---|---|---|---|---|---|---|---|---|---|
| LAMOST DR6 | Low-resolution optical spectra for nearly 6 million AFGK stars | LAMOST stellar parameter pipeline: https://ui.adsabs.harvard.edu/abs/2015MNRAS.448..822X/abstract **T_eff**: min=3460K, max=8500K, med=5600K, med_err=85K, med_frac_err=1.5% **log(g)**: min=-0.162, max=4.9, med=4.192, med_err=0.139, med_frac_err=3.6% **[Fe/H]**: min=-2.5, max=1, med=-0.202, med_err=0.082, med_frac_err=37% | 3690-9100 A | 5.77 million | Magnitude limited | Spectroscopic | down to r~19 in single exposure | AFGK | SNR > 10 in g and i bands | -10 to 60 degrees dec | http://dr6.lamost.org/catalogue | https://ui.adsabs.harvard.edu/abs/2012RAA....12.1197C/abstract | ~1800 | [Fe/H] | 2020, live | Search interface or full fits/csv catalog download | Search: http://dr6.lamost.org/v2/search Download full catalog: http://dr6.lamost.org/v2/catalogue |
| LAMOST DR6 | Low-resolution optical spectra for M stars | LAMOST stellar parameter pipeline: https://ui.adsabs.harvard.edu/abs/2015MNRAS.448..822X/abstract **T_eff**: min=2300K, max=4300K, med=3721K, med_err=88K, med_frac_err=2.39% **log(g)**: min=0, max=6, med=5, med_err=0.145, med_frac_err=2.94% **[m/H]**: min=-2.5, max=0.5, med=-0.323, med_err=0.1053, med_frac_err=34% | 3690-9100 A | 635,285 | Magnitude limited | Spectroscopic | down to r~19 in single exposure | M | SNR > 10 in g and i bands | -10 to 60 degrees dec | http://dr6.lamost.org/catalogue | https://ui.adsabs.harvard.edu/abs/2019ApJS..244...8Z/abstract | ~1800 | [m/H], TiO, CaH, CaOH, H-alpha, and Na spectroscopic indices (not abundances) | 2020, live | Search interface or full fits/csv catalog download | Search: http://dr6.lamost.org/v2/search Download full catalog: http://dr6.lamost.org/v2/catalogue |
| LAMOST DR6 Medium Resolution | Medium-resolution optical spectra for 341,631 stars | LAMOST stellar parameter pipeline: https://ui.adsabs.harvard.edu/abs/2015MNRAS.448..822X/abstract **T_eff**: min=3100K, max=8500K, med=5626K, | Blue 4950-5350 A & Red 6300-6800 A | 341,631 | Magnitude limited | Spectroscopic | Down to G=15 | AFGKM | SNR > 10 | -10 to 60 degrees dec | http://dr6.lamost.org/catalogue | https://ui.adsabs.harvard.edu/abs/2020arXiv200507210L/abstract | ~7500 | [Fe/H], [α/Fe], Elemental abundances relative to iron: C, N, O, Mg, Al, Si, S, | 2020, live | Search interface or full fits/csv catalog download | Search: http://dr6.lamost.org/v2/search Download full catalog: http://dr6.lamost.org/v2/catalogue |
| | | med_err=45K, med_frac_err=0.8% **log(g)**: min=-0.232, max=5.043, med=4.058, med_err=0.052, med_frac_err=1.33% **[Fe/H]**: min=-2.5, max=1, med=-0.167, med_err=0.031, med_frac_err=16.9% **RV (km/s)**: min=-589, max=604, med=-2.92, med_err=1.36, med_frac_err=6.28% | | | | | | | | | | | | Ca, Ti, Cr, Ni, Cu | | | |
| LAMOST DR5 Abundances | Abundances for 16 elements for 6 million stars | https://ui.adsabs.harvard.edu/abs/2019ApJS..245...34X/abstract "for S/Npixel ≥ 50, the typical internal abundance precision is 0.03-0.1 dex for the majority of these elements, with 0.2-0.3 dex for Cu and Ba, and the internal precision of T eff and log g is better than 30 K and 0.07 dex, respectively. Abundance systematics at the ~0.1 dex level are present in these estimates but are inherited from the high-resolution surveys' training labels" | 3690-9100 A | 6 million | Magnitude limited | Spectroscopic | down to r~19 in single exposure | AFGKM | SNR > 10 in g and i bands | -10 to 60 degrees dec | http://dr5.lamost.org/v3/doc/vac | https://ui.adsabs.harvard.edu/abs/2019ApJS..245..34X/abstract | ~1800 | C, N, O, Na, Mg, Al, Si, Ca, Ti, Cr, Mn, Fe, Co, Ni, Cu, and Ba | 2019, final | fits file | Download full catalog: http://dr5.lamost.org/v3/doc/vac |
| APOGEE | Near-IR spectroscopic catalog of stellar parameters and abundances from SDSS/APO. Based on whole-spectrum fitting with ASPCAP, which works best for stars with 4000 < T < 5500 K. | APOGEE dr16 https://ui.adsabs.harvard.edu/abs/2020AJ....160..120J/abstract Uncertainties as functions of TEFF, S/N and [M/H]. Fitting models are split between giant and dwarf pops, here are catalog stats: **T_eff:** min=3100K, max=9800K, med=4768K, med_err=96K, med_frac_err=2.02% **log(g):** min=-0.58, max=6.54, med=2.69, med_err=0.015, med_frac_err=2.31% **[m/H]:** min=-2.47, max=0.84, med=-0.16, med_err=0.0046, med_frac_err=6.27% **RV (km/s):** min=0.0, max=±1660, med=±30.2, med_err=0.00016, med_frac_err=0.06% abundances uncertainties varied. See here | 1.51-1.70 µm | > 430,000 | Magnitude-limited. | Spectroscopy w/ APO. | Depends on number of visits, with single-visit targets between 7.0 < H < 11.0 and 24-visit targets between 12.8 < H < 13.8. | Primarily red giants, with ancillary programs for KOIs, K2, M Dwarfs, and more. | Typically SNR > 100. | full sky | https://www.sdss.org/dr16/irspec/spectro_data/ | APOGEE Overview; DR16 ASPCAP | ~ 22,500 | [M/H], [Fe/H], [alpha/M]. Elemental abundances relative to iron: C, Cl, N, O, Na, Mg, Al, Si, P, S, K, Ca, Ti, Ti II, V, Cr, Mn, Co, Ni, Cu. Also have Ge, Rb, Ce, Nd, Yb for very few (if any) stars. | dr16 12/2019, dr17 06/2021 | SQL/CasJobs, full .fits file download | Data Access Overview (contains use-case instructions, tools and tutorials): https://www.sdss.org/dr16/data_access/ Download parameter summary catalog allStar (dr16): https://www.sdss.org/dr16/irspec/spectro_data/ SDSS CasJobs: https://skyserver.sdss.org/casjobs/ |
| GALAH DR3 | Southern sky high-resolution spectroscopic catalog | Use Spectroscopy Made Easy for parameter and abundance info: https://docs.datacentral.org.au/galah/dr3/data-reduction-and-analysis/<br>**T_eff**: min=3000K, max=7933K, med=5542K, med_err=96K, med_frac_err=1.8%<br>**log(g)**: min=-0.45, max=5.18, med=3.985, med_err=0.192, med_frac_err=4.8%<br>**[Fe/H]**: min=-4.78, max=1.0, med=-0.161, med_err=0.088, med_frac_err=24% | 471-789 nm | 588,571 | ~31179pc | Spectroscopic | 12 < V < 14 | Mostly FGK | Typically SNR > 25 in at least one channel | \|b\| > 10, dec<+10 | https://docs.datacentral.org.au/galah/ | https://ui.adsabs.harvard.edu/abs/2018MNRAS.478.4513B/abstract | ~28,000 | Li, C, O, Na, Mg, Al, Si, K, Ca, Sc, Ti, V, Cr, Mn, Fe, Co, Ni, Cu, Zn, Rb, Sr, Y, Zr, Ru, Ba, La, Ce, Nd, and Eu | 2020, live | Download full catalog or SQL/ADQL query | Download catalog: https://cloud.datacentral.org.au/teamdata/GALAH/public/GALAH_DR3/<br>SQL/ADQL query: https://datacentral.org.au/services/query/ with sample queries (from DR2): https://docs.datacentral.org.au/galah/dr2/sample-queries/ |
| Gaia DR2 | Astrometry and photometry of 1.3 billion stars with ESA's Gaia spacecraft. | Gaia DR2: https://gea.esac.esa.int/archive/documentation/GDR2/<br>**Parallax:** number = 1.3 billion, uncertainty = 0.02-0.04mas (G<15) - 2.0ma (G =21)<br>**T_eff:** number = 161 million, min = 3,000 K, max = 10,000 K, uncertainty = 324 K<br>**Radius:** number = 77 million, uncertainty = 10%<br>**Luminosity:** number = 77 million, uncertainty = 15% | G, G_BP, G_RP | Sky position + parallax + proper motions: > 1.3 billion; RVs: > 7.2 million | Magnitude-limited. | Astrometry (parallax); photometry (T_eff, radius, luminosity). | Sky position + parallax + proper motions: 3 < G < 21 (most complete for 12 < G < 17); RVs: 4 < G < 13. | all (RVs only for 3550 < Teff < 6900) | N/A (?) | full sky | http://cdn.gea.esac.esa.int/Gaia/gdr2/ | Gaia mission; Gaia DR2 | N/A | N/A | 04/2018, final. | ADQL, MAST, CasJobs. Full download impractical, but partial download possible. | ADQL: https://gea.esac.esa.int/archive/<br>MAST: https://mast.stsci.edu/portal/Mashup/Clients/Mast/Portal.html<br>CasJobs: http://mastweb.stsci.edu/mcasjobs/default.aspx<br>Downloads: http://cdn.gea.esac.esa.int/Gaia/<br>Gaia Archive FAQ: https://www.cosmos.esa.int/web/gaia-users/archive/faq<br>Gaia Data Access Tutorials: https://www.cosmos.esa.int/web/gaia-users/archive |
| Gaia DR3 | Improve astrometry and photometry. Release splits in EDR3 (Dec 3, 2020) and full DR3 (in 2022). Uses improved RUWE to assess multiplicity and spurious sources (Alessandro Sozetti) | Improved astrometric accuracy as compared with DR2. Improved T_eff, radii only in full DR3.<br>**parallaxes** improve by 20%, **proper motions** by a factor of 2. | optical | 1.5 billion + 0.3 billion without full astrometric solution | magnitude-limited | astrometry, photometry | G: 3-21 mag | all | n.a. | full sky | Dec 3, 2020 | Dec 3, 2020 | n.a. | n.a. | 2020 | TAP or ADQL query, VizieR query. Full download impractical, but partial download possible. | Dec 3, 2020 |
| RAVE | Spectroscopic (R~7,500) catalog, 451,783 stars | Section 7.1<br>https://ui.adsabs.harvard.edu/abs/2020AJ....160...83S/abstract<br>BDASP for T_eff, log(g), MADERA for [m/H], GAUGUIN for Fe, Al, Ni, and [α/Fe]<br>**T_eff**: min=2248K, max=15412K, med=4881K, med_err=98K, med_frac_err=2%<br>**log(g)**: min=-1.1, max=5.3, med=2.97, med_err=0.07, med_frac_err=2.4%<br>**[m/H]**: min=-4.7, max=0.95, med=-0.18, med_err=0.09, med_frac_err=23.7%<br>**[Fe/H]**: min=-8.3, max=3.88, med=-0.17, med_err=0.16, med_frac_err=36%<br>**[Al/H]**: min=-2.2, max=3.83, med=-0.05, med_err=0.14, med_frac_err=29.2%<br>**[Ni/H]:** min=-2.96, max=3.78, med=-0.2, med_err=0.24, med_frac_err=64.4%<br>**[α/Fe]**: min=-0.49, max=0.89, med=0.1, med_err=0.17, med_frac_err=63% | 8410-8795 A | 451,783 | | spectroscopy with HIRES | 9 < I < 12 | all | SNR > 10, ~40 for I=10-11 | dec < 0 degrees | http://rave-survey.org/ | https://ui.adsabs.harvard.edu/abs/2020AJ....160...83S/abstract | 7500 | [M/H], Fe, Al, Ni, [α/Fe] | 2020, final | SQL/ADQL Query | https://www.rave-survey.org/query/ |
| SPOCS | Visible spectra from Keck/HIRES for nearby FGKM stars used to measure Teff, logg, vsini and [M/H]. | Stellar parameters methodology described in S4, elemental abundance methods described in S5, Error analysis described in S7<br>**T_eff precision:** 25K<br>**log(g) precision:** 0.028 dex<br>**[m/H] precision:** 0.01 dex<br>**vsini precision:** 0.5 km/s | 364-799 nm | 1617 | 200 pc, most within 150 pc | spectroscopy with HIRES | Mostly V<12, but some fainter targets for targeted follow up | FGK dwarfs | generally SNR > 200, but lower for faint stars. See distinct break in performance at SNR = 100, so sample is divided into two bins above/ below SNR 100 | Mostly Northern hemisphere | https://cdsarc.unistra.fr/viz-bin/cat/J/ApJS/225/32 | https://ui.adsabs.harvard.edu/abs/2016ApJS..225...32B/abstract | ~70,000 (HIRES data) | [Fe/H], C, N, O, Na, Mg, Al, Si, Ca, Ti, V, Cr, Mn, Fe, Ni, & Y | 2016, final | Vizier search/table | |
| Cool Dwarfs Catalog | Late K and M dwarf input catalog for TESS | T_eff from T_eff-color relations (Mann+ 2015,2016)<br>**T_eff**: min=2500K, max=4243K, med=3421K, med_err=100K, med_frac_err=3.1%<br>R* from M_K-R* relations (Mann+2015)<br>**R***: min=0.112, max=0.727, med=0.345, med_err=0.067, med_frac_err=19.2%<br>M* from M_K-M* relations (Benedict+ 2016 or Delfosse+ 2000)<br>**M***: min=0.078, max=0.718, med=0.348, med_err=0.084, med_frac_err=23.6% | optical & IR photometry | 1.14 million | 1000 pc | phot | 1.35 < V < 20.92, median 18.23 | late-K and M dwarfs | N/A | full sky | https://vizier.u-strasbg.fr/viz-bin/VizieR?-source=J/AJ/155/180 | https://ui.adsabs.harvard.edu/abs/2018AJ....155..180M/abstract | N/A | N/A | 2018, final | Vizier search/table | https://vizier.u-strasbg.fr/viz-bin/VizieR?-source=J/AJ/155/180 |
| Gaia Ultracool Database | "M8-T6 with accurate Gaia coordinates, proper motions, and parallaxes [combined] with published spectral types and photometry from large area optical and infrared sky surveys" | pi, mu, G, BP, RP, JHKs, SpT... Ongoing: massive spectral follow-up and parameter determination by big international team (Smart, Jones, Burgasser, Faherty...) | Mostly 0.8-2.4 μm | ~1000 stars and brown dwarfs | ~100 pc | both (published photometry and some spectroscopy) | 16 < G < 21 | late-M, L and a few T dwarfs | N/A | full sky | Upon request to richard.smart@inaf.it | https://ui.adsabs.harvard.edu/abs/2017MNRAS.469..401S/abstract | | N/A | | | |
| HERMES-TESS | Bulk parameters for stars in the TESS CVZ-South | Teff, logg, Rstar, Vsini, Fe/H | blue 4718–4903 Å green 5649–5873 Å red 6481–6739 Å IR 7590–7890 Å | 16,000 | | multifiber spectroscopy | 10 < V< 13 | early F to early K, dwarfs and subgiants, with some giants | SNR > 100 per resolution element | TESS southern CVZ, everything south of -78 in ecliptic latitude | http://www.physics.usyd.edu.au/tess-hermes/ | https://ui.adsabs.harvard.edu/abs/2018MNRAS.473.2004S/abstract | 28,000 | Fe/H | 2017, supposedly not final | Full catalog downloadable at catalog site: http://www.physics.usyd.edu.au/tess-hermes/ | Can be queried through MAST: https://archive.stsci.edu/prepds/tess-hermes/ |
| Schweitzer 2019 | Determines radii and masses of 293 nearby, bright M dwarfs using CARMENES spectra. | **radii:** 2-3% accuracy for 0.1<R[R_Sun]<0.6 **masses:** 3-5% accuracy for 0.1 < M[M_Sun]<0.6 | 520-1710nm | 293 | magnitude limited | spectroscopy with CARMENES | J: 4-11.5 | M dwarfs (M0 - M9.5) | ~150 per pixel in the I band | RA: all, Dec > −23◦ | https://vizier.u-strasbg.fr/viz-bin/VizieR?-source=J/A+A/625/A68 | https://ui.adsabs.harvard.edu/abs/2019A%26A...625A..68S/abstract | 94000 (optical), 80000 (IR) | [Fe/H] in Passegger et al. (2018, 2019) | 2019, should be updated when CARMENCITA catalog is released in 2021 | Vizier search/table | |
| RECONS | Stellar neighborhood measurements of trigonometric parallax, orbital parameters and photometry. | **pi**: 1.5 mas **mu**: <1 mas/yr **orbital parameters**: ~10% or better depending on data quality | optical | ~3000 | 25 pc | astrometry, photometry | V:11-23 | All nearby stars with an emphasis on M dwarfs | parallax error <1.5 mas | Visible from Chile | http://www.recons.org/ | http://www.recons.org/ | | | | Vizier search/table | |
| Terrien M dwarfs | Catalog of NIR spectra from IRTF/SpeX for nearby M dwarf stars used to measure Teff, Rstar, Lstar, [Fe/H], [M/H], and systematic RV. | Description of all parameters presented in Table 4 **T_eff:** 73K (from Newton 2015 calibration) **[m/H] precision**: 0.1 dex **[Fe/H] precision**: 0.1 dex **Systemic RV precision**: 10 km/s **R_star:** 0.027 R_Sun (from Newton 2015 calibration) **L_star**: 0.049 dex (from Newton 2015 calibration) | 0.8-2.4 μm | 886 | 100 pc, but many stars without parallax entries | spectroscopy with SpeX on IRTF | J: 4-12.4 | M dwarfs (M0-M8) | > 100-150 | RA: all. Dec: +67 to -32◦ | https://vizier.u-strasbg.fr/viz-bin/VizieR?-source=J/ApJS/220/16 | https://ui.adsabs.harvard.edu/abs/2015ApJS..220..16T/abstract | R~2000 | [Fe/H] and [M/H] | 2015, final | Vizier search/table | |
| Newton M dwarfs | M dwarf (IR) spectroscopic catalog that uses 24 M dwarfs with interferometrically measured parameters to calibrate empirical relationships. Those relationships are then applied to the Mearth M dwarf sample and the Kepler cool dwarfs sample. | Description of using results interferometric sample (N=25) to derive empirical calibrations is in Section 3. **T_eff:** 73K<br>**[m/H] precision**: 0.1 dex<br>**[Fe/H] precision**: 0.1 dex<br>**Systemic RV precision**: 10 km/s<br>**R_star:** 0.027 R_Sun<br>**L_star**: 0.049 dex | 0.8-2.4 µm | 564 | 3000 light years (applied to Kepler stars) | spectroscopy with SpeX on IRTF | | calibrations are valid from mid K to mid M dwarf stars, roughly 3100 to 4800 K. | | | https://vizier.u-strasbg.fr/viz-bin/VizieR?-source=J/ApJ/800/85 | https://ui.adsabs.harvard.edu/abs/2015ApJ...800...85N/abstract | R~2000 | | 2015, final | Vizier search/table | |
| Mann M dwarfs | M dwarf (Vis/NIR) spectroscopic catalog, using stars selected from the CONCHSHELL and Lupine&Gaidos 2011 target lists | Parameter derivations described in Section 4.<br>**Teff precision**: 2-5%<br>**R star precision**: 2-5%<br>**Systemic RV precision**: 10 km/s<br>**R_star:** 0.027 R_Sun<br>**L_star**: 0.049 dex | 0.3-2.4 µm | 183 | 38pc | spectroscopy. Via spec from SNIFS (3200-9700A, R~1000) and NIR spec from SpeX (0.8-2.4 micron, R~2000) | J: 3.5 - 11 | K7 - M7 dwarfs | Vis: S/N > 100 per resolution element in the red arm of SNIFS. NIR: S/N > 100 (and often > 150) in the H and K band region of SpeX | Visible from Mauna Kea | https://vizier.u-strasbg.fr/viz-bin/VizieR?-source=J/ApJ/804/64 | https://ui.adsabs.harvard.edu/abs/2015ApJ...804...64M/abstract | R~1000 in Vis, R~2000 in NIR | [Fe/H] | 2015, final | Vizier search/table | |
| CONCH shell | All-sky catalogue of nearby (d < 50 pc), bright M- or late K-type dwarf stars. Visible spectra are used to determine Teff and [Fe/H], empirical relations are used to determine stellar radii, masses, and luminosities. | **T_eff**: min=1941K, max=4803K, med=3735K, med_err=77K, med_frac_err=2%<br>**Lum**: min=0.003, max=0.292, med=0.043, med_err=0.009, med_frac_err=<br>**Radius**: min=0.19, max=0.8, med=0.5, med_err=0.05, med_frac_err=<br>**Mass**: min=0.14, max=0.78, med=0.54, med_err=0.07,<br>**[Fe/H]**: | range of instruments, but all visible | 2970 | | spec | J < 9 | Late K - M stars | | all sky | https://vizier.u-strasbg.fr/viz-bin/VizieR?-source=J/MNRAS/443/2561 | https://ui.adsabs.harvard.edu/abs/2014MNRAS.443.2561G/abstract | R~1000 | | 2014, final | Vizier search/table | |
| California Kepler Survey | Spectroscopic (R~60,000) catalog, 2025 Kepler stars | Parameters from SpecMatch<br>**T_eff**: min=4619K, max=6651K, med=5691K, med_err=65K, med_frac_err=1.14%<br>**logg**: min=2.732, max=4.656, med=4.354, med_err=0.023, med_frac_err=1.89%<br>**[Fe/H]**: min=-0.59, max=0.433, med=0.041, med_err=0.042, med_frac_err=17.72%<br>**R***: min=0.614, max=10.4, med=1.11, med_err=0.018, med_frac_err=10.18%<br>**M***: min=0.624, max=2.119, med=1.001, med_err=0.02, med_frac_err=3.8%<br>**Age**: min=1.05Gyr, max=14.49Gyr, med=5.99Gyr, med_err=2.04Gyr, med_frac_err=33.36% | 3640-7990 A | 2025 | 10 pc | HIRES spectroscopy | 6.8 < J < 15.5 | FGK stars | SNR > 45 per pixel | Kepler field | https://california-planet-search.github.io/cks-website/ | https://ui.adsabs.harvard.edu/abs/2017AJ....154..107P/abstract | R~60,000 | [Fe/H] | 2017, final | csv download | csv: http://www.astro.caltech.edu/~howard/cks/cks_physical_merged.csv<br>column description: http://www.astro.caltech.edu/~howard/cks/column-definitions.txt |
| SDSS SEGUE | Spectroscopic (R~1800) catalog, ~350,000 stars | T_eff, log(g), [Fe/H] | 3850-9200 A | 347085 | ~100k pc | spec | g < 19 | 4000 K-10 000K | SNR > 10 | all sky | https://www.sdss.org/dr16/tutorials/segue_sqlcookbook/ | https://ui.adsabs.harvard.edu/abs/2009AJ....137.4377Y/abstract | R~1800 | [Fe/H], [alpha/Fe] | 2009, final | SQL Query | https://www.sdss.org/dr16/tutorials/segue_sqlcookbook/ |
| Starhorse | Photo-astrometric parameters for 265 million stars brighter than G=18 | Sect. 5, Fig. 14+15 in Anders et al. 2019<br>**dist**: 5%<br>**extinction**: 0.2 mag in V-band<br>**T_eff**: 245 K for G<14 mag<br>**log(g)**: < 0.1 dex (but some systematics)<br>**[Fe/H]**: ~0.27 dex<br>**mass**: 0.10-0.15 M_Sun | optical +IR (Gaia, PanSTARRS-1, 2MASS, and AllWISE) | 265.6 million (137 million with most accurate parameters) | - | based on astrometry and photometry | G < 18 mag | all | - | full sky | http://vizier.u-strasbg.fr/viz-bin/VizieR-3?-source=I/349. or https://data.aip.de/projects/starhorse2019.html | https://ui.adsabs.harvard.edu/abs/2019A%26A...628A..94A/abstract | - | [Fe/H] | 2019, final (for now) | query (VizieR) or download (choice of hdf5, fits or csv) | VizieR column description: https://cdsarc.unistra.fr/viz-bin/ReadMe/I/349?format=html&tex=true<br>download: https://data.aip.de/projects/starhorse2019.html |
| Winters et al. 2020 | Volume-Complete Sample of M Dwarfs with Masses from 0.1 | | | 512 M dwarfs | | | | | | | | https://ui.adsabs.harvard.edu/abs/2020ar | | | | | |
| | - 0.3 M_sun within 15 Parsecs | | | | | | | | | | | Xiv201109409W/abstract | | | | | |
| **Stellar Multiplicity** | | | | | | | | | | | | | | | | | |
| SB9 - The 9th Catalogue of Spectroscopic Binary Orbits | bright stars checked for evidence of spectroscopic binaries | SB1, SB2, orbital parameters of spectroscopic binaries | optical | 4004 (version of 2019) | - | radial velocities | m_phot ~9 | all | - | full sky | http://vizier.u-strasbg.fr/viz-bin/VizieR-3?-source=B/sb9 or https://sb9.astro.ulb.ac.be/mainform.cgi | https://ui.adsabs.harvard.edu/abs/2004A%26A...424..727P/abstract | - | - | 2019, live since 2004 | query (VizieR) or download zipped text file | VizieR column description: https://cdsarc.unistra.fr/viz-bin/ReadMe/B/sb9?format=html&tex=true download: https://sb9.astro.ulb.ac.be/mainform.cgi |
| WDS - Washington Double Star Catalog | Washington Double Star catalog | visual physical and optical companions, accuracy depends | usually optical and IR | 153002 components (number of systems not available) | - | visual, speckle, interferometry etc. | V ~ 24 | all | - | full sky | http://www.astro.gsu.edu/wds/ or https://vizier.u-strasbg.fr/viz-bin/VizieR?-source=B/wds | https://ui.adsabs.harvard.edu/abs/2001AJ....122.3466M/abstract | - | - | 2020, live | query (VizieR) or download text file | VizieR column description: https://cdsarc.unistra.fr/viz-bin/ReadMe/B/wds?format=html&tex=true download: http://www.astro.gsu.edu/wds/ |
| Speckle | Completed with a variety of surveys including Zorro/Alopeke on the Gemini's, HRCam on SOAR, and NESSIE on WIYN 3.5 meter. Some are speckle imaging and others are in interferometry mode. (Steve Howell, David Ciardi, Carl Ziegler) | separation, PA, magnitude | Infrared and Optical | ~1000 | ~100pc | speckle, astrometry, photometry | R<13 | all | DelR~4 at 1" for NESSI | full sky | No central catalog but observations on Exo-FOP | https://ui.adsabs.harvard.edu/abs/2011AJ....142...19H/abstract | - | - | on going | target catalogs and multiplicity data on cds | https://vizier.u-strasbg.fr/viz-bin/VizieR?-source=J/AJ/142/19 |
| Robo-AO | Three surveys completed at Kitt | separation, PA, magnitude | Infrared and | ~20,000 | ~150 pc | adaptive optics, | K<9 | all inclu | DelK~8 at 1" | full sky | No centr | https://arxiv.org/abs/ | - | - | on going | multiple catalogs on vizier | https://cdsarc.unistra.fr/viz-bin/cat/ |
| | Peak 2.1 meter, Palomar 60 inch and Hawaii 88 inch. Surveys completed in optical and infrared. All use adaptive optics with a UV LGS. (Carl Ziegler, Christoph Baranec, Nick Law) | | Optical | | | astrometry, photometry | | ding Kepler and TESS targets and nearby M dwarfs | for Robo-AO2 at MK 2.2m, sensitivity varies for three different locations | | al catalog | [1703.08867](1703.08867) | | | | | [J/ApJ/804/30](J/ApJ/804/30) |
| AO surveys | Multiple surveys in multiple publications. Some use the coronagraph and some do not. Most are at infrared wavelengths (exception Magellan VisAO). | separation, PA, magnitude | Infrared with some optical | >3000 with some overlap (Bowler talk) | ~200 pc | adaptive optics, astrometry, photometry | Depends on survey, H<9 early 200x's, H<12 late 200x's | all with special attention to early types | SNR > 5 sigma considered a companion candidate detection | full sky | No central catalog | [https://ui.adsabs.harvard.edu/abs/2020AJ....159...63B/abstract](https://ui.adsabs.harvard.edu/abs/2020AJ....159...63B/abstract) | - | - | on going | multiple catalogs with target lists and maybe contrast information in publications. Many surveys do not publish non-detections. Some raw data in archives. | |

## Compiled Resources

| Name | Link | Description | Basic Stellar Parameters | Wavelength | Sky Coverage |
|---|---|---|---|---|---|
| Hypatia Catalog | https://www.hypatiacatalog.com/ | Compiled catalog of abundances for 9434 stars | Compiled [Fe/H]+abundances | - | |
| TIC V8.1 | https://mast.stsci.edu/portal/Mashup/Clients/Mast/Portal.html | Compiled catalog of stellar parameters for 1.7 billion stars | Compiled | - | all |
| NASA Exoplanet Archive | http://exoplanetarchive.ipac.caltech.edu/ | Properties of known exoplanet hosts | Compiled | mostly visible, when applicable | all |
| Starchive | in prep | Properties of nearest stars, young stars, stars with disks, WDs, planet hosts | all | all | all |
| The extrasolar planets encyclopaedia | http://exoplanet.eu/ | Properties of known exoplanet hosts | Compiled | mostly visible, when applicable | |
| EXOCAT | | | | | |
| PASTEL | https://ui.adsabs.harvard.edu/abs/2016A%26A...591A.118S/abstract | | | | |
| SDSS MaSTAR | https://www.sdss.org/dr16/mastar/mastar-catalogs/ | High quality, R~1800 spectra for 3321 stars | No derivation of their own stellar parameters, parameters compiled from APOGEE, LAMOST, SEGUE where available. | 3622-10354 A | northern hemisphere |

**Broadband Photometry**

| Name | Link | Description | Bandpass (include units) | Sky Coverage | Magnitude range (include units) | Ongoing/ complete | Is there a homogeneous catalog? |
|---|---|---|---|---|---|---|---|
| 2MASS | https://irsa.ipac.caltech.edu/Missions/2mass.html | Infrared all-sky survey | J, H, Ks | All sky | ~15 mag | Complete (2001) | yes |
| WISE | https://irsa.ipac.caltech.edu/Missions/wise.html | Infrared survey mission that observed the entire sky in four bands (3.4, 4.6, 12, and 22 um) with angular resolutions of angular resolution of 6.1", 6.4", 6.5", and 12.0", respectively. | 3.4, 4.6, 12 and 22 microns | All sky | Limiting Vega magnitudes of 16.5, 15.5, 11.2, and 7.9 in bands W1 - W4. Saturation affects photometry for sources brighter than approximately V=8.1, 6.7, 3.8 and -0.4 mag at 3.4, 4.6, 12 and 22 µm, respectively. | complete | yes |
| DENIS | http://cdsarc.u-strasbg.fr/viz-bin/cat/B/denis | The Deep Near Infrared Southern sky survey. It was a dedicated survey with the ESO 1m telescope. | I, J, Ks | declination range +03 to -72 degrees. | limiting magnitudes of 16.5, 14 and 18.5 for J, Ks and Gunn-i. | complete | yes |
| CMC15 | http://svo2.cab.inta-csic.es/vocats/cmc15/ | CMC15 is an astrometric and photometric catalogue of more than 122.7 million stars in the magnitude range 9 < r (SDSS) < 17. With a positional accuracy to about 35 mas, the catalogue covers the declination range -40deg to 50deg | sloan r' | declination range −40º to +50º | magnitude range of 9 to 17 | complete | yes |
| APASS | https://www.aavso.org/apass | AAVSO optical photometric catalog | B, V, u', g', r', i', z_s, Z | all sky | 7 to 17 | ongoing | yes |
| Pan-STARRS | https://panstarrs.stsci.edu/ | Northern sky optical photometric catalog | g, r, i, z, y | Dec > -30 deg | mag~13-23 | Ongoing | yes |
| Gaia | https://www.cosmos.esa.int/web/gaia/early-data-release-3 | Full-sky astrometric/photometric mission | G_BP (blue), G_RP (red), G (optical) | All | G~3-20 | Ongoing | yes |
| GALEX | https://archive.stsci.edu/missions-and-data/galex-1/ | Orbital imaging telescope with near-UV (NUV) and far-UV (FUV) bands and a 'grism' for low-res (90 and 200, respectively) spectroscopy. Completed an All-Sky and Medium surveys, as well as deep-imaging and nearby galaxy surveys. | FUV: 1344-1786 Å effective: 1538.6 Å NUV: 1771-2831 Å effective: 2315.7 Å | All sky & medium sky | All-sky, NUV: down to 20.5 mag Medium-sky, NUV: down to 23.5 mag FUV: down to 19 mag | Complete (2013) | Yes? |
| SkyMapper | http://skymapper.anu.edu.au/ | Southern sky optical photometry | u, v, g, r, i, z | 21,000 sq. deg, Southern sky | ~10-20 mag | Ongoing | yes |
| SDSS | https://www.sdss.org/dr16/imaging/ | Northern sky optical photometry | u, g, r, i, z | 14,000 sq. deg | Down to g~23 | Complete (2009) | yes |
| 2RXS (The 2nd ROSAT All-Sky Survey Bright Source Catalog) | https://www.aanda.org/articles/aa/full_html/2016/04/aa25648-15/aa25648-15.html | second publicly released ROSAT catalogue of point-like sources obtained from the ROSAT all-sky survey (RASS) observations performed with the position-sensitive proportional counter (PSPC) between June 1990 and August 1991, and is an extended and revised version of the bright and faint source catalogues. | 0.1−2.4 keV | All | | Complete | Yes |
| eRASS - future successor of 2RXS | | eROSITA's all-sky survey. In the 0.3-2 keV band, it is expected to be 25 times more sensitive than the pioneering ROSAT mission of the 1990s, and will effectively supersede it. Expected to detect 700,000 stars in the Milky Way | 0.3-10 keV | All | | Ongoing, 7 year planned survey from 2019-2026 | there will be, but not clear if incremental catalogs will be released ahead of the final version |
| Akari | https://heasarc.gsfc.nasa.gov/W3Browse/all/akaripsc.html | The AKARI/IRC Point Source Catalogue, Version 1.0 provides positions and fluxes for 870,973 sources observed with the InfraRed Camera (IRC): 844,649 sources in the S9W filter, and 194,551 sources in the L18W filter. | 9 and 18 microns | all | 50mJy at 9 micron and 120 mJy at 18 micron [5{sigma}] | complete | http://cdsarc.u-strasbg.fr/ftp/cats/II/297/ |

**Specialized Output Catalogs**

| Name | Link to Paper | Link to File | Description | Basic Stellar Parameters |
|---|---|---|---|---|
| Berger et al. (2020) | https://ui.adsabs.harvard.edu/abs/2020AJ....159..280B/abstract | https://cdsarc.unistra.fr/viz-bin/cat/J/AJ/159/280 | Updated homogeneous catalog for 186,000 Kepler stars based on Gaia DR2 parallaxes, spectra, and photometry | Teff, logg, [Fe/H], R*, M*, L*, Age, Distance |
| Hardegree-Ullman et al. (2020) | https://ui.adsabs.harvard.edu/abs/2020ApJS..247...28H/abstract | https://cdsarc.unistra.fr/viz-bin/cat?J/ApJS/247/28 | Updated homogeneous catalog for 222,000 AFGKM K2 stars based on Gaia DR2 parallaxes, LAMOST spectra, and photometry | Teff, logg, [Fe/H], R*, M*, Distance |

## Appendix E: Task Force 1 -- Mission Observables & Deliverables Final Report

# Exoplanet-related capabilities of current and planned space telescopes

*Leads: David Wilson (U. Texas at Austin) & Karan Molaverdikhani*

This list is intended as a guide for the work required to complete the objectives of SAG22 Task Force 1. Please add your initials next to points that you have a particular expertise and/or you would like to work on for our report. Also feel free to add more details, or edit incorrect information on the list.

Note: Missions with no NASA involvement are still relevant for discussion but should be treated as low priority.

People:
Dimitri Veras = DV
Devanshu Jha = DJ
Karan Molaverdikhani =KM
David Wilson =DJW
Navya Nagananda = NN
Kimberly Smith = KSm
Karl Stapelfeldt = KSt
Lokesh Mishra = LM
Chris Stark = CS
Joshua Pepper = JP

# Missions with direct relevance to exoplanets

## Current

- **Hubble Space Telescope (HST)**
    - Planetary atmospheric spectroscopy
        - **Science goals:**
            - Robust observations of exoplanetary atmospheres and recovery of spectral continuum using transmission and emission spectroscopy; commonly with the Wide Field Camera-3 (WFC3) https://iopscience.iop.org/article/10.3847/2041-8213/aa8e40/meta
            - https://www.spacetelescope.org/about/general/instruments/
            - https://hubblesite.org/mission-and-telescope/instruments

- https://www.researchgate.net/figure/Upper-panel-transmission-curves-of-the-HST-ACS-and-HST-WFC3-filters-used-in-the-CLASH_fig2_256199450
        - **Anticipated Data Products:**
            - UV wavelength spectroscopy of planets and stars for majority of systems using the Cosmic Origins Spectrograph (COS) (815-3200 Å, R~1,550–24,000, single/multiple epochs)
                - https://www.stsci.edu/hst/instrumentation/cos
                - https://en.wikipedia.org/wiki/Cosmic_Origins_Spectrograph
            - UV using Space Telescope Imaging Spectrograph (STIS) with three detector arrays (200 nm to 1030 nm; 160 nm and 310 nm; 115 nm and 170 nm; R vary depending on the detector and mode)
                - https://www.stsci.edu/hst/instrumentation/stis/performance/spectral-resolution
                - https://en.wikipedia.org/wiki/Space_Telescope_Imaging_Spectrograph
            - Visible wavelength spectroscopy of planets and stars for majority of systems (STIS: 200 nm to 1030 nm)
            - NIR/IR wavelength spectroscopy of planets and stars for majority of systems (STIS: 200 nm to 1030 nm, NICMOS: 800 to 2500 nm)
                - https://www.spacetelescope.org/about/general/instruments/nicmos/
    - Transit/Planetary atmospheric photometry
        - **Science goals:**
            - Precision aperture photometry
            - https://iopscience.iop.org/article/10.1086/444553
            - https://iopscience.iop.org/article/10.1086/320580
            - https://academic.oup.com/mnras/article/450/2/2043/981103
            - https://arxiv.org/abs/1807.10652
        - **Anticipated Data Products:**
            - Light curves with precisions on the order of ~100 ppm per ~10 sec exposure time
            - For transiting planets, radius, inclination, and possible dayside temperature and hot-spot offset, and in some occasions active regions can be characterized (their location on the disk and/or their temperature)
            - Phase curves can be also extracted upon good precision and stability of the instrument during observations
    - Exoplanet Detection in Survey mode
        - **Science goals:**
            - Discovery of new exoplanets through transit method, e.g. SWEEP
            - https://en.wikipedia.org/wiki/Sagittarius_Window_Eclipsing_Extrasolar_Planet_Search
        - **Anticipated Data Products:**

- Deep field images and corresponding light curves (ACS/WFC: 350–1100 nm, field-of-view of 202″×202″)
- Confirmation of candidate transiting planets (low p)
    - **Science goals:**
        - Confirmation of candidate transiting planets
        - Measurement of possible TTVs
        - More precise ephemeris for the follow-ups (to less extend because observations are expensive and makes more sense to use ground-based observations)
    - **Anticipated Data Products:**
        - Precise light- and phase-curves (usually better than 100 ppm for bright systems)
        - Precise ephemeris (usually order of minutes at the present epoch)
- Characterisation of host stars ultraviolet SEDs and activity (high p)
    - **Science goals:**
        - ~30 targets with xray to IR, publicly available (or planned to be) SEDs.
        - Mostly done on an ad-hoc basis as planets discovered.
        - A few deliberate surveys, e.g. (Mega) MUSCLES (https://cos.colorado.edu/~kevinf/muscles.html), HAZMAT (https://arxiv.org/pdf/1407.1344.pdf)
        - Expensive as most planet hosts are UV faint→ small samples.
    - **Anticipated Data Products:**
        - Full SEDs for a representative sample of planet hosts.
- Direct imaging of planets and debris discs (~high p)
    - **Science goals:**
        - **Incomplete.**
    - **Anticipated Data Products:**
        - **Incomplete.**
- Chemistry of planetary debris at white dwarfs
    - **Science goals:**
        - UV spectroscopy of white dwarfs to detect metal absorption lines from planetary debris.
        - Time-series spectroscopy of systems with variable debris discs and/or transiting debris to search for variation in absorption strength and line-of-sight gas absorption.
        - 100-200 white dwarfs observed with COS. Most easy targets done, future HST observations will focus on detailed observations of known systems: https://arxiv.org/abs/1904.04839
    - **Anticipated Data Products:**
        - Bulk chemistry of ~25 rocky objects,
        - Limits on prevalence of “exotic” chemistry e.g. carbon planets.
        - Absorption spectroscopy of a few (<10) debris discs.

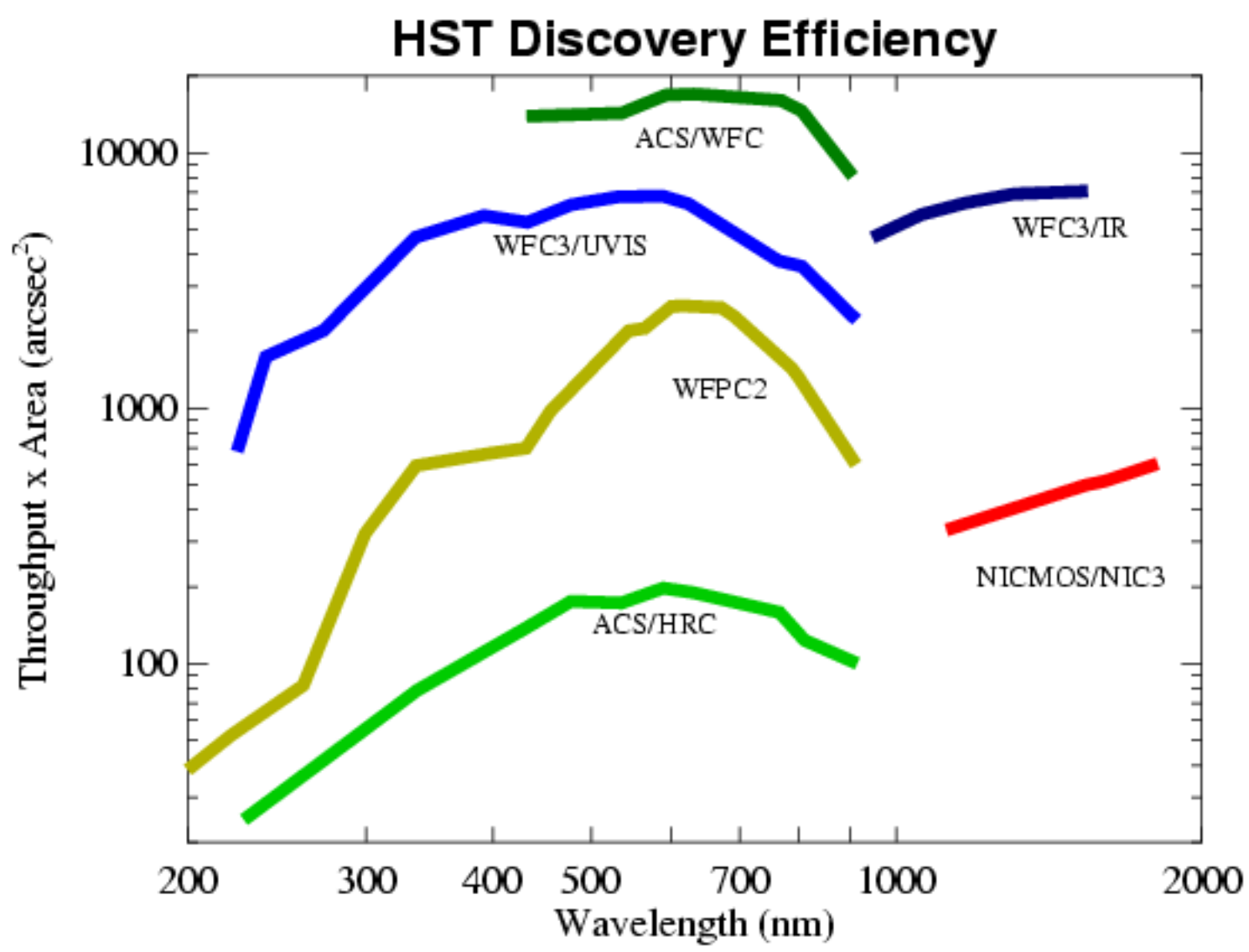

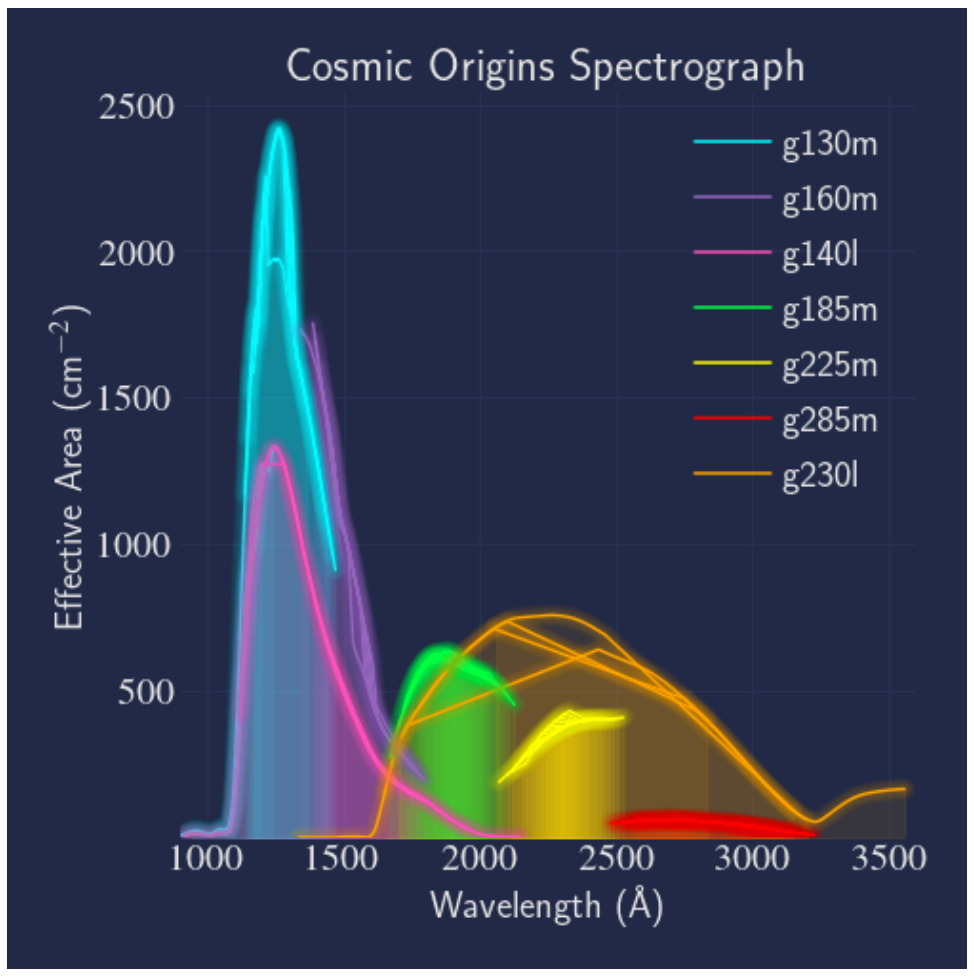

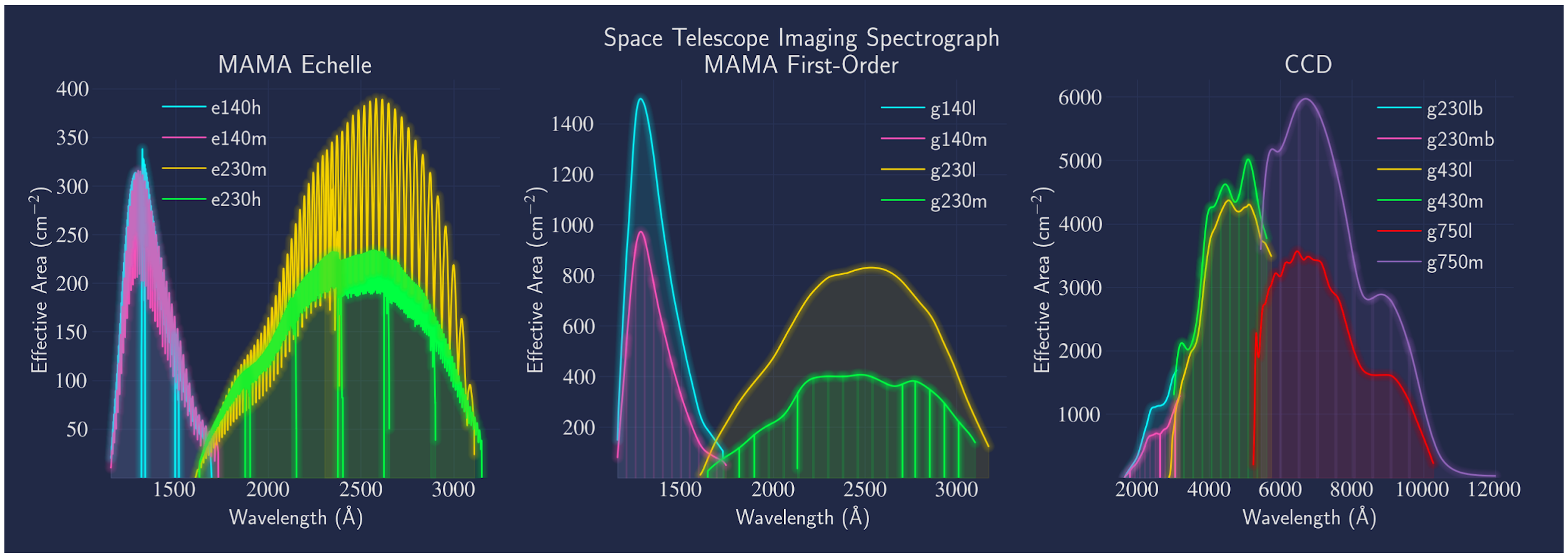

- **TESS (low p)**
  - Transit detection
    - Predicted planet yields: https://arxiv.org/abs/1807.11129
    - **Science goals:**
      - Planet physical and orbital properties
      - TTV and system architecture
    - **Anticipated Data Products:**
      - Planet radii (precision usually better than 30% for rocky planets and as good as stellar radii precision for massive planets)
      - Planet orbital periods (precision mostly depends on the transit depth, host star's magnitude, and the number of observed transits)
      - Planet inclination

- Phase-curve (that in turn provides information on the temperature brightness and heat redistribution)
            - Transit asymmetry: the degree of non-spherical geometry of the planet/star
            - Stellar characterization
        - **Science goals:**
            - Asteroseismology component of stars
            - Stellar rotation
            - Activity characterization
        - **Anticipated Data products**
            - Stellar masses and ages from asteroseismology
            - Rotational periods
            - Flare rates
            - Characterization of spots/faculae through both occulted and unocculted configurations

- **Chandra X-ray observatory (Chandra) (low p)**
    - Characterisation of host stars X-ray SEDs and activity - see above, mostly done in the context of HST UV observations.
        - **Incomplete.**
    - Searches for x-ray transits of clouds from evaporation planet have been made but with no detections thus far https://arxiv.org/abs/1804.11124
        - **Incomplete.**
    - Transits of planets in other galaxies https://arxiv.org/abs/2009.08987
        - **Science goals:**
            - Search for transits around bright X-ray sources in other galaxies.
            - Constrain planet formation/survival around X-ray binaries.
        - **Anticipated data products:**
            - Unclear: discovery paper finds one planet candidate in a survey of three galaxies, future yield depends on amount and quality of x-ray observations of different galaxies.

- **NASA's Stratospheric Observatory for Infrared Astronomy (SOFIA) (low p)**
    - https://ui.adsabs.harvard.edu/abs/2018haex.bookE.191A/abstract
    - https://iopscience.iop.org/article/10.1086/656386
    - https://exoplanets.nasa.gov/news/1435/sofia-confirms-nearby-planetary-system-is-similar-to-our-own/
    - Planet confirmation
        - https://exoplanets.nasa.gov/news/1435/sofia-confirms-nearby-planetary-system-is-similar-to-our-own/
        - **Science goals:**
            - Confirmation of candidate transiting planets (particularly around nearby and bright stars)
            - Measurement of possible TTVs

- More precise ephemeris for the follow-ups (to less extend because observations are expensive and makes more sense to use ground-based observations)
        - **Anticipated Data Products:**
            - More precise measurements of planet radii, inclination, etc
            - Precise light- and phase-curves (additionally for possible TTV, hot-spot offset, heat redistribution, etc)
    - Atmospheric spectrophotometry
        - **Science goals:**
            - Absolute optical photometry
            - Optical and NIR spectrophotometry (in particular with the HIPO-FLITECAM and FPI+ optical and near-infrared instruments)
        - **Anticipated Data Products:**
            - Emission and transmission spectra at low resolution (0.3 - 1,600 micrometer)
            - Multi-band phase-curve photometry

# Future

- **James Webb Space Telescope (JWST, 2021)**
    - **Incomplete.**
    - JWST can cover a variety of exoplanet sciences similar to that of HST, except where shorter wavelengths are needed, for instance, to study some atomic lines, or characterize stellar radiance at shorter wavelengths than that of JWST's.
    - Planetary atmospheric spectroscopy
        - https://arxiv.org/abs/1803.04985
        - https://www.stsci.edu/jwst/observing-programs/approved-ers-programs/program-1366
        - https://www.stsci.edu/jwst/observing-programs/program-information?id=1185
        - https://www.stsci.edu/jwst/observing-programs/program-information?id=1177
        - https://www.stsci.edu/jwst/observing-programs/program-information?id=1201
        - https://www.stsci.edu/jwst/observing-programs/program-information?id=1312
        - https://www.stsci.edu/jwst/observing-programs/program-information?id=1331
        - https://www.stsci.edu/jwst/observing-programs/program-information?id=1353
        - **Science goals:**
            - Characterization of planetary atmospheres
        - **Anticipated Data Products:**
            - Mid-resolution spectroscopy of planetary atmospheres through transmission, emission, and phase-curve by using MIRI (Low Resolution Spectroscopy), NIRCam (Grism Time Series), NIRISS

(Single-object Slitless Spectroscopy), and NIRSpec (Bright Object Time Series)

- Direct imaging of planets & debris disks (NIRCam & MIRI)
  - https://jwst-docs.stsci.edu/near-infrared-camera/nircam-observing-modes/nircam-coronagraphic-imaging
  - https://jwst-docs.stsci.edu/mid-infrared-instrument/miri-observing-modes/miri-coronagraphic-imaging
  - https://www.stsci.edu/jwst/observing-programs/program-information?id=1188
  - https://www.stsci.edu/jwst/observing-programs/program-information?id=1193
  - **Science goals:** Directly image known self-luminous giant exoplanets around nearby stars, detect new self-luminous giant planets, image known debris disks at wavelengths transitioning from scattered light to thermal emission, constrain disk composition and size distribution
  - **Anticipated Data Products:**
    - NIR planet photometry & disk images (1.8-5 microns, IWA > 0.4”, contrast better than 10^-5 beyond IWA, narrow and broadband)
    - MIR planet photometry & disk images (10-25 microns, IWA > 0.3”, broadband >~ 6%)
- Spectroscopy of debris discs around WDs
  - **Science goals:**
    - **Incomplete.**
  - **Anticipated Data Products:**
    - **Incomplete.**

- **Habitable Exoplanet Observatory (HabEx, 2030s if selected)**
  - https://arxiv.org/abs/2001.06683
  - Direct imaging & spectroscopy of exoplanets
    - **Science goals:** Directly image & spectrally characterize exoplanets around nearby stars, “map out” planetary systems by detecting planets and measuring orbits, explore planet diversity, discover ~< 10 potentially Earth-like plans and search for signs of habitability & biosignatures
    - **Anticipated Data Products:**
      - All orbital parameters for nearly all detected planets (semi-major axis better than 25%, eccentricity better than 0.1)
      - Planet radii constrained to w/in a factor of 2
      - Visible wavelength photometry for nearly all detected planets (~0.5 microns, broadband (20%), multi-epoch, multi-band)
      - Visible wavelength spectroscopy of planets and stars for majority of systems (0.3-1.0 microns, R~140, single epoch)
      - UV and NIR spectroscopy of planets and stars for a few systems (~0.2-1.8 microns, multi-epoch)
  - Direct imaging & spectroscopy of debris disks

- **Science goals:** “map out” planetary systems by measuring debris disk distribution, investigate disk composition and volatile delivery, measure exozodi structures & brightness
        - **Anticipated Data Products:**
            - Visible wavelength broadband (~20%) images of inner regions of disks for all systems (OWA ~ 0.8”)
            - Visible wavelength ultra-broadband (~100%) images of outer regions of disks for a few systems (OWA ~ 6.0”, multi-epoch)
            - Spatially-resolved visible wavelength spectroscopy of inner debris disks for majority of systems (0.3-1.0 microns, R~140, single epoch, 1” OWA)
            - Spatially-resolved UV and NIR spectroscopy of disks for a few systems (~0.2-1.8 microns, multi-epoch)
    - Transit spectroscopy
        - https://www.jpl.nasa.gov/habex/pdf/HabEx_Interim_Report.pdf (section 3.6)
        - **Science goals:**
            - Transmission and emission spectroscopy of transiting planets (GO program)
        - **Anticipated Data Products:**
            - UV-NIR Spectra (0.2-1.8 microns, High SNR in a single transit, Low R, and for limited number of exoplanets because the advantage of HabEx is mainly to explore other corners of parameter-space relevant to characterization of planetary systems through direct imaging)

- **Large UV/Optical/IR Surveyor (LUVOIR, 2030s/40s if selected)**
    - https://arxiv.org/abs/1912.06219
    - https://asd.gsfc.nasa.gov/luvoir/reports/
    - Direct imaging & spectroscopy of exoplanets
        - **Science goals:** Directly image & spectrally characterize exoplanets around nearby stars, “map out” planetary systems by detecting planets and measuring orbits, explore planet diversity, discover ~< 55 potentially Earth-like plans and search for signs of habitability & biosignatures, constrain how common Earth-like conditions/life are/is
        - **Anticipated Data Products:**
            - All orbital parameters for nearly all detected planets (to precision of a few percent)
            - Planet radii constrained to w/in a factor of 2
            - UV + VIS photometry for nearly all detected planets (broadband (10-20%), multi-epoch, multi-band)
            - UV + VIS spectroscopy of planets and stars for ~50 systems (10-20% bandpasses, R<150, single epoch)

- NIR spectroscopy of planets and stars for a few systems (<1.8 microns, 20% bandpass, single epoch)
- Direct imaging & spectroscopy of debris disks
    - **Science goals:** "map out" planetary systems by measuring debris disk distribution, investigate disk composition and volatile delivery, measure exozodi structures & brightness
    - **Anticipated Data Products:**
        - UV + VIS broadband (~10-20%) images of inner regions of disks for all systems (OWA ~0.16/0.5" for UV/VIS)
        - Spatially-resolved UV + VIS spectroscopy of inner debris disks for ~50 systems (~10-20% bandpasses, single epoch, OWA ~0.16/0.5" for UV/VIS)
        - Spatially-resolved NIR spectroscopy of inner debris disks for a few systems (<1.8 microns, 20% bandpass, single epoch, OWA~1")
- Astrometric observations of direct imaging targets
    - See section 4.2.3.1 of final report
    - **Science goals:**
        - Measure masses of directly imaged planets
    - **Anticipated Data Products:**
        - V band widefield (3' x 2') images of direct imaging target stars via High Definition Imager (HDI)
        - Astrometric positions (< 1 micro-arcsec precision per observation, < 0.1 micro-arcsec final precision) of all stars that host directly imaged exoEarth candidates to measure exoEarth mass.
- Direct imaging & spectroscopy of known RV planets
    - **Science goals:**
        - Understanding how atmospheric composition varies with planet mass and stellar properties.
    - **Anticipated Data Products:**
        - Reflected light spectra of gas and ice giants, down to sub-Neptune/super-Earth sizes.
        - Spectral range for coronagraphic spectroscopy provides the ability to detect and measure the abundances of a large array of possible gases, including $H_2O$, $CH_4$, $O_2$, $O_3$, and $CO_2$.
        - See table from LUVIOR final report for number of targets.
- Transit spectroscopy
    - **Science goals:**
        - Characterizing the atmospheres of potentially habitable planets to constrain the chemical compositions and search for signs of life.
    - **Anticipated Data Products:**
        - 0.2-2.5um transit spectroscopy of rocky planets around M dwarfs
        - Abundances of (among others) $H_2O$, $CO_2$, $CH_4$, $O_2$, $O_3$.
        - Targets from TESS sample, ground-based surveys and TRAPPIST-1.

**Table B-8.** *Science time and overheads required for each component of the LUVOIR-A non-earth direct spectroscopy program. The program requires 424 hours to investigate the compositions of 30 planets*

| Mass | $T_{eq} \leq 180$ K | $180 < T_{eq} \leq 300$ K | $T_{eq} > 300$ K |
|---|---|---|---|
| $M_p > 0.15\ M_J$ | **5 planets**<br>science: 132 hrs<br>overhead: 38 hrs<br>***total: 171 hrs***<br>Model atmosphere = "jupiter" | **11 planets**<br>science: 65 hrs<br>overhead: 62 hrs<br>***total: 127 hrs***<br>Model atmosphere = "jup2" (warm Jupiter at 2 AU from Cahoy et al. 2010) | **3 planets**<br>science: 18 hrs<br>overhead: 17 hrs<br>***total: 35 hrs***<br>Model atmospheres= "jup08" (warm Jupiter at 0.8 AU from Cahoy et al. 2010) |
| $M_p \leq 0.15\ M_J$ | **0 planets**<br>science: 0 hrs<br>overhead: 0 hrs<br>***total: 0 hrs***<br>Model atmosphere = "neptune" | **6 planets**<br>science: 18 hrs<br>overhead: 17 hrs<br>***total: 35 hrs***<br>Model atmosphere = "nep2" (warm Neptune at 2 AU from Hu & Seager 2014) | **5 planets**<br>science: 7 hrs<br>overhead: 26 hrs<br>***total: 33 hrs***<br>Model atmosphere = "cloudy_nep1" (cloudy warm Neptune at 1 AU from Hu & Seager 2014) |

- Spectroscopy of planetary-polluted WDs (Also a goal with HabEx)
  - https://arxiv.org/abs/1904.04839
  - **Science goals:**
    - Obtain detailed statistics on the bulk chemistry of rocky extrasolar planets, analogous to abundance statistics from Solar system meteorites.
  - **Anticipated Data Products:**
    - UV spectroscopy of >1000 targets, ~25% of which will have pollution from planetary debris. Atmosphere model fits and metal abundances for those spectra.

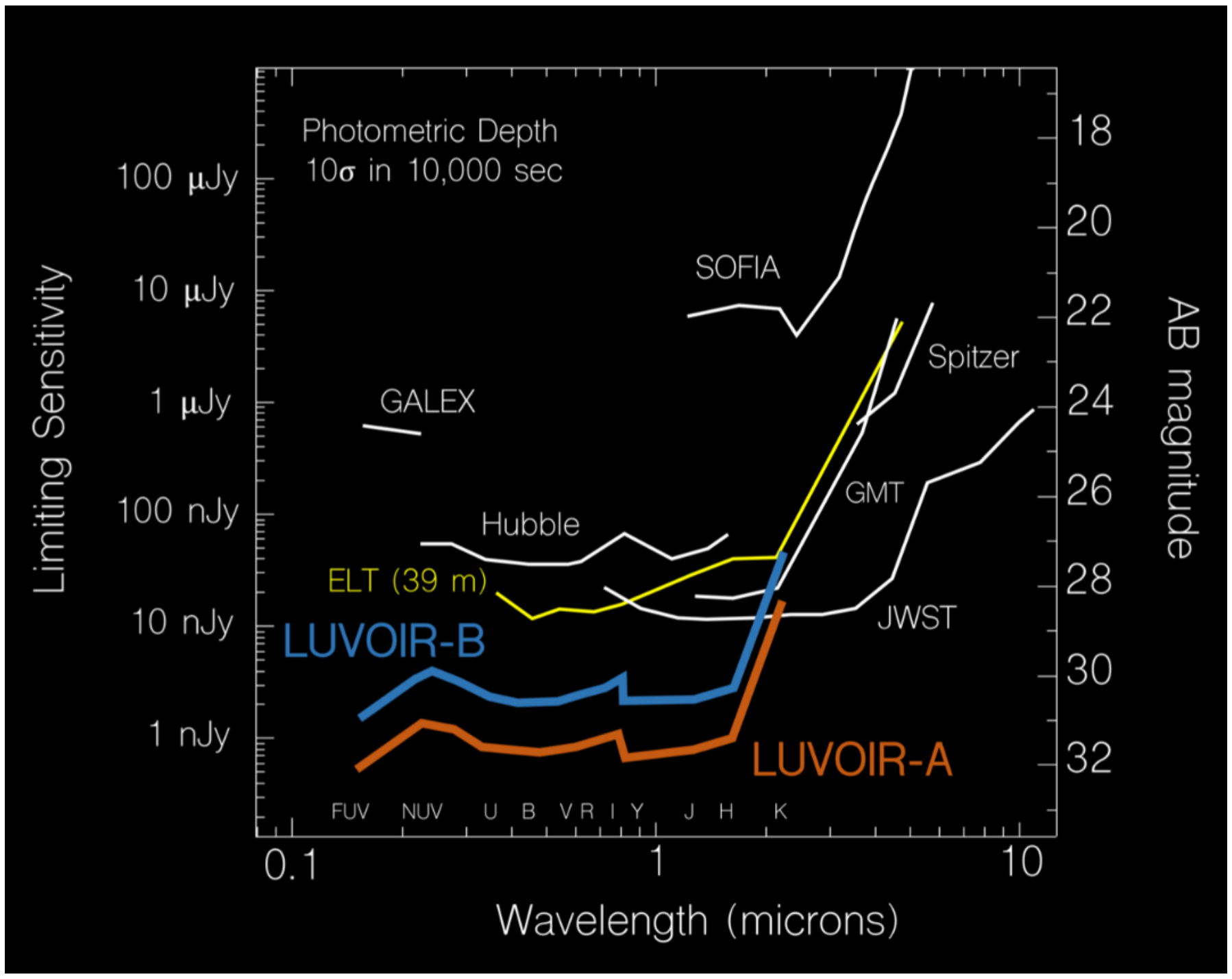

- **Origins Space Telescope (OST, 2030s if selected)**
    - https://asd.gsfc.nasa.gov/firs/docs/OriginsVolume1MissionConceptStudyReport25Aug2020.pdf
    - Planetary atmospheric spectroscopy
        - https://www.spiedigitallibrary.org/conference-proceedings-of-spie/10698/1069815/The-Origins-Space-Telescope-mission-concept-overview/10.1117/12.2313823.full?SSO=1
        - https://www.cosmos.esa.int/documents/1866264/3219248/WiednerM_OriginsSpaceTelescope-Wiedner.pdf/6bd4b85d-7dea-baa8-8758-a007e4c62a4e?t=1565184755214
        - https://arxiv.org/abs/2012.06549
        - **Science goals:**
            - How do the conditions for habitability develop during the process of planet formation?
            - What role does water play in the formation and evolution of habitable planets?
            - How and when do planets form?
            - How were water and life's ingredients delivered to Earth and to exoplanets?
            - Do planets orbiting M--dwarf stars support life?
        - **Anticipated Data Products:**
            - Imaging of young disks, detection of water in disks
            - Measurement of the evolution of the gas content in disks
            - Detection of solar system comets to measure D/H ratio
            - 2.8–20 μm transmission spectra to detect nature of the atmosphere (clear, cloudy, or tenuous). A subset of the targets with enough signal to get atmospheric composition and biosignatures
            - Thermal emission of planets (not clear if is secondary eclipse or direct imaging)

- **Atmospheric Remote-sensing Infrared Exoplanet Large-survey /Contribution to ARIEL Spectroscopy of Exoplanets (ARIEL/CASE, 2028?, non-NASA, with NASA contribution)**
    - Planetary atmospheric spectroscopy
        - https://www.cosmos.esa.int/documents/1783156/3267291/Ariel_RedBook_Nov2020.pdf/30a9501c-8b63-227b-bcaf-b7f544c3628e?t=1604684048651
        - **Science goals:**
            - Detect and determine the composition and structure of a large number of planetary atmospheres
            - Constrain planetary interiors by removing degeneracies in the interpretation of mass-radius diagrams

- Constrain planetary formation and evolution models through measurements of the elemental composition (evidence for migration)
            - Determine the energy budget of planetary atmospheres (e.g. albedo, vertical and horizontal temperature structure, weather/temporal variations)
            - Identify and constrain chemical processes at work (e.g thermochemistry, photochemistry, transport, quenching)
            - Constrain the properties of clouds (e.g. cloud type, particle size, distribution, patchiness)
            - Investigate the impact of stellar and planetary environment on exoplanet properties
            - Identification of different populations of planets and atmospheres (e.g. through colour- colour diagrams)
            - Capacity to do a population study and go into a detailed study of select planets
        - **Anticipated Data Products:**
            - Survey of ~1000 transiting exoplanets from gas giants to rocky planets, in the hot to temperate zones of A to M-type host stars
            - Target selection before launch based on ESA science team and community inputs
            - Delivery of a homogeneous catalogue of planetary spectra, yielding refined molecular abundances, chemical gradients and atmospheric structure; diurnal and seasonal variations; presence of clouds and measurement of albedo

- **<u>Roman Space Telescope (2025)</u>**
    - Microlensing ( Wide Field Instrument)
        - https://arxiv.org/pdf/1808.02490.pdf
        - **Science goals:**
            - Occurrence rates of terrestrial to giant planets beyond 1 AU (mostly G stars & later), constraints on planet formation
        - **Anticipated Data Products:**
            - NIR (~1.5 micron) wide field images of 2.8 sq deg of the galactic bulge (15 min cadence) for a little over a year of total observing time, 2 years wall clock time, split into 6 72 days seasons
            - NIR magnification light curves for all sources
    - Transits ( Wide Field Instrument) -- parallel project w/ microlensing
        - https://arxiv.org/pdf/1610.03067.pdf
        - **Science goals:**
            - Detect ~100k transiting planets around stars w/ known distances w/ orbital periods < 2 mo and radius > few R_Earth.
            - Improvements to the Kepler occurrence rates
        - **Anticipated Data Products:**

- NIR (~1.5 micron) wide field images of 2.8 sq deg of the galactic bulge (15 min cadence) for a little over a year of total observing time, 2 years wall clock time, split into 6 72 days seasons
        - NIR light curves for all sources
  - Direct imaging of planets (coronagraph)
    - https://science.nasa.gov/science-pink/s3fs-public/atoms/files/Kruk_WFIRST_update_APAC_March2020.pdf
    - https://www.nationalacademies.org/documents/embed/link/LF2255DA3DD1C41C0A42D3BEF0989ACAECE3053A6A9B/file/D656AEF4631BDA7A4F24DC482B6AC6A2696847F06216
    - **Science goals:** image and spectrally characterize known RV giant planets in reflected light
    - **Anticipated Data Products:**
        - VIS (575 nm) broadband (10%) photometry & polarimetry of “warm” giant planets (0.14”-0.45”)
        - NIR (825 nm) broadband (10%) photometry & polarimetry of “cold” giant planets (0.45”-1.4”)
        - VIS (730 nm) spectroscopy of planets (R~50, 15% bandwidth)
        - Constrained orbits by combining direct imaging w/ RV
  - Direct imaging of debris disks (coronagraph)
    - https://www.nationalacademies.org/documents/embed/link/LF2255DA3DD1C41C0A42D3BEF0989ACAECE3053A6A9B/file/D656AEF4631BDA7A4F24DC482B6AC6A2696847F06216
    - **Science goals:** image debris disks in scattered light, constrain composition and dust transport mechanisms, measure exozodiacal dust brightness
    - **Anticipated Data Products:**
        - VIS (575 nm) broadband (10%) photometry & polarimetry of “warm” debris disks (0.14”-0.45”)
        - NIR (825 nm) broadband (10%) photometry & polarimetry of “cold” debris disks (0.45”-1.4”)

- **ELTs (TMT, GMT: possible NASA participation)**
  ***The following information is restricted to the first generation of instruments at these observatories.***
  GMT Science Book: ttps://www.gmto.org/wp-content/uploads/GMTScienceBook2018.pdf
  TMT Science Book: https://www.tmt.org/download/Document/10/original
  G-CLEF: https://www.gmto.org/SPIE_2016/SPIE_2016_9908-76.pdf

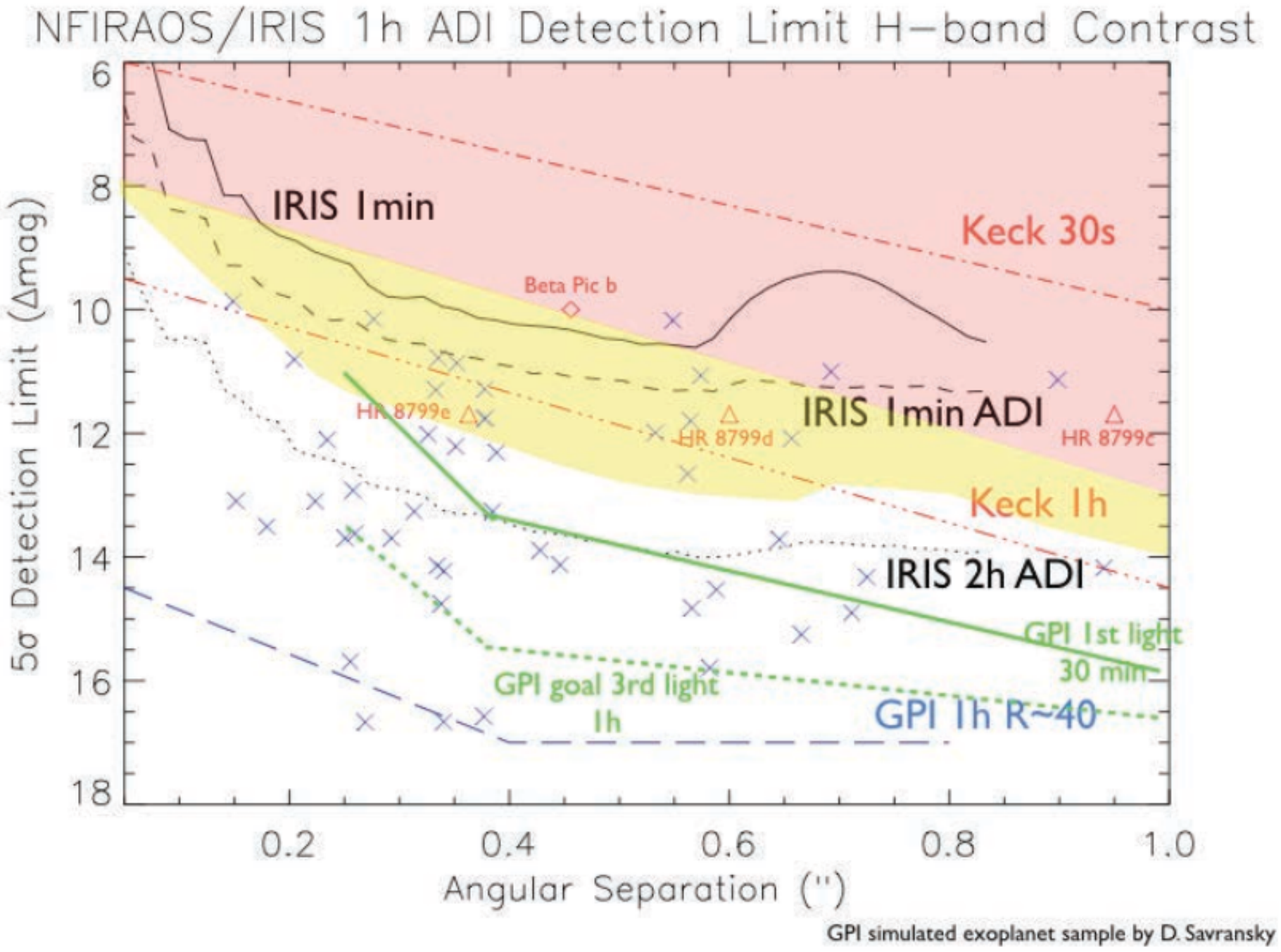

Abridged Caption: Exoplanet imaging science with TMT/IRIS. Top: Predicted contrast of TMT/IRIS using angular differential imaging (ADI) compared to contrasts for Gemini/GPI and Keck/NIRC2 and a simulated population of 50 super-Jovian exoplanets imaged around a sample of 600 stars (crosses; Marois et al. 2012).

- Direct imaging
  - **Science goals:**
    - Direct direction of reflected and thermal emission from exoplanets
    - Albedo measurements, follow-up spectroscopy and astrometry of imaged planets
    - Measure molecular abundances in exoplanet atmospheres (particular targets are the known potentially habitable terrestrial planets: YZ Cet d, HD 85512 b, GJ 3323 b, Wolf 1061 c, GJ 273 b, Proxima b, and GJ 667C f)
    - Follow-up observations of gas-giant planets  found by GPI, SPHERE and SCExAO.
    - Detect sub-jupiter mass planets around low-mass M dwarfs
  - **Anticipated Data Products:**
    - 60-150 planets directly imaged by GMT (in reflected light); 15 of these planets are expected to be rocky (R < 1.6 Rearth); 20 planets, between 1.6 to 4 Rearth, in the habitable zone.
    - Direct imaging of planetary companions in open clusters (Pleiades, Hyades) or in regions ~ 150 pc (Taurus, Sco-Cen, ρ Ophiuchus).
  - Radial Velocity
    - **Science goals:**

- Measure mass of Earth-sized planets (transit follow-up)
- Determine eccentricity of planetary orbits
- Search nearest quietest stars for non-transiting planets (candidates for direct imaging)

■ **Anticipated Data Products** (based on G-CLEF on GMT)**:**

- PRV measurements: Goal ~ 10 cm/s, Requirement <= 50 cm/s, Wavelength range: 400-700 nm, Resolution: 108,000

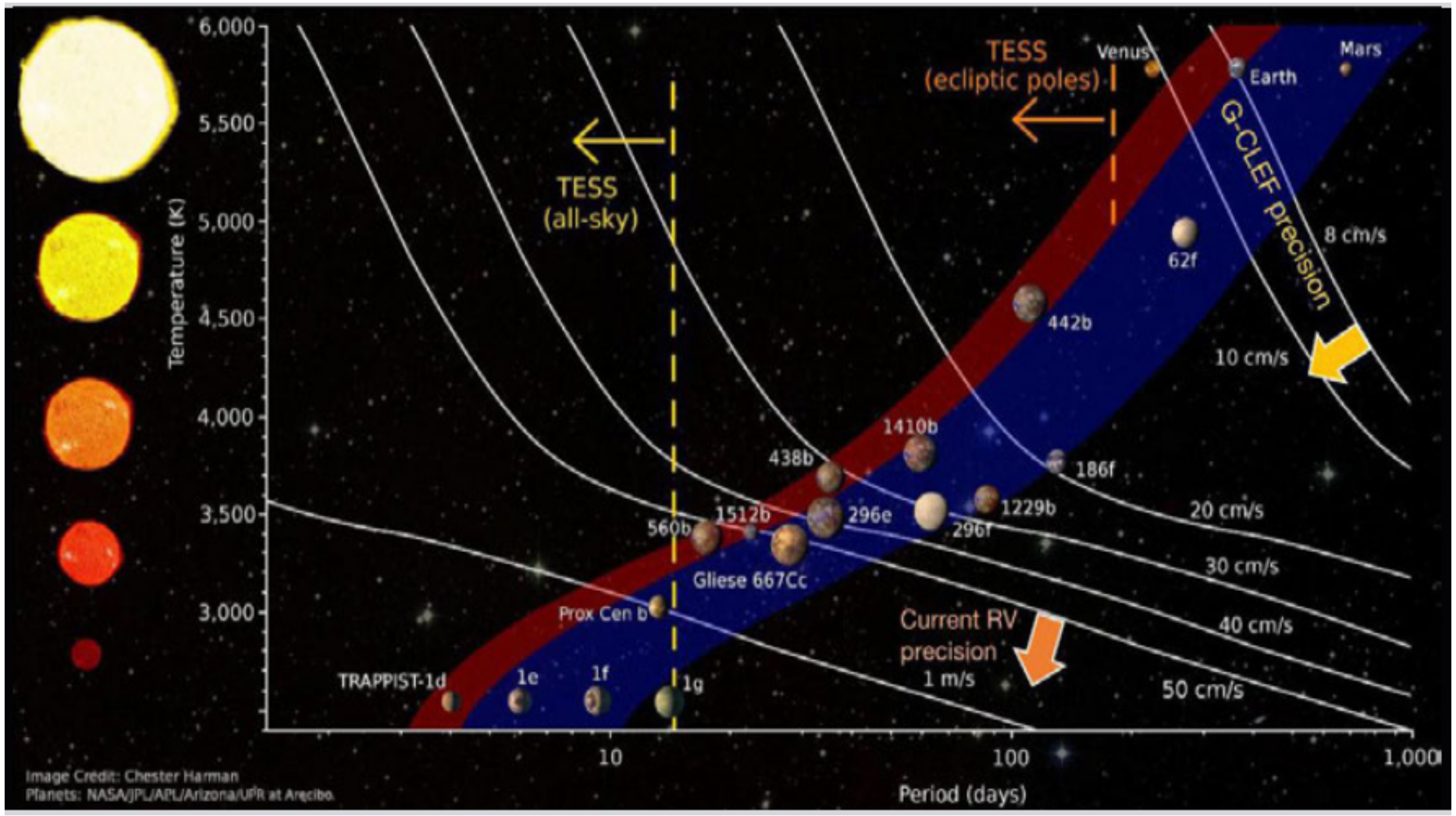

Abridged Caption: Figure 2-1 The GMT will determine the masses and characterize the atmospheres of temperate rocky planets in the habitable zone. This illustration from Chester Harman shows most of the currently known exoplanets that are smaller than 1.6 times the size of Earth (as of March 2017), around moderately bright stars, and in the “habitable zone”, along with the GMT’s radial velocity capability. It explains how precision (10 cm/s) measurements of radial velocity with the G-CLEF instrument will revolutionize the study of Earth-like worlds, through the determination of mass and density.

○ Transits and Transmission Spectroscopy

■ **Science goals:**

- Measure planet orbital alignment via Rossiter McLaughin effect
- Characterize atmosphere and search for biomarkers
- Especially, 3-sigma detection of $O_2$ in exoplanet atmospheres: G-CLEF targets are expected to be dwarfs of class M3 to M8 within 20 pc of the Sun

■ **Anticipated Data Products:**

- Wavelength range: 350-1000 nm, R>100,000 (Snellen et. al. 2013)

# Indirectly relevant missions (just the type of data)

(e.g., surveys or missions related to stellar astrophysics with implications for host stars.)

Current

- **Swift**
    - **Incomplete.**
    - Characterisation of host stars UV/X-ray SEDs and activity

Non-NASA Relevant Missions

- **Gaia**
    - **Incomplete.**
    - Host star distances
    - Detection of giant planets via astrometry
- **XMM-Newton (XMM)**
    - **Incomplete.**
    - Contribution similar to Chandra above, mainly provides supplementary data to observations with e.g. HST.
- **CHaracterising ExOPlanets Satellite (CHEOPS, non-NASA)**
    - **Incomplete.**
    - Confirmation of candidate transiting planets
    - Phase-curve measurements of transiting planets

Future

- **Lynx X-ray Observatory (Lynx, 2030s if selected)**
    - **Incomplete.**
    - X-ray transits: None detected with Chandra/XMM, but possible with Lynx: https://arxiv.org/abs/1804.11124
    - Characterisation of host stars X-ray SEDs and activity https://wwwastro.msfc.nasa.gov/lynx/docs/LynxConceptStudy.pdf
        - **Science Goals:**
            - Infer EUV emission from high-resolution x-ray spectroscopy.
            - Measure stellar winds from K-shell emission
            - Find coronal mass ejections from line doppler shifts.
        - **Anticipated data products:**
            - EUV measurements for stars with a range of ages and rotation periods.
            - Solar-strength winds detectable at all stars <10pc.
            - Systematic detection of CME events.
- **PLAnetary Transits and Oscillations of stars (PLATO, 2027, non-NASA)**
    - **Incomplete.**
    - Detection of transiting planets
    - Host star optical activity

## Appendix F: Task Force 2 -- Target Lists and Target Final Report

*Leads: Ilaria Pascucci (U. Arizona) & David Ciardi (Caltech/IPAC-NExScI)*

**Goals**: The primary goal of the SAG22 Task Force 2 (TF2) is to assemble a primary set of criteria that would be the basis of a target catalog associated with the SAG22 effort. The Phase 1 task for TF2 is to collate target lists that have been assembled for various NASA and related non-NASA exoplanet missions and projects, or the individual criteria being considered. The primary goal is not a detailed target list but rather to elucidate our understanding of the overlap in targets for the various missions and of the multi-dimensional stellar parameter space necessary to characterize the targets.

**Participants:** Ilaria Pascucci, David Ciardi, Katia Cuhna, Lynne Hillenbrand, Wei-Chun Jao, Eric Mamajek, Lokesh Mishra, Josh Pepper, Chris Stark, Keivan Stassun, Jennifer Winters

**Deliverable:** Federated list of targets collected from the literature, project websites, and personal communications that includes the

- Target Name: as listed in the original target list
- 2MASS Designation (or equivalent if no 2MASS source is available)
- Equatorial coordinates J2000
- V magnitude if available
- Gaia G-magnitude if available
- 2MASS J magnitude if available
- 2MASS K-short magnitude if available
- Parallax of the target
- Spectral Type of the target
- Identification of which target lists the source is registered
- Working notes on different targets and catalogs is a Google document
- The Federated Target List is a txt file in the TF2 Google drive

**Categories of Targets:** Atmospheric Characterization is the unifying scientific theme for the targets in this. Atmospheric characterization can be accomplished via direct imaging/spectroscopy and/or transmission/eclipse spectroscopy, both of which require good determinations of the masses and orbits of the planets typically through precision radial velocity observations. We have identified the following Categories of Targets and associated relevant missions or projects.

- Direct Imaging
    - Roman CGI
    - JWST
    - LUVOIR

- HabEx
    - Origins
    - GPI and SPHERE
    - US ELT Programs
- Transmission and Emission Spectroscopy
    - JWST
    - ARIEL
- Precision Radial Velocity
    - NEID WIYN Survey
    - Keck HIRES - California Planet Search
    - VLT ESPRESSO
    - Calar Alto - CARMENES
    - Maroon-X
    - US ELT Programs

**Caveats on Collected Target Lists:** There are a series of caveats with regards to the collection of targets that the group have identified and have enumerated here.

1. The collection of targets identified in this effort is not expected to be complete or comprehensive. Exoplanet host stars are continuously being discovered by the community, and target lists, especially for ongoing and future missions, are constantly evolving. Further, not all related programs or missions have published their target lists (e.g., JWST Cycle 1 targets are still to be determined).
2. The properties of the targets in this list are not intended to be complete or comprehensive as one of the goals of the entire SAG is to determine what information should be collected. Rather the information in the target lists (e.g., position, color, and distance) is intended to be able to provide a general overview of the physical properties and distribution of the targets.
3. Many of the lists contain targets that are fairly bright and as a result survey information about the brightest targets may not be the most accurate. At this point in the SAG process, effort was not made to provide the best or most-reliable properties of all of these stars - rather uniformity and overall breadth of targets was chosen as more important than depth or detail on any one target.

**Process:** Individual target groups were assigned to TF2 participants where the participant expertise was most effective. The participants utilized the literature and/or project resources to identify the appropriate target list for that mission or project. Once a target list for a given mission/project was identified, the source list was matched against the 2MASS catalog - if an appropriate 2MASS identifier could not be found, a reasonable other catalog was used (e.g., WISE). The 2MASS identifiers were then used to federate all of the lists together such that a single target only appears once and the missions/projects associated with that target are recorded. If a specific target list for a mission/project could not be identified, the overall “target-demographics” were recorded. Below are specific notes on the individual targets collected.

**Notes on Individual Target Lists:**

- Roman CGI: Target lists taken from Dmitry Savransky’s page and is the most up-to-date list of exoplanet targets for Roman (confirmed by chat w/ Dmitry). https://plandb.sioslab.com/
- JWST: Target lists were taken from the JWST Guaranteed Time Observations and the Early Release Observation programs. The approved programs are available online at the JWST Approved Programs page. Note that only the Extrasolar Planets category was used. If there are other exoplanet targets that overlap with non-exoplanet JWST categories (e.g., Brown Dwarfs, Debris Disks), those targets have not been captured in this list.
- LUVOIR and HabEX: Target lists were taken from the LUVOIR and HabEx studies. The targets were marked as LUVOIR-A, LUVOIR-B, and HabEx
- Origins: Targets from the mission study need to be collected (this are primarily transit and emission spectroscopy targets)
- GPI and SPHERE: The GPI targets are from the GPIES survey published in Nielsen+2019. The SPHERE targets are from the SHINE survey published in Vigan+2020.
- US ELT: The US ELT programs did not specify a particular target list but there are a series of white papers describing the primary target types. For direct imaging, the primary targets are nearby M-dwarfs (Guyon et al. 2019); for searches for biosignatures, the targets are also primarily nearby M-dwarf (Lopez-Morales et al. 2019), and for high precision radial velocity observations the targets are primarily bright nearby FGKM stars (Ciardi et al. 2019).
- ARIEL: ARIEL targets were taken from the reference mission paper (Edwards et al. 2019). Note that this target list is incomplete as it will be revised heavily in the coming years as ARIEL approaches launch and operations.
- NEID/CPS/ESPRESSO/Carmenes/Maroon-X: Precision radial velocity targets were acquired through private communications to the individual groups except for the ESPRESSO GTO target list which is published on the ESO website.
    - See also: Target Prioritization and Observing Strategies for the NEID Earth Twin Survey - https://arxiv.org/abs/2101.11689

# Appendix G: Task Force 3 -- Interdisciplinary Use Cases Final Report

*Leads: Ravi Kopparapu (NASA GSFC) & Jacob Lustig-Yaeger (JHU APL)*

**Members:** Jake Lustig-Yaeger, Ravi Kopparapu, Tyler Richey-Yowell, Emily Safron, Carl Melis, Noah Tuchow, Shawn Domagal-Goldman, Chuanfei Dong, Vladimir Airapetian, Kim Smith, John Ahlers, Natalie Hinkel

## Introduction

The goal of the Exoplanet Study Analysis Group (SAG) 22 is to identify key information for a target star archive for exoplanet science which will be used to both plan exoplanet missions and interpret the data. Ideally, the proposed archive would serve a broad user base, from those designing future missions to those conducting observations and analyzing observations. In this Task Force, we have specifically focused on defining and categorizing the different interdisciplinary science use cases that we could expect such a future archive to serve. Meeting this goal, which includes specifying the observables/properties necessary for utilization of the archive by a broad range of disciplines, requires discussion with experts from a variety of disciplines to understand the major lines of related scientific inquiry. For example,

- *How would a geologist/(astro)biologist/planetary scientist/theorist/astronomer want to interact with the catalog and what data products would they want?*
- *Are there specific stellar or planetary system characteristics that would help prioritize targets for observation and/or lead to more robust individual-planet analysis?*

In order to answer these questions, we contacted hundreds of our colleagues from a variety of disciplines with the following set of questions:

1. What are your stellar or planetary science questions?
2. What stellar or planetary information will you need to interpret your data/models/output (including primary and secondary use case)?
3. What is ideal vs. threshold precision?
4. What wavelength range might these observations cover?
5. How have you used stellar databases/information in the past? What was missing?

We received a total of 66 responses. Based on the content of each response, we categorized each respondent into the single disciplinary groups, shown in Fig. 1. Since the responses were long form (including many meandering narratives), we processed some of the answers to create more easily parsable and uniform short form phrases. For example, we converted “chemical species in the atmospheres of planets would be a great start” to “atmospheric composition.” In addition, where the primary science case relied on derived quantities, we added the stellar or planetary observables necessary to perform the implicit inference.

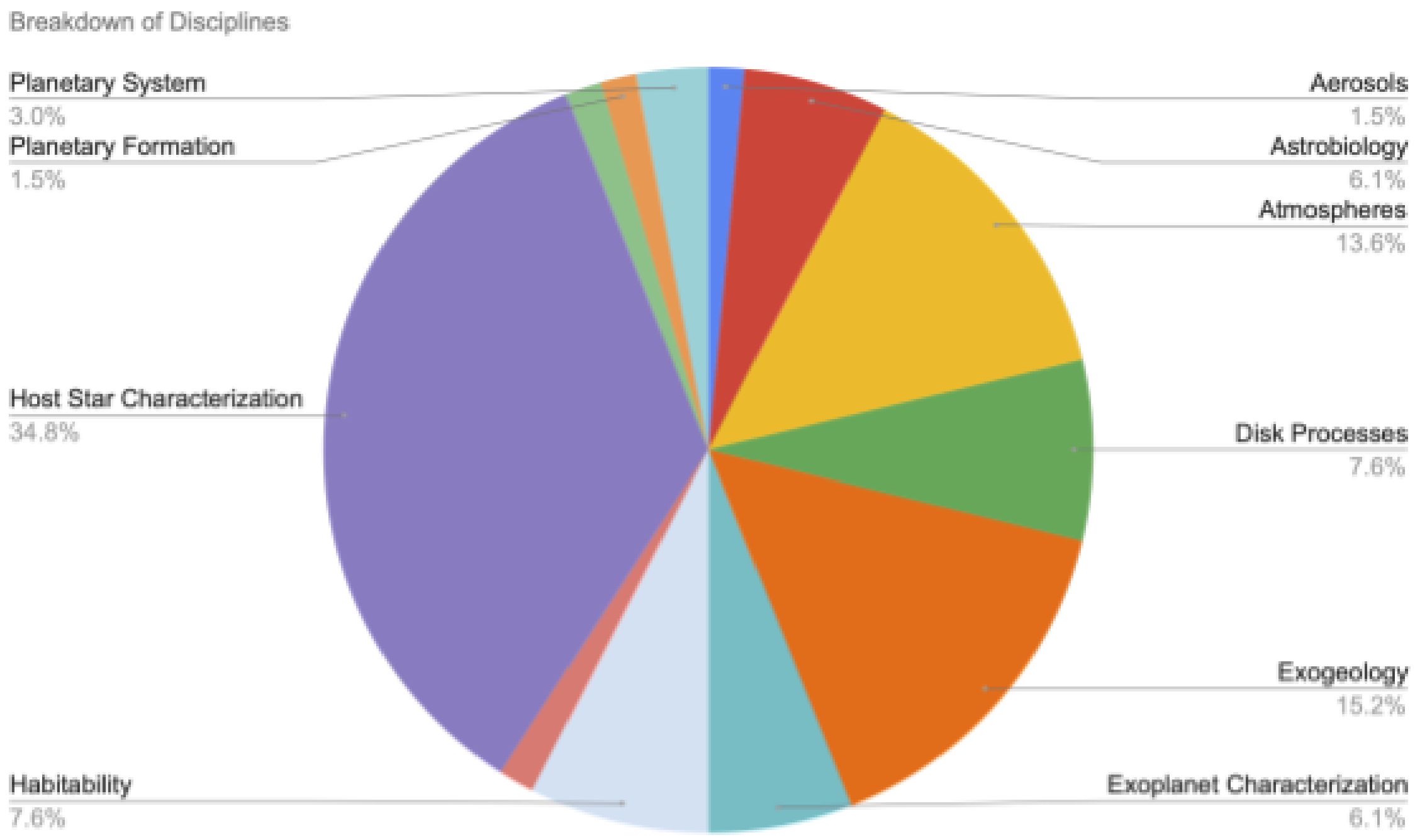

Fig 1: Pie chart showing the breakdown of the different discipline responses as labeled by Task Force 3 (i.e. labels were not assigned by those who responded to the set of questions).

## Findings

In this section, we provide a stove-piped summary of our findings in terms of the following sub-disciplines: host stars, planetary system, exogeology, and habitability and astrobiology. Then we take a step back and examine the interdisciplinary overlap.

### 1. Sub-Disciplines

#### A. Host Stars

Among the different applications for a catalog of exoplanet host star properties, perhaps the most straightforward case is for the study of the stars themselves. In the study of stellar astrophysics, many areas of research could greatly benefit from more precise measurements of a variety of stellar properties. Many of these applications directly flow into the study of exoplanets around these stars, since in most cases, having a better knowledge of a host star increases one's knowledge of the planets around it.

One such science case is understanding the interior structure and evolution of stars. Among the most important stellar properties inferred from such studies are the mass and age of a star, which can be inferred via fitting stellar models to observed properties such as the luminosity, effective temperature, surface gravity, and composition/metallicity of a given star. Measurements of a given star's mass, age, and metallicity allow one to reconstruct the past evolution of the star in terms of, for example, temperature and luminosity. Respondents reported the need for precisions to be better than ~20% for stellar mass and ~10% for stellar age. Knowledge of a star's luminosity and temperature evolution informs how its habitable zone changes over time, which is important for the study of the long term habitability of planets. For studying the UV evolution of stars, in addition to masses, ages, and effective temperatures, stellar rotation periods and EUV models would be useful. Stellar rotation periods (ideally within ~10%) can also be used to infer the age of a target star and serve as a model independent constraint on stellar ages. Measurements of asteroseismic modes of oscillation can serve to provide another set of constraints on model fitting for masses and ages, and can be used to infer the interior structure of a star, such as its convective zone radius. Additionally, measurements of stellar oblateness along with radii, luminosities, and surface gravities would be useful for the study of stellar gravity darkening.

Another major family of use cases deal with the study of stellar activity, variability, and winds. For studies of stellar activity, it would be useful to have high resolution (R > ~30,000) optical spectra to infer stellar metallicities, effective temperatures, and spectral types. The study of stellar granulation can be done with high resolution measurements of jitter from spectroscopy and flicker from photometry. Spectrophotometry measurements in UV and X-ray wavelengths allow for the study of a wide variety of areas, such as the radiation environment around a star, coronal heating, stellar activity, and activity cycles. To study the magnetic activity around other stars, one would like to have measurements of magnetic field strengths and topologies. To obtain magnetic field topologies would require spectropolarimetry measurements, such as those from Zeeman Doppler Imaging. Stellar flaring is another area of study, and one would like to obtain measurements of stellar flare rates, the spectra of flares, and their energy distribution. The study of stellar winds would benefit from measurements of rotation and magnetic field strengths along with direct inferences of stellar winds. Furthermore, we would like to know how EUV radiation and coronal mass ejections (CMEs) affect habitable zone planets and how this changes as a function of stellar mass and age. CMEs are not yet directly measurable around other stars, and are instead inferred from relations for flares and CMEs from the Sun. Indirectly constraining CME rates and properties will require short-cadence lightcurves in multiple bandpasses and possibly associated spectroscopic coverage.

Access to additional properties of exoplanet host stars would help advance numerous avenues of stellar and planetary science. Proxies for stellar age (other than rotation periods) include magnetic activity indicators (flare rates, H alpha emission, Ca H & K lines, UV and X-ray fluxes), Li abundances, and 3D space velocities. For understanding exoplanet host star formation and dynamics and evolution of systems, including multiple star systems, high-resolution imaging in the optical and IR as well as astrometric measurements such as those from *Gaia* are required. Planet detection around binary stars requires spectroscopic and photometric observations of the host stars, including precision radial velocity measurements and high-contrast imaging and archival time-series photometry. Stellar elemental abundances are correlated with the availability of those elements on the surface and interior of a planet. Therefore, it is necessary to have measurements of elements important to planetary and biological formation, especially those bio-essential elements, such as N, F, P, Cl, and K, that are not well or often measured in stars.

### B. Planetary System

Debris disk characterization can inform us about planetary system architectures, volatile delivery and eventual (interior and atmospheric) composition of a planet. However, all of these depend upon the host star characteristics such as stellar spectrum, variability, mass, luminosity and chemical abundances. Observations of debris disks in the UV to far-IR, at distances ranging from 0.5-50 AU from the host star, could help in cataloging sources that could provide a relation between the general planet formation process and host star characteristics. Furthermore, with existing catalogs, cross-referencing sources between catalogs is needed so that a best estimate of stellar host parameters can be obtained. Stellar age and chemical abundances are important for disk fractionation processes. Therefore, a ~10% precision in both optical disk photometry and stellar age is needed, with a ~5% precision in chemical abundances (Mg/Si, C/O).

Exoplanet atmospheric characterization can provide information about the bulk and trace molecular species in an atmosphere. For rocky exoplanets, it can give crucial data on the potential habitability of the planet. Currently, transmission spectra and reflection/emission spectra are the two most widely used methods for exoplanet atmospheric characterization. However, transmission spectra contamination from stellar activity and spot coverage can potentially mask or mimic planetary signals and the corresponding molecular features from their atmospheres. Obtaining information about stellar spectra (~10% precision) from UV-NIR (0.2 - 25μm), spot temperatures (+/- 100K precision), and stellar masses (~20%) could be useful in accurately characterizing exoplanet atmospheres.

Additional observations of low-mass stars (which are one of the more active types of stars) in radio (< 100 MHz) with a precision of ~10mJy may provide information about the strength of the host star's magnetic fields, which in turn can tell us about the magnitude of the stellar activity that could affect atmospheric retention of a planet. While stellar activity may be detrimental to retaining a planet's atmosphere, a protective planetary magnetic field may potentially limit the atmospheric escape. However, it's not clear whether the net effect of planetary magnetic fields will encourage planetary atmospheric loss or retention.

Identifying the presence and estimating the mass of planetesimal belts that may orbit with planets in a system would enable us to assess the interactions with the planetesimal belt which could drive planetary migration (e.g., the Nice model for the solar system) and/or planetesimal delivery to inner planetary systems (which might be important for volatile delivery). The inner and outer radii of debris disks with a precision of ~10% could provide constraints on the extent of planetesimal belts.

### C. Exogeology

The fields of planetary and geological sciences are strongly informed by characteristics that we are able to measure for planets (and other bodies) in the Solar System. Within the context of exoplanetary systems, the scientific topics of these communities typically require high-precision measurements of the mass and composition of rocky planets, along with assorted stellar and greater planetary system characterization.

Planetary masses are desired at the <10% level generally, ideally approaching a few percent in precision and accuracy. This in turn requires tightly constrained orbital parameters, for which values like semi-major axis/period/eccentricity are needed at the level of a few percent for secondary science topics like planetary system architectures. Completing planetary system architectures necessitates the identification and characterization of planetesimal belts including their mass and radial extent. Combining all of these measurements for a planetary system—and the host star mass (at the level of 1 to a few %) and age (to within 10%)—sets the stage for a current epoch dynamical assessment of the system and possibly a look at its dynamical history. This information could inform planet migration episodes or volatile delivery to the inner planetary system.

Planetary composition is critical in assessing the structure and habitability of a rocky exoplanet. This includes important processes like outgassing of secondary atmospheres, plate tectonics, and activity that might produce volcanism, as well as heat flux through a planet that might maintain these processes and core magnetic field generation. To inform on these topics for exoplanetary systems, where *in situ* measurements are not possible, it is essential to have highly precise planetary radii (at the 1% level) and stellar abundance measurements (<10%). Major constituents of rocky material (Fe, Mg, Si, O) are especially in need of high precision, while for some trace elements and biologically important species (e.g. CHNOPS) any measurement is of value, and precisions at the level of 50% may be sufficient. The ideal is of course attaining precision and accuracy comparable to that for the Sun.

Significant work remains to be done on using stellar abundance measurements to estimate or constrain rocky planet interior compositions and structure. Verification for models developed in this regard are of critical importance. In addition, the use of polluted white dwarf stars, which have accreted rocky planetary system material that are observable in their spectra, to better understand the composition of rocky planets should be extensively explored. The assessment of planetary atmospheric composition and physical state for planets is necessary for habitability studies, and need to be combined with planet surface composition constraints and stellar input. Stellar effective temperature needs to be constrained to the few percent levels, while robust measurements or estimates for spectral energy distributions from the host star (and associated variability) are needed at the 5% level or better.

The measurements of interest to inform the above science topics typically are collected within the UV, optical, or IR wavelength ranges. Sub-millimeter/millimeter wavelength observations are required for planetesimal belt characterization.

### D. Habitability and Astrobiology

There are no single observables that, on their own, meet the science requirements to assess the habitability of and presence of life on exoplanets. These science cases are broad and interdisciplinary in scope, and will ultimately culminate at the intersection of exhaustive astronomical observations and cutting-edge theoretical modeling efforts.

Community needs for the study of exoplanet habitability and astrobiology tend to revolve around the composition of target exoplanet atmospheres and surfaces as gleaned through observed planetary spectra, and as interpreted within the full context of the stellar environment. To answer their top-level science questions, most respondents with science cases linked to constraining planetary habitability and astrobiology/biosignatures report the need for detailed information about exoplanet atmospheric compositions and surface conditions. Furthermore, constraining the composition of exoplanet atmospheres will fundamentally require spectroscopic observations of planets in order to make such inferences using atmospheric retrieval models. Atmospheric retrieval experts report a need for more precise stellar flux measurements (to better than 10% in the NIR and MIR). In this way, the study of exoplanet habitability and biosignatures flows directly from exoplanet atmospheric characterization, which itself flows directly from the spectroscopy of exoplanets and their host stars.

Respondents interested in spectroscopic observations of exoplanets express a strong preference towards IR (IR/NIR/MIR) and optical wavelengths, while UV was typically mentioned in reference to the need for better stellar data with which to constrain photochemical models and aerosol production in exoplanet atmospheres, as well as a need for UV laboratory measurements of gas and aerosol optical properties. That is, UV measurements were often requested to help provide crucial inputs (stellar and atmospheric) required to perform exoplanetary atmospheric modeling.

Science cases investigating exoplanet habitability tended to be uniquely interested in atmospheric and surface conditions, such as the surface temperature (+/-30 K precision), pressure, albedo, composition, and ultimately the presence of water. These habitability indicators are required to provide converging lines of evidence needed to report a planet as habitable. Astrobiology themed science cases involving biosignatures and their detection all relied upon some combination of surface/atmospheric compositional information, such as characterizing gaseous and surface biosignatures, how much H2 can be generated through Fe photo-oxidation on anoxic planets with oceans, and observing properties of organic aerosols and clouds. These biosignatures must not only be uniquely identified on the planet, but also interpreted within the context of the planetary environment to assess whether or not they are biogenic in origin.

### 2. Interdisciplinary Overlap

While the earlier sections discussed the data product needs of different sub-disciplines, common themes can be identified that would be useful for an interdisciplinary investigation; i.e, information/data collected from one discipline could flow into another discipline. **The most common data products that all disciplines need seem to be stellar effective temperature and luminosity, planetary mass and radius, with spectroscopy (either stellar or planetary) as a major priority (see Fig. 2).**

This is not entirely surprising considering that fundamental stellar and planetary is crucial for studying planetary system properties such as disk formation, volatile/ice-line location, atmospheric retention, and habitability. These topics span all the disciplines discussed above indicating a common need. Obtaining information about stellar spectra (~10% precision) from UV-NIR (0.2 - 25mu). UV and X-ray wavelengths allow for the study of a wide variety of areas, such as the radiation environment around a star, stellar activity, and activity cycles that could impact a planet’s atmospheric composition and habitability.

Stellar and planetary masses (~10% precision), in addition to stellar (or system’s) age (< ~10%) have applications across disciplines from mapping dynamical histories of system architectures to evolutionary histories of interactions between stars and planets.

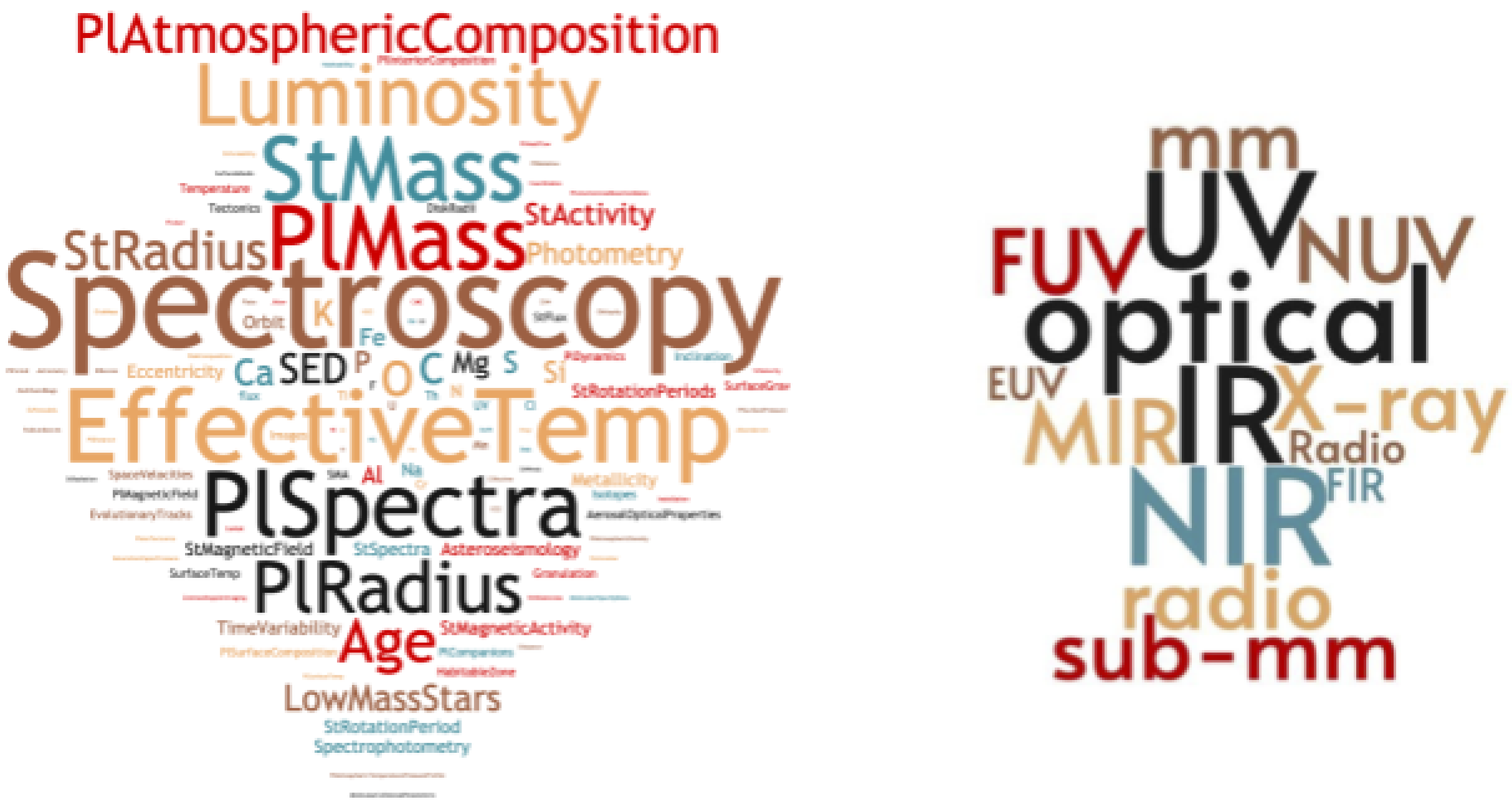

Fig 2: Word clouds showing the properties most important to all disciplines (left) as well as the wavelength regimes (right). Note that "Pl" is short for "planet" and "St" is short for "star" as a way to qualify the properties without impacting the results of the word cloud.

The properties most important to the individual sub-disciplines are given in Fig. 3 (left column), with the applicable wavelength regimes necessary for the science in the right column. The most noticeable difference between the properties important to the four sub-disciplines is the variation -- namely, characterizing host stars and planetary systems require a variety of properties, while exogeology, habitability, and astrobiology are fairly limited to the physical properties of the planet (mass, radius, and spectra) as well as the chemical composition (as typically inferred from the stellar abundances). In addition, the study of habitability and astrobiology requires a much more diverse set of characteristics, as noted by the presence of a large number of properties with smaller fonts. Also, while it may be useful for the host star and planetary system communities to observe their targets across a wide range of wavelengths, exogeology and habitability/astrobiology are preferentially interested in the UV, optical, and IR bandpasses. This is caveated by realizing responses may have been catered or biased towards specific upcoming instruments/telescopes, such as HabEx or LUVOIR. Finally, we note that there are properties in every sub-disciplines that are standalone because they don't quite fit; however, this really illuminates the point that exoplanet science is very interdisciplinary and that these sub-disciplines "silos" don't work across the board.

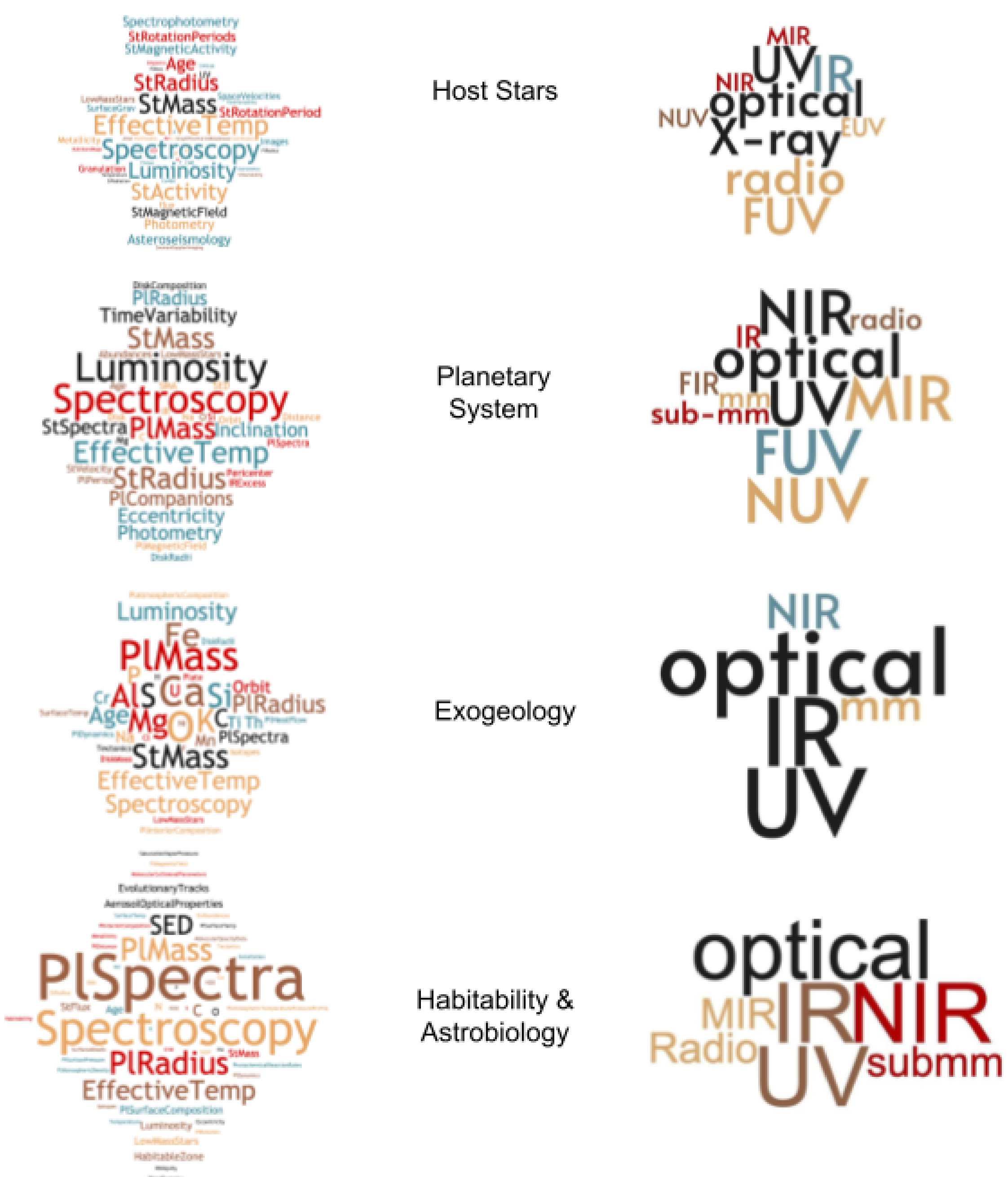

Fig 3: Similar to Fig 2, word clouds for individual discipline properties (left) and wavelengths (right), as listed in the center column.

## Conclusion

As mentioned in the introduction, within our set of questions we asked interdisciplinary scientists to identify what was missing within currently available databases or archives. Fig 4 shows a word cloud of their abbreviated responses. The most common concern across the sub-disciplines was that it was difficult to cross-correlate target data from different resources, whether active databases or telescope archives. There are also a variety of properties that are necessary for exoplanet science but are lacking in current databases, either because they have not been amalgamated or because they do not exist: spectroscopy of atmospheric gases (especially high resolution), broadband spectroscopy, stellar abundances for elements important to planet formation, high precision UV data, and information on stellar activity.

Members of the planetary and geological sciences communities have by and large used astrophysical databases in the course of their exoplanet-related research. In their experience, they find that stellar abundances for the major components of planets are not always available and that a list of habitable zone planets would be helpful. Easily accessible cross-correlated data for potential exoplanet targets and any associated planet and star data for each system would be of value.

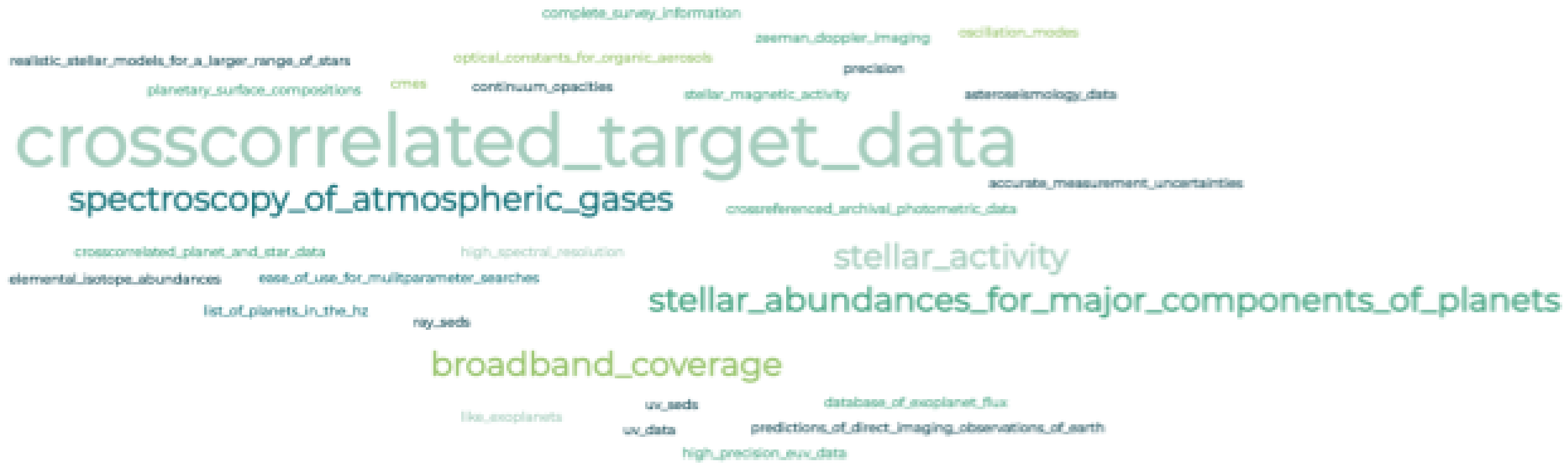

Fig 4: Word cloud of properties missing from current databases.

Many people expressed concern about how to compare stellar and planetary data, such as determining variations between different groups or methodologies and how to filter out information that may be of low(er) quality or not useful for certain analyses. It is also unclear how to compare uncertainties in measurements that utilized different approaches. Instead, it was stated that summaries of the likelihood and/or the posterior distribution computed with a well-chosen interim prior with broad support would be useful for determining true uncertainties. Finally, complete survey information, i.e. about the entire sample surveyed, as well as what observations were taken for each object and how it was decided to take those observations (namely target of interest vs homogenous survey), would help understand biases within the data.

Overall, the comments made by members of multiple interdisciplinary fields made it clear that a comprehensive target star archive would be highly valued within the community. In addition, there are a number of gaps in our current understanding of stellar and planetary systems that should be addressed as quickly as possible, in order to capitalize on upcoming mission data and more fully characterize confirmed exoplanets. Finally, ease of use and parameter searching, as well as cross-matching, are important for database usefulness such that scientists can access information in an intuitive manner.

## Appendix H: Task Force 4 -- Existing Catalogs Final Report

*Leads: Jennifer Burt (JPL) & Kevin Hardegree-Ullman (U. Arizona)*

**Members:** Jennifer Burt, Kevin Hardegree-Ullman, Galen Bergsten, Angelle Tanner, Srija Chakraborty, Jacob Luhn, Josh Pepper, Dave Latham, Sabine Reffert

### Outline of charge from SAG

Task force four was charged with providing a list of existing catalogs of stellar properties, with a breakdown of the catalog contents, including estimates of the range of values of each parameter and the typical precision. It was requested that we note which catalogs are publicly available through open online interfaces, how often they are updated, and indicate the agency, institution, or person responsible for maintaining the catalog with contact information.

### Catalog Selection Criteria

The list of stellar catalogs was compiled by task force members based on their knowledge of the field. Catalogs were included if stellar parameters were derived using observational data and existing pipelines. These catalogs were required to be publicly available and the data have gone through the referee process.

A majority of the included catalogs are based on spectroscopic surveys because they generally provide a large set of stellar parameters. We included broadband photometric catalogs separately, but note that these catalogs can be incomplete for some stars of interest to future NASA missions because they are often incomplete for very bright stars, and in some cases also incomplete for high proper motion, nearby late-type stars. Also, while stellar models are important for pipelines to compute stellar parameters, we viewed the compilation of such models as out of scope for our work.

We made a limited effort to assemble a list of compiled catalogs of stellar parameters. These catalogs typically contain data from numerous sources which may have been inhomogeneously derived and contain a wide range of parameter quality. Observing archives (such as NASA's Exoplanet Archive and exoplanet.eu) typically identify a "top" set of stellar parameters from the literature for their exoplanet catalogs. Other catalogs like the TIC are built for broad reliability across the whole sky, rather than highly-reliable information for a small number of top targets. Some catalogs like the Hypatia Catalog and the Starchive are highly curated, but draw their contents from a highly heterogeneous set of primary sources, and are not complete for all stars of exoplanet characterization interest.

**Description of Existing Catalogs Table**
The existing catalogs table includes information which task force four members decided would be most useful to the community. The following is a list of columns:

| | | |
|---|---|---|
| Catalog name | Basic catalog description | Description of stellar parameters |
| Wavelength range | Number of stars | Observational method |
| Magnitude range | Spectral types | SNR parameters |
| Sky coverage | Link to catalog | Link to seminal paper |
| Spectral Resolution | Chemical abundances | Year of most recent update |
| Interface to data | Link to info on how to query catalog | |

The information provided by task force four is mostly complete, but due to the differing nature of how stellar catalogs are compiled and delivered some information might not be easily accessible or available.

**Auxiliary tabs for other catalogs**
We have additionally provided lists of broadband photometry catalogs, compiled catalogs (as described above), and specialized subset catalogs in separate spreadsheet tabs. The specialized subsets label refers to catalogs that are derived from a specific subset of targets from a larger catalog (Kepler, LAMOST, etc) that have their stellar parameters derived in a homogeneous way. All additional tabs include less detail than the main ‘existing catalogs’ tab, but provide a useful starting point for future efforts to assemble detailed lists in these areas.

**Notes on any resources/data/etc that we found to be missing**
While compiling existing catalogs for this compendium, we noted certain target types and/or stellar parameters that were not readily available in existing catalogs. We make note of them here and suggest that future efforts focus on identifying existing resources or funding the observations/data reduction necessary to fill these gaps.

- Types of data:
    - UV data and corresponding stellar parameters and activity indicators
    - Volume limited samples of stars -- new Gaia DR3 addresses this (volume limited out to 100pc) but Gaia is incomplete for very bright, very faint, and late-type nearby stars. Future efforts should try to assess gaps left by Gaia.
    - Exozodi measurements

- Presence of circumstellar disks
    - Broadband photometry for very bright (V<5) nearby stars
    - Broadband photometry for nearby, late type M dwarf stars
    - Spectral types (as distinct from Teff and general luminosity class) are rarely compiled. The value of this information for future exoplanet missions is unclear.
- Types of data products/measurements:
    - Stellar activity metrics (R'HK, MWS, H-alpha, etc) -- lack of comprehensive, ongoing spectroscopic time series that allow for tracking of both rotation modulation and magnetic activity cycles. There are some photometric facilities doing this (TESS, EVRYSCOPE) that can be used to get at rotation signals and flares. Should look into the exact level of spot coverage that is likely to show up in photometric monitoring.
    - Stellar inclinations obtained by combining vsini measurements with photometric rotation periods.
    - Planetary inclination -- for non-transiting planets this is necessary to calculate actual masses and not minimum masses using EPRV observations. Gaia astrometric orbits will be available for a subset of known planets.
    - Stellar ages -- PLATO supposed to get asteroseismological ages for 100,000s of stars, so if mission proceeds on time (launch ~2026) then this could help solve the lack of age information
    - Stellar associations and cluster membership -- Many individual papers, but no modern catalog compilation. Existing WEBDA website is heterogeneous and reliability is unclear. Sky coverage is also not well characterized.

**Notes on improved catalog utility**

During the catalog compilation effort we found that the most useful resources were those where the data access process is well documented. This is especially true for large data sets where access is not simply handled via querying vizier or downloading key tables. In these cases, we recommend that the authors provide tutorials that walk users through the data access/query/download process.